\pdfoutput=1
\documentclass[10pt]{article}
\usepackage[utf8]{inputenc}
\usepackage{cite}
\usepackage{hyperref}
\hypersetup{colorlinks=true,linkcolor=bluecyan,citecolor=orange3,urlcolor=green3,pdfencoding=auto,linktocpage}
\usepackage{amsmath,amssymb,amsbsy,amstext,amsthm,simplewick,amsfonts}
\usepackage{mathrsfs}
\usepackage{graphicx}
\usepackage{wrapfig}
\usepackage{upgreek}
\usepackage{bm} 
\usepackage{framed}
\usepackage{bbm}
\usepackage{textcomp}
\usepackage{adjustbox}
\usepackage{makecell}
\usepackage{tcolorbox}
\usepackage{empheq}
\usepackage[normalem]{ulem}
\usepackage{enumitem}
\usepackage{braket} 
\usepackage{array}
\usepackage{dsfont}
\usepackage{physics}
\usepackage{ulem}
\usepackage{xcolor}
\usepackage{caption}
\usepackage{subcaption}
\usepackage{tocloft}
\usepackage{tikz}

\usepackage[framemethod=default]{mdframed}

\newmdenv[backgroundcolor=gray!15,%
skipabove=5pt,%
skipbelow=5pt,%
leftmargin=2pt,%
rightmargin=2pt,%
innertopmargin=-6pt,%
innerbottommargin=5pt,%
innerleftmargin=5pt,%
innerrightmargin=5pt,%
splittopskip=0pt,%
splitbottomskip=0pt,%
linewidth=0pt,%
nobreak=true]%
{keyeqn}

\newmdenv[backgroundcolor=gray!15,%
skipabove=5pt,%
skipbelow=5pt,%
leftmargin=2pt,%
rightmargin=2pt,%
innertopmargin=-2pt,%
innerbottommargin=5pt,%
innerleftmargin=5pt,%
innerrightmargin=5pt,%
splittopskip=0pt,%
splitbottomskip=0pt,%
linewidth=0pt,%
nobreak=true]%
{keythrm}

\graphicspath{{Fig/}}

\definecolor{lightgreen}{cmyk}{0.2, 0, 0.2, 0.2}
\definecolor{lightgray}{cmyk}{0.1,0.2,0,0.1}
\definecolor{lightgray2}{cmyk}{0.1,0.1,0,0.1}
\definecolor{bluecyan}{RGB}{0, 100, 200}
\definecolor{blue3}{RGB}{31,119,180}
\definecolor{red3}{RGB}{214,39,40}
\definecolor{orange3}{RGB}{255,127,14}
\definecolor{green3}{RGB}{44,160,44}

\definecolor{red2}{RGB}{255,0,0}
\definecolor{green2}{RGB}{0,170,0}
\definecolor{blue2}{RGB}{0,128,255}
\definecolor{magenta2}{RGB}{191,64,191}
\definecolor{purple2}{RGB}{112,48,160}
\definecolor{orange2}{RGB}{255,192,0}

\newtheorem{theorem}{Theorem}[section]
\newtheorem{corollary}{Corollary}[section]

\def\bfk{\textbf{k}}
\def\bfx{\textbf{x}}
\def\Res{\mathrm{Res}\,}
\def\Re{\mathrm{Re}\,}
\def\Im{\mathrm{Im}\,}

\renewcommand{\star}{*}

\numberwithin{equation}{section}

\allowdisplaybreaks[1]
\begin{document}

\begin{titlepage}
	\setcounter{page}{1} \baselineskip=15.5pt 
	\thispagestyle{empty}

    \begin{center}
		{\fontsize{18}{18}\centering {\bf{Cosmological Correlators Through the Looking Glass:}} \\ \vspace{0.2cm} {\textit{Reality, Parity, and Factorisation}}\;}\\
	\end{center}
 
	\vskip 18pt
	\begin{center}
		\noindent
		{\fontsize{12}{18}\selectfont David Stefanyszyn\footnote{\tt david.stefanyszyn@nottingham.ac.uk}$^{,a}$, Xi Tong  \footnote{\tt xt246@cam.ac.uk}$^{,b,c}$ and Yuhang Zhu \footnote{\tt yhzhu@ibs.re.kr}$^{,d,c}$}
	\end{center}
	
	\begin{center}
		\vskip 8pt
		$a$ \textit{School of Mathematical Sciences \& School of Physics and Astronomy,
			University of Nottingham, University Park, Nottingham, NG7 2RD, UK} \\
        $b$ \textit{Department of Applied Mathematics and Theoretical Physics, University of Cambridge,\\Wilberforce Road, Cambridge, CB3 0WA, UK} \\
		$c$ \textit{Department of Physics, The Hong Kong University of Science and Technology,\\Clear Water Bay, Kowloon, Hong Kong, P.R. China} \\
        %
        $d$ \textit{Cosmology, Gravity and Astroparticle Physics Group,\\
        Center for Theoretical Physics of the Universe,\\
        Institute for Basic Science, Daejeon 34126, Korea}
	\end{center}
	
	
	\noindent\rule{\textwidth}{0.4pt}
	\noindent \textbf{Abstract} ~~ We consider the evolution of quantum fields during inflation, and show that the total-energy singularities appearing in the perturbative expansion of the late-time Wavefunction of the Universe are purely real when the external states are massless scalars and massless gravitons. Our proof relies on the tree-level approximation, Bunch-Davies initial conditions, and exact scale invariance (IR-convergence), but without any assumptions on invariance under de Sitter boosts. We consider all $n$-point functions and allow for the exchange of additional states of any mass and integer spin. Our proof makes use of a decomposition of the inflationary bulk-bulk propagator of massive spinning fields which preserves UV-convergence and ensures that the time-ordered contributions are purely real after we rotate to Euclidean time. We use this reality property to show that the \textit{maximally-connected} parts of wavefunction coefficients, from which total-energy singularities originate, are purely real. In a theory where all states are in the complementary series, this reality extends to the full wavefunction coefficient. We then use our reality theorem to show that parity-odd correlators (correlators that are mirror asymmetric) are \textit{factorised} and do not diverge when the total-energy is conserved. We pay special attention to the parity-odd four-point function (trispectrum) of inflationary curvature perturbations and use our reality/factorisation theorems to show that this observable is factorised into a product of cubic diagrams thereby enabling us to derive exact shapes. We present examples of couplings between the inflaton and massive spin-$1$ and spin-$2$ fields, with the parity-violation in the trispectrum driven by Chern-Simons corrections to the spinning field two-point function, or from parity-violating cubic interactions which we build within the Effective Field Theory of Inflation. In addition, we present a first-of-its-kind example of a parity-violating trispectrum, generated at tree-level, that arises in a purely scalar theory where the inflaton mixes linearly with an additional massive scalar field.
	
	\noindent\rule{\textwidth}{0.4pt}
	
	
\end{titlepage} 


\newpage
\setcounter{page}{2}
{
	\tableofcontents
}

\newpage


\section{Introduction}

The fundamental observables of inflationary cosmology are late-time cosmological correlators, namely expectation values of quantum fields evaluated at the end of inflation. In the simplest models of inflation we are interested in correlations between the two massless fields that survive the rapid expansion of the background spacetime: the Goldstone boson of broken time translations, and the graviton. These correlators provide the initial conditions for Hot Big Bang cosmology, and observations of the Cosmic Microwave Background and Large Scale Structure then allow us to in principle distinguish between different models of the very early universe by measuring these spatial correlations. Given the very high energies that could characterise inflation, additional massive particles can be produced from the vacuum and decay into the light states that make it to the end of inflation. Such heavy states leave distinctive imprints on late-time correlators that encode their masses (through time evolution) and spins (through kinematics), thereby leading to the tantalising prospect of using the early universe as a very energetic \textit{cosmological collider} \cite{Chen:2009zp,Baumann:2011nk,Noumi:2012vr,Arkani-Hamed:2015bza,Lee:2016vti}.

Traditionally, such correlators are computed in de Sitter space perturbation theory using the in-in/Schwinger-Keldysh formalism (see e.g. \cite{Chen:2010xka,Wang:2013zva,Chen:2017ryl} for reviews), or the Wavefunction of the Universe. Such computations quickly become very complicated due to the background time dependence which, in contrast to flat-space, means that the time integrals one must compute are very non-trivial. Further complications arise from the fact that the mode functions of massive fields in de Sitter space are usually characterised by Hankel functions (or similar functions) which are harder to integrate compared to the plane wave solutions which are familiar in flat-space. Such complications provide a stumbling block in ones quest to understand the late-time effects of massive fields during inflation, and more generally to understand the fundamental properties of cosmological correlators. 

This has motivated the \textit{cosmological bootstrap programme} which aims to develop new computational techniques from which one can compute cosmological correlators while avoiding the unobservable time evolution of quantum fluctuations and the associated complicated nested time integrals. The idea is to use general physical principles such as symmetries, locality and unitarity to directly fix the structure of boundary correlators. This programme is motivated by the highly successful S-matrix bootstrap programme where scattering amplitudes are computed while avoiding the complications of Feynman diagrams \cite{TASI,Benincasa:2007xk,Elvang:2013cua,NAH,Kruczenski:2022lot,Dixon:1996wi}. Much progress on understanding the structure of cosmological correlators has been made in recent years focusing on correlators in exact de Sitter space \cite{WFCtoCorrelators1, CosmoBootstrap1,CosmoBootstrap2,CosmoBootstrap3,Bzowski:2013sza}, correlators arising from interactions with broken de Sitter boosts \cite{BBBB,Green:2020ebl}, the role of analyticity and unitarity in these late-time observables \cite{COT,Melville:2021lst,Goodhew:2021oqg,DiPietro:2021sjt,Baumann:2021fxj,Albayrak:2023hie,Salcedo:2022aal,Cespedes:2020xqq,Tong:2021wai,Meltzer:2021zin,Meltzer:2020qbr}, constraints from bulk locality \cite{MLT}, the effects of additional degrees of freedom during inflation \cite{Liu:2019fag,Jazayeri:2022kjy,Pimentel:2022fsc,Wang:2022eop,Cabass:2022rhr,Tong:2022cdz,Qin:2023ejc,Jazayeri:2023xcj,Kumar:2019ebj,Werth:2023pfl}, loop effects \cite{Lee:2023jby,Cohen:2020php,Xianyu:2022jwk,Gorbenko:2019rza,Agui-Salcedo:2023wlq,Wang:2021qez,Baumgart:2019clc,Qin:2023bjk,Qin:2023nhv}, graviton correlation functions \cite{Cabass:2021fnw,Bonifacio:2022vwa,Cabass:2022jda,WFCtoCorrelators2}, double-copy structures \cite{Armstrong:2023phb,Mei:2023jkb,Lee:2022fgr}, non-linear realisations \cite{Armstrong:2022vgl,Hui:2022dnm,Bittermann:2022nfh,Bonifacio:2021mrf}, scattering equations \cite{Armstrong:2022csc,Gomez:2021qfd}, the connections between scattering amplitudes and correlators \cite{Maldacena:2011nz,Raju:2012zr,Bonifacio:2021azc,GJS,PSS}, Mellin space representations \cite{Sleight:2019hfp,Sleight:2019mgd,Sleight:2021plv,Qin:2022lva,Qin:2022fbv,Qin:2023nhv,Qin:2023bjk}, differential representations \cite{Hillman:2021bnk,De:2023xue}, non-perturbative effects \cite{Hogervorst:2021uvp,Penedones:2023uqc}, relations to geometry \cite{Arkani-Hamed:2017fdk,Arkani-Hamed:2018bjr,Benincasa:2019vqr,Hillman:2019wgh}, and with much more fun to be had in the coming years. For reviews see \cite{Baumann:2022jpr,Benincasa:2022omn,Benincasa:2022gtd}. 

In this work we focus on the Wavefunction of the Universe and the associated wavefunction coefficients which contain the dynamical information about the evolution of quantum fields in de Sitter space. It is now well-known that for fields that satisfy Bunch-Davies initial conditions, i.e. for fields that have Minkowski-like behaviour in the far past, wavefunction coefficients have a restricted set of singularities. Indeed, wavefunction coefficients are only singular when the total-energy of the external states vanishes, or when the total-energy entering a sub-graph vanishes.\footnote{As usual in cosmology here we use the term ``energy" to describe the magnitude of a spatial momentum vector, even though strictly speaking there is no notion of energy in cosmology given that time translations are broken.} For physical configurations i.e. for real momenta, such singularities cannot be reached, however much of our understanding of wavefunction coefficients and cosmological correlators stems from analytically continuing away from real momenta in which case energies can become negative and singularities can be probed. For example, the leading total-energy singularity of $n$-point functions allows us to probe the corresponding flat-space (boost-breaking) scattering amplitude \cite{Maldacena:2011nz,Raju:2012zr,COT,PSS} which provides a non-trivial link between the cosmological bootstrap and the S-matrix bootstrap, while at four-points the additional singularities are partial-energy ones on which wavefunction coefficients factorise into a product of a lower-point wavefunction coefficient and a scattering amplitude, see e.g. \cite{CosmoBootstrap3}. When wavefunction coefficients are rational, which is often the case for massless scalars and gravitons \cite{Goodhew:2022ayb}, the residues of these partial-energy poles can be fixed by unitarity and energy shifts \cite{MLT,Baumann:2021fxj}.

In general, wavefunction coefficients in momentum space are complex functions of the external kinematics and cosmological correlators correspond to taking the real or imaginary parts, at least at tree-level. \textit{For parity-even interactions it is the real part that contributes to correlators, while for parity-odd interactions it is the imaginary part} \cite{Cabass:2022rhr}. While it is known that the tree-level wavefunction coefficients in simple massless theories are purely real \cite{Liu:2019fag} (thereby making parity-odd correlators a neat probe of exotic inflationary physics), in this paper we greatly extend this concept of reality. More precisely, we show that tree-level wavefunction coefficients with massless scalar external states have purely real total-energy poles (both leading and sub-leading) as long as the bulk interactions are IR-convergent. Our proof is valid for all $n$-point functions at tree-level and we allow for Feynman diagrams corresponding to the exchange of massive spinning fields of any mass and integer spin. More concretely, we show that contributions to wavefunction coefficients from the maximally-connected parts of Feynman diagrams are purely real, where ``maximally-connected" corresponds to the contributions to the integrands with the maximal number of $\theta$-functions (and this is where total-energy singularities come from). Our proof makes use of Wick rotations of the time integrals that compute these wavefunction coefficients and a simple property of the bulk-bulk propagator of massive fields in de Sitter space: the time-ordered part is purely real after the time variables are both Wick rotated by $90^{\circ}$ in the complex plane. This property follows as a simple consequence of the differential equation that the bulk-bulk propagator must satisfy given that it is a Green's function. We provide more explicit and detailed proofs within two different set-ups for describing massive spinning fields during inflation: \textit{cosmological condensed matter physics} (CCM) and \textit{cosmological collider physics} (CC). 

The former was introduced in \cite{Bordin:2018pca} and requires a sizeable coupling between the new massive degrees of freedom and the time-dependent inflaton. In this set-up fields are classified with respect to how they transform under the unbroken group of symmetries which for cosmology is spatial rotations. The theory should in addition have all of the symmetries of the Effective Field Theory of Inflation (EFToI) \cite{Cheung:2007st} and this can be guaranteed thanks to particular couplings with the inflaton which can be built straightforwardly using the building blocks of the EFToI \cite{Bordin:2018pca}. From the point of view of spontaneous symmetry breaking and the coset construction, such new degrees of freedom are classified as matter fields which can couple to the Goldstone boson of broken time translations. The masses of these fields are not restricted by the Higuchi bound \cite{Higuchi:1986py}, which illustrates that they cannot exist in an exactly de Sitter invariant theory. This set-up is somewhat similar to condensed matter systems where linearly realised Lorentz boosts are not used to construct the effective theory (and hence the name). The latter is perhaps more familiar to the reader and corresponds to describing massive degrees of freedom as representations of the de Sitter group. The masses of these fields must adhere to a lower bound in order for the theory to remain unitary, which is the Higuchi bound \cite{Higuchi:1986py}. The Lagrangians for such fields are known \cite{spinslagrangian1,spinslagrangian2,Zinoviev:2001dt}, but quickly become very complicated due to the need to include auxiliary fields (which ultimately enforce the transverse and traceless conditions on the fields as required by the degrees of freedom counting). One can instead work directly with the equations of motion which take a simpler form \cite{Deser:2003gw}. Although the free theories for these new degrees of freedom are de Sitter invariant, de Sitter boosts can be broken when we come to couple these fields to the inflaton. This can again be done within the language of the EFToI as done in \cite{Lee:2016vti}. We will review both set-ups in Section \ref{sec:massivefields}.

In each case we consider light and heavy fields i.e. those in the complementary and principle series, respectively. We concentrate on the reality properties of the Wick-rotated bulk-bulk propagators which, in cosmology, are composed of the usual time-ordered Feynman propagator and a factorised term necessitated by the boundary conditions. For light fields, we show that the full bulk-bulk propagator is purely real after Wick rotation. For heavy fields, we show that although the full bulk-bulk propagator is complex in general, we are able to add and subtract factorised contributions to the full bulk-bulk propagator in such a way that we cancel the imaginary parts of the Wick-rotated Feynman propagator. The new time-ordered part, which enjoys reality after rotation, and which is now a sum of the Feynman propagator and a factorised contribution, is referred to as the \textit{connected propagator}. This decomposition is diagrammatically represented in \eqref{DecompositionFigureIntro}. In addition, this connected propagator enjoys the crucial property that it vanishes in the far past, which ensures that Feynman diagrams constructed using this propagator maintain UV convergence of the associated time integrals. The detailed treatment of the propagator realities differs from one set-up to another. In the CCM set-up, we allow for parity violation in the free theory of the massive spinning field coming from a chemical potential term with a single spatial derivative in the action \cite{Wang:2019gbi,Wang:2020ioa,Sou:2021juh,Tong:2022cdz}. This splits the helicities and changes the mode functions from Hankel functions to Whittaker functions. The propagator realities then require cancellations once we sum over the helicities. In the limit of a vanishing chemical potential with a parity-conserving propagator, reality holds for each helicity mode separately with the proof for light fields already appearing in \cite{Cabass:2022rhr}. In the CC set-up we maintain parity in the free theory of the massive spinning field (except for spin-1 which is essentially identical to the CCM case) and again show that for light fields the Wick-rotated propagator is purely real, for each helicity mode, while for heavy fields it is again only the connected part that is purely real. In order to arrive at this conclusion we use the fact that the transverse and traceless conditions for these fields relate modes with the same helicity via differential operators that have simple properties under Wick rotation. These proofs make up Section \ref{PropagatorPropertySection}.

In Section \ref{RandFSection} we then use these general properties of Wick-rotated bulk-bulk propagators to prove that total-energy poles of wavefunction coefficients with external massless scalars are purely real under the assumptions of IR-convergence, scale invariance and the tree-level approximation. Our proof does not rely on de Sitter boosts and is therefore directly applicable to inflationary correlators. We also point out that our proof applies to external gravitons, and to wavefunction coefficients with an even number of conformally coupled scalars. Our approach is to extract the connected part of the full bulk-bulk propagator and show that diagrams that only involve connected propagators (maximally-connected diagrams) are purely real, and indeed total-energy singularities come from such diagrams. This is a general result, but if we restrict ourselves to the exchange of light fields only, then the full wavefunction coefficient is real. In Appendix \ref{CuttingComparison}, we offer a complementary proof of the reality of total-energy poles using the \textit{Hermitian analyticity} properties of the external bulk-boundary propagators and the internal bulk-bulk ones. This property combined with exact scale invariance allows us to draw the same conclusions we arrived at using Wick rotations. Hermitian analyticity of bulk-bulk propagators in the context of the CCM scenario has been established in \cite{Goodhew:2021oqg}, and in Appendix \ref{CuttingComparison} we extend the analysis to the CC scenario.

\begin{figure}[ht!]
\centering
\includegraphics[width=0.48\textwidth]{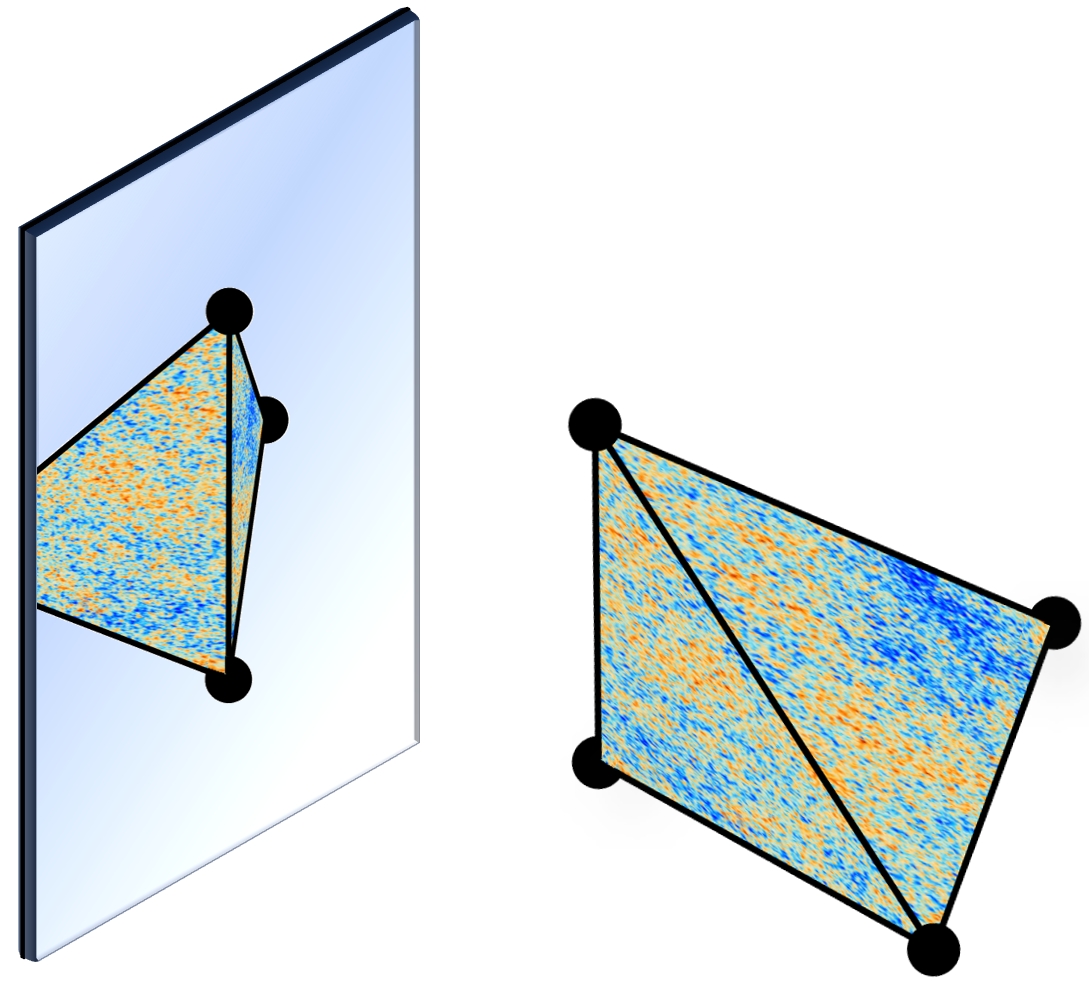}\\
    \caption{Cosmological correlators through the looking glass.}\label{LookingGlass}
\end{figure}

As a proof of the usefulness of this observation, we consider the parity-odd scalar trispectrum (see Figure \ref{LookingGlass} for a cartoon illustration) in Section \ref{ExactTrispectraSection}, which has recently gained some attention \cite{Philcox:2022hkh,Hou:2022wfj,Cabass:2022oap,Niu:2022fki,Creque-Sarbinowski:2023wmb,Cabass:2022rhr,Philcox:2023ypl,Coulton:2023oug,Jazayeri:2023kji}. This correlator is fixed by the imaginary parts of wavefunction coefficients and given that total-energy poles are real, they do not contribute to the parity-odd trispectrum (unless there is some IR divergence as in \cite{Creque-Sarbinowski:2023wmb}, or if they come from loop diagrams as in \cite{Lee:2023jby}). This implies that this observable is in fact factorised at tree-level, and can only have partial-energy poles. In computing cosmological correlators the primary difficulties arise when one computes the nested time integrals coming from the time-ordered propagators. Since here these contributions are purely real, computing the parity-odd trispectrum due to the exchange of massive spinning fields reduces to computing lower-order, factorised time integrals which have known closed-form solutions. We present a number of \textit{exact} parity-odd trispectra for both the CCM and CC descriptions of massive spin-$S$ fields focusing primarily on $S\leq 2$. We consider different sources of parity-violation: parity-violating bulk interactions and parity violation in the free theory describing the massive field. The latter case is usually studied in the context of cosmological chemical potentials, where it is known that the chemical potential can assist particle production \cite{Adshead:2015kza,Sou:2021juh} and boost the cosmological collider signal \cite{Chen:2018xck,Hook:2019zxa,Kumar:2019ebj,Wang:2019gbi,Hook:2019vcn,Wang:2020ioa,Tong:2022cdz,Chen:2023txq,Tong:2023krn}. In all cases we also allow for a general speed of sound for both the inflaton and the exchanged field which can also boost the signal if the inflaton moves more slowly \cite{Jazayeri:2022kjy}. We consider both light and heavy fields, and show that the final correlators for these two cases can be converted into each other by analytic continuation. We pay special attention to the spin-$1$ case with a parity-violating propagator (this corresponds to taking the Proca action and adding a Chern-Simons term that mixes the massive spin-$1$ field with the inflaton), and compare our exact result for the corresponding parity-odd trispectrum with that recently derived in \cite{Jazayeri:2023kji} using a non-local EFT approach. We find that the result from the non-local EFT provides a good approximation to our exact result for small chemical potential, but deviations start to appear as the magnitude of the chemical potential is increased (as expected from numerics \cite{Jazayeri:2023kji}). For comparison, the exact result we present here holds for all values of the chemical potential within the perturbative regime. As a final example, we show how a parity-odd trispectrum can be generated in a purely scalar theory at tree-level. This can arise due to a parity-odd quartic vertex that mixes the inflaton with another massive scalar field, in addition to a linear mixing term. We present an exact analytic result for the corresponding trispectrum.

Our results do not just imply that parity-odd trisepctra are factorised; rather any $n$-point correlator of massless scalars and massless gravitons that requires one to take the imaginary part of wavefunction coefficients will be factorised and therefore lend itself to a simpler computation. This includes all $n$-point parity-odd correlators.

We conclude and discuss further directions in Section \ref{Conclusions}. We also include a number of appendices. In Appendix \ref{GeneralDeltaGhAppendix} we provide full details on how to realise the reality property of the time-ordered part of the Wick-rotated propagator. In Appendix \ref{InIn} we provide a proof of our reality theorems using the in-in formalism which might be more familiar to the reader. Appendix \ref{CuttingComparison} includes a complementary proof of our results using Hermitian analyticity that we have already mentioned above, and in Appendix \ref{MoreRealities} we extend our results to other Friedmann–Lemaître–Robertson–Walker (FLRW) spacetimes.

\newpage

\begin{figure}[h!]
	\centering
	\begin{tikzpicture}
	\node at (0,0) {\includegraphics[width=0.8\textwidth]{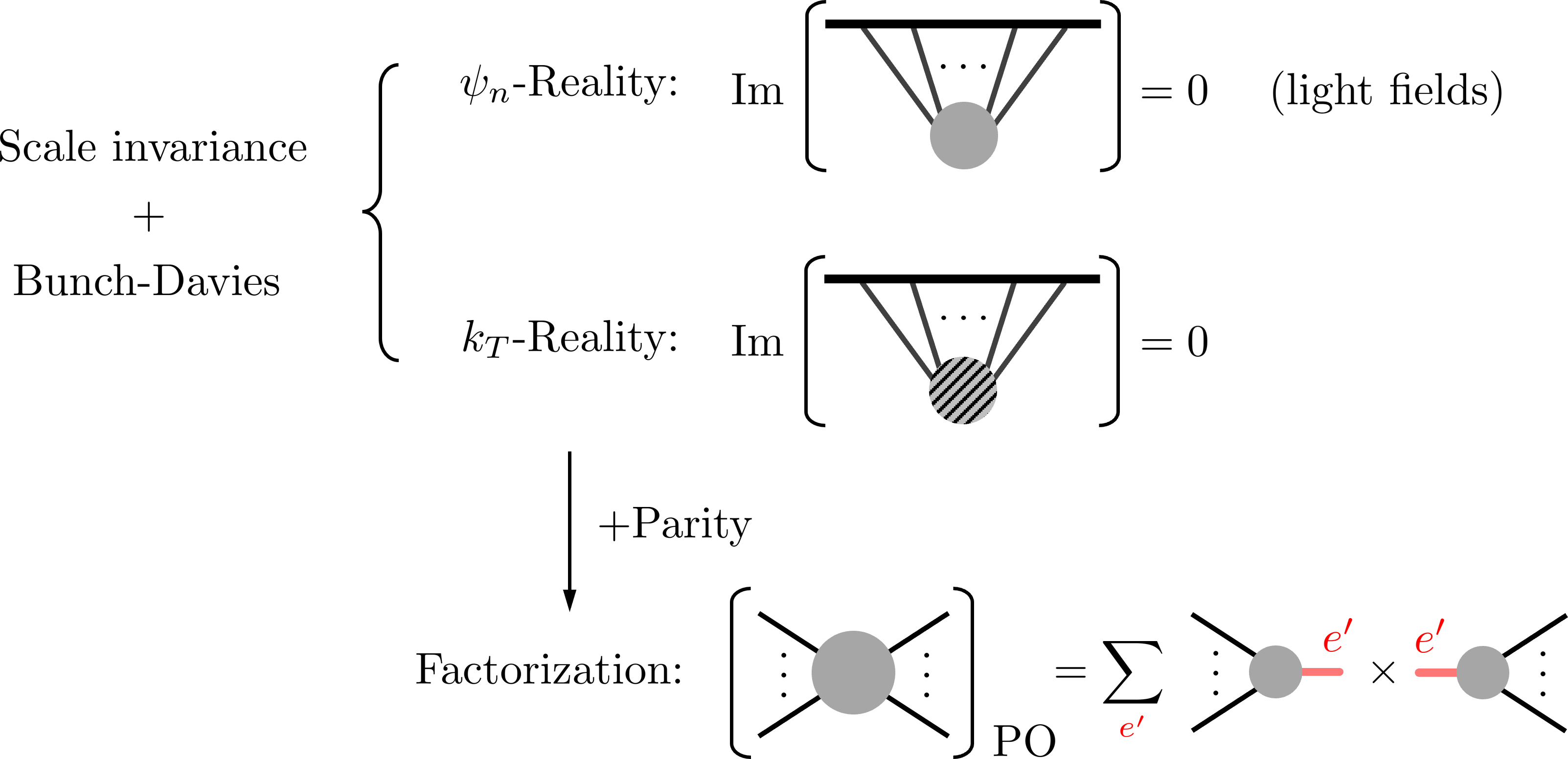}};
	\node at (-1.77,1.9) {\eqref{psiNRealityTheorem}};
	\node at (-1.77,-0.2) {\eqref{kTRealityTheorem}};
	\node at (-1.77,-2.95) {\eqref{POFactorizationTheorem}};
	\end{tikzpicture}
	\caption{Scale invariance plus a Bunch-Davies vacuum implies that the maximally-connected parts of wavefunction coefficients are purely real where the maximally-connected part is given by replacing all bulk-bulk propagators $G$ with the connected propagators $C$. Such a maximally-connected part is depicted as a hatched grey blob in this figure. We refer to this result as the $k_{T}$-reality since all total-energy singularities come from the maximally-connected wavefunction coefficients. This is a general result that applies for the exchange of fields of any mass and integer spin. If the exchanged fields are light i.e. they are in the complementary series, then the full wavefunction coefficient is real (not just the maximally-connected part) since in this case $C$ coincides with $G$. If we further consider parity-odd correlation functions, which on general grounds must be imaginary if we consider correlators of a massless scalar, then the wavefunction reality implies that such $n$-point correlators are factorised since only the imaginary part of maximally-connected wavefunction coefficients contribute to these observables and the imaginary parts vanish.}\label{summaryFig}
\end{figure}
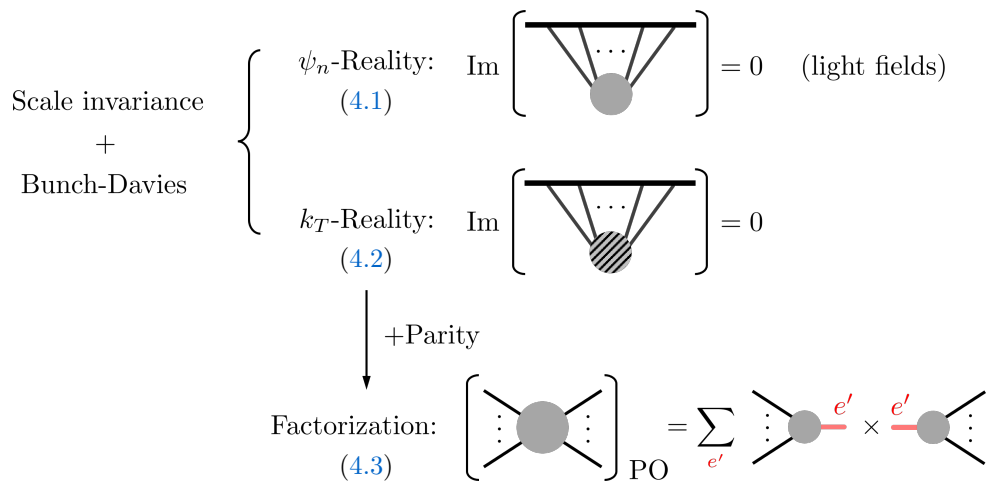

\begin{figure}[htp]
	\includegraphics[width=0.75\textwidth]{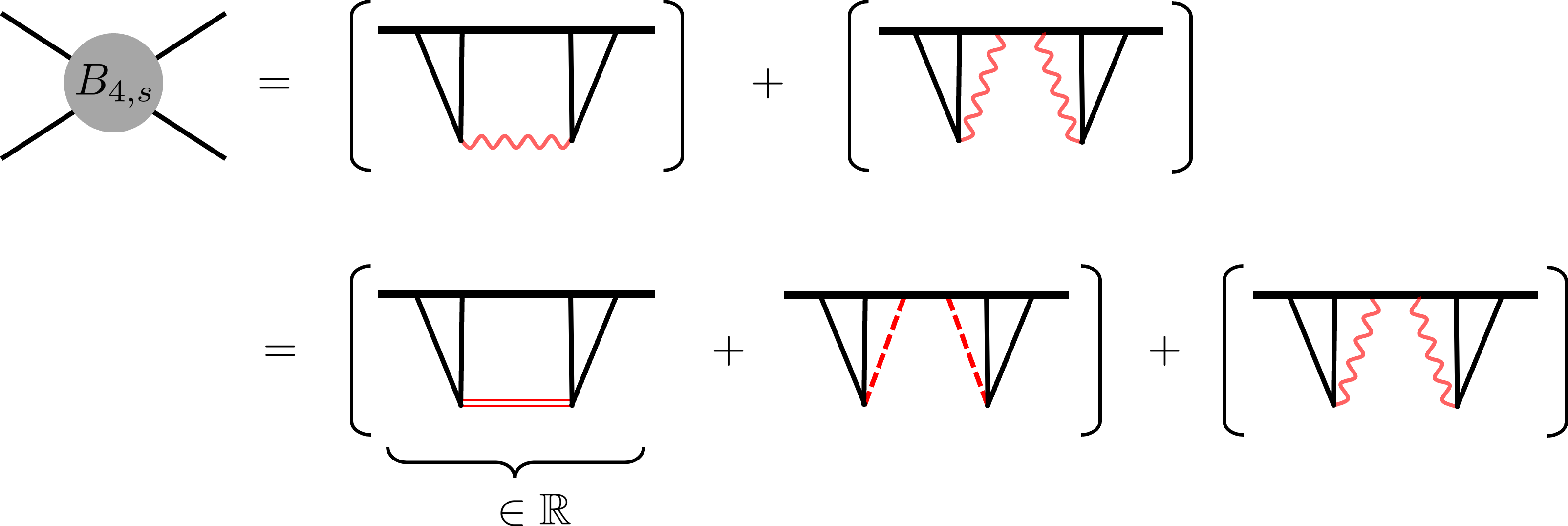}
  \caption{The first line is the usual relationship between ($s$-channel) four-point correlators and wavefunction coefficients. In the second line we have decomposed the full bulk-bulk propagator into the connected and factorised parts. The $k_{T}$-reality tells us that the maximally-connected part is purely real which in turn implies that the parity-odd part of the four-point function is factorised (as above). }\label{fourpointcorrealtorfigure}
\end{figure}

\paragraph{Summary of main results} Before moving to the main body of the paper, let us first summarise our main results for the convenience of the reader. 
\begin{itemize}
 \item Throughout this paper we consider the following decomposition of the wavefunction bulk-bulk propagator $G$:
\begin{keyeqn}
	\begin{align} \label{DecompositionFigureIntro}
		\begin{gathered}
			\includegraphics[width=0.4\textwidth]{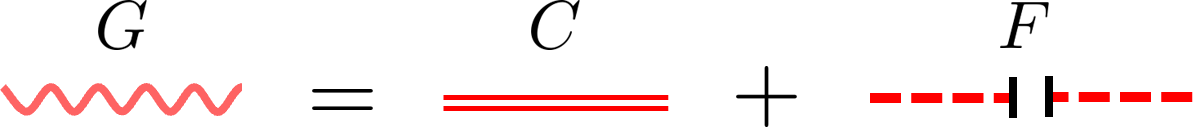}
		\end{gathered}
	\end{align}
\end{keyeqn}
where we refer to $C$ as the connected part and $F$ as the factorised part. The connected part is defined as containing the time-ordered parts of the full propagator, along with the crucial property that it is purely real after both time variables are Wick rotated by $90^{\circ}$ in the complex plane. As we explain in detail in Section \ref{PropagatorPropertySection}, $C$ is \textit{not} equal to the Feynman propagator; they differ by a factorised contribution. For fields in the complementary series $C$ coincides with $G$ i.e. $F=0$, however for fields in the principle series $F \neq 0$. In all cases both $C$ and $F$ vanish in the far past, while only the sum satisfies the future Dirichlet boundary condition with all $\eta_0$ dependence (where $\eta_0$ is the late-time cut-off) contained in $F$.

\item Using this decomposition and the reality properties of $C$, we derive reality properties of wavefunction coefficients of massless scalars and gravitons with our main results depicted in Figure \ref{summaryFig}.

\item If we specialise to four-point correlators of inflationary curvature perturbations then the relationship between such a correlator and wavefunction coefficients is depicted in Figure \ref{fourpointcorrealtorfigure}.

\item We used our factorisation theorem to compute some exact parity-odd trispectra of curvature perturbations. We consider a number of examples in Section \ref{ExactTrispectraSection} but in all cases the final trispectrum takes the following schematic form:
\begin{keyeqn}
\begin{align}\label{B4Schematic}
\resizebox{0.9\textwidth}{!}{$B_{4}^{\text{PO}} = \sum  \text{constants} \times \text{kinematics} \times (\text{hypergeometric function})_{L} \times (\text{hypergeometric function})_{R}$} 
\end{align}
\end{keyeqn}
where $(L,R)$ correspond to kinematic structures with partial-energy ($E_{L,R}$) singularities. There are no total-energy singularities. In all cases the time evolution is characterised by a product of hypergeometric functions.
\end{itemize}

\paragraph{Notations and conventions} Throughout this paper, we adopt natural units $c=\hbar=1$ and work with the $(-+++)$ metric sign convention. Fourier transforms are defined by
\begin{align}
	f(\mathbf{x})=\int\frac{d^3k}{(2\pi)^3}e^{i\mathbf{k}\cdot \mathbf{x}}f(\mathbf{k})\equiv\int_{\mathbf{k}}e^{i\mathbf{k}\cdot \mathbf{x}}f(\mathbf{k})~.
\end{align}
We will always adhere to exact translational and rotational invariance. In fact, we shall work in the Poincaré patch of de Sitter space with the metric
\begin{align}
	ds^2=a^2(\eta)(-d\eta^2+d\mathbf{x}^2)~,\quad a(\eta)=-\frac{1}{H\eta}~,
\end{align}
as an approximation for an inflationary spacetime. We will mainly work with the wavefunction formalism and define the wavefunction coefficients by the expansion
\begin{align} 
	\label{WFU}
	\Psi[\varphi] = \exp\left[+\sum_{n=2}^{\infty}\frac{1}{n!} \int_{\mathbf{k}_{1} \cdots \mathbf{k}_{n}} \psi_{n}(\{ k \}, \{ \mathbf{k} \}) (2 \pi)^3 \delta^3 \bigg(\sum_{{\sf a}=1}^n {\mathbf{k}}_{\sf a} \bigg)\varphi(\mathbf{k}_{1}) \cdots \varphi(\mathbf{k}_{n}) \right] \,\,,
\end{align}
where $\varphi$ denotes a general field with indices suppressed. The bulk-boundary and bulk-bulk propagators in the wavefunction formalism are given by
\begin{align}
	K(\eta,k)&=\frac{\varphi^*(k,\eta)}{\varphi^*(k,\eta_{0})}~,\\
	G(\eta_1,\eta_2,k)&=P(k)\left[\Big(K^*(\eta_1,k)K(\eta_2,k)\theta(\eta_1-\eta_2)+(\eta_1\leftrightarrow\eta_2)\Big)-K(\eta_1,k)K(\eta_2,k)\right]~,\label{GdefIntro}
\end{align}
with $P(k)=\varphi(\eta_0,k) \varphi^*(\eta_0,k)$ denoting the power spectrum at $\eta=\eta_0$. When computing the wavefunction coefficients, we adopt the amplitude Feynman rules by including a factor of $i$ for every vertex (our bulk-bulk propagator therefore differs from that in \cite{COT} by a factor of $i$). Wick rotation turns out to be extremely crucial in this paper. Hence for clarity, we adopt $\chi>0$ to denote the Wick-rotated conformal time, defined by
\begin{align} \label{WickRotation}
	\eta=i e^{i\epsilon}\chi~,\quad \epsilon\to0^+~.
\end{align}
Under this transformation, the propagators are dressed with tildes to indicate Wick rotation,
\begin{align}
	\tilde{K}(\chi,k)&=K(\eta,k)~,\\
	\tilde{G}(\chi_1,\chi_2,k)&=G(\eta_1,\eta_2,k)~.\label{GPropWickRotDef}
\end{align}
We will pay much attention to the total energy of a diagram with external momenta $\{\mathbf{k}_1,\cdots, \mathbf{k}_n\}$ given by
\begin{align}
	k_T=k_1+\cdots+k_n~,
\end{align}
where $k_{\sf a}=|\mathbf{k}_{\sf a}|,{\sf a}=1,\cdots,n$ are the external energy variables. Note that correlators with a prime denotes the removal of an overall momentum-conserving $\delta$-function, e.g.
\begin{align}
	\left\langle\varphi(\mathbf{k}_1)\cdots\varphi(\mathbf{k}_n)\right\rangle=(2\pi)^3\delta^3 \left(\sum_{{\sf a}=1}^n \mathbf{k}_{\sf a} \right)\left\langle\varphi(\mathbf{k}_1)\cdots\varphi(\mathbf{k}_n)\right\rangle'=(2\pi)^3\delta^3 \left(\sum_{{\sf a}=1}^n \mathbf{k}_{\sf a} \right) B_n^\varphi(\mathbf{k}_1,\cdots,\mathbf{k}_n)~.
\end{align}
In the case of 4-point correlation functions, we evoke the Mandelstam-like variables
\begin{align}
	\nonumber\mathbf{s}&=\mathbf{k}_1+\mathbf{k}_2~, &\mathbf{t}&=\mathbf{k}_1+\mathbf{k}_3~, &\mathbf{u}&=\mathbf{k}_1+\mathbf{k}_4~,\\
	s&=|\mathbf{k}_1+\mathbf{k}_2|~,& t&=|\mathbf{k}_1+\mathbf{k}_3|~,& u&=|\mathbf{k}_1+\mathbf{k}_4|~,
\end{align}
which satisfy the non-linear relation
\begin{align}
	k_{1}^2+k_{2}^2+k_{3}^2+k_{4}^2 = s^2+t^2+u^2,
\end{align}
by momentum conservation. We define a dimensionless curvature trispectrum $\mathcal{T}$ by
\begin{align}
	B_4^\zeta(\mathbf{k}_1,\mathbf{k}_2,\mathbf{k}_3,\mathbf{k}_4)=(2\pi)^6\Delta_\zeta^6\frac{(k_T/4)^3}{(k_1 k_2 k_3 k_4)^3}\mathcal{T}(\mathbf{k}_1,\mathbf{k}_2,\mathbf{k}_3,\mathbf{k}_4)\,.\label{dimlessTDef}
\end{align}
For spinning fields, the CCM and CC set-ups historically chose different conventions for polarisation tensors. We choose to respect these distinct conventions and use different fonts to avoid confusion:
\begin{align}
	\nonumber\text{Cosmological Condensed Matter Scenario: }& \mathrm{e}_{i_1\cdots i_S}^{(h)}~,\\
	\text{Cosmological Collider Scenario: }& {\mathfrak e}_{i_1\cdots i_S}^{(h)}~.
\end{align}
The different properties of these tensors will be explained in Section \ref{sec:massivefields}. Throughout this paper, we make use of a number of mathematical formulae for Hankel and Whittaker functions, which can be found in \href{https://dlmf.nist.gov/10}{Chapter 10} and \href{https://dlmf.nist.gov/13}{Chapter 13} of NIST \cite{NIST:DLMF}.


\section{Massive spinning fields during inflation} \label{sec:massivefields}

As we explained in the introduction, in this work we are interested in $n$-point functions of massless scalar fields (and gravitons) that can be generated at tree-level due to interactions with massive spinning fields. In this section we introduce these massive spinning fields and derive their mode functions. We consider the two different cases of interest, cosmological condensed matter physics and cosmological collider physics, separately. This section does not contain any new results, so readers familiar with these descriptions of massive spinning fields (including parity-violating corrections from the chemical potential) can skip to Section \ref{PropagatorPropertySection}.


\subsection{Cosmological condensed matter physics} \label{CCMmodefunctions}

We begin with the description of massive spinning fields during inflation advocated in \cite{Bordin:2018pca}. The idea is to classify states with respect to the unbroken rotational symmetries, rather than as representations of the full de Sitter group. Fields therefore only have spatial indices, and we denote a field of spin $S$ by $\Sigma^{i_{1} \cdots i_{S}}$. For this field to carry $2S+1$ degrees of freedom, it should be traceless but not transverse. To ensure that the symmetries of the EFToI are respected, this field is promoted to $\Sigma^{\mu_{1} \cdots \mu_{S}}$ where the new temporal components depend on the Goldstone of broken time translation $\pi$. For example, for $S=1$ we have \cite{Bordin:2018pca} 
\begin{align}
	\Sigma^{0}(\pi, \Sigma^{i}) = -\frac{\partial_{i} \pi \Sigma^{i}}{1 + \dot{\pi}}~,
\end{align}
while for $S=2$ we have
\begin{align}
	\Sigma^{00}(\pi, \Sigma^{ij}) = \frac{\partial_{i} \pi \partial_{j} \pi \Sigma^{ij}}{(1 + \dot{\pi})^2}~, \qquad \Sigma^{0j}(\pi, \Sigma^{ij}) = -\frac{\partial_{i} \pi \Sigma^{ij}}{1 + \dot{\pi}}~.
\end{align}
We then write down a quadratic action for $\Sigma^{\mu_{1} \cdots \mu_{S}}$ which will introduce quadratic terms for the spatial components, and interactions between the spatial components and the inflaton with the same coefficients. This illustrates the fact that such a theory cannot exist in the absence of the inflaton. The general action with at most two derivatives is 
\begin{align}
	S_2=\frac{1}{2 S!}\int d^4x\sqrt{-g} \Bigg[&\left(1-c^2\right)n^{\mu}n^{\lambda}\nabla_\mu\Sigma^{\nu_1 \cdots\nu_S}\nabla_{\lambda}\Sigma_{\nu_1 \cdots \nu_S}-c^2\nabla_\mu \Sigma^{\nu_1 \cdots \nu_S}\nabla^{\mu}\Sigma_{\nu_1\cdots\nu_S} \nonumber \\ &-\delta
	c^2\nabla_{\mu}\Sigma^{\mu\nu_2\cdots\nu_S}\nabla_\lambda\Sigma^{\lambda}_{~\nu_2\cdots\nu_S}
	-\left(m^2+S c^2H^2\right)\Sigma^{\nu_1\cdots\nu_S}\Sigma_{\nu_1\cdots\nu_S} \nonumber \\ &-2S\kappa\, n^{\mu}\epsilon_{\mu\rho\gamma\lambda}\Sigma^{\rho\nu_2\cdots \nu_S}\,\nabla^{\gamma}\Sigma^{\lambda}_{~\nu_2\cdots \nu_S}\Bigg]~,
\end{align}
where $n^\mu n_\mu=-1$ is a timelike unit vector that defines a preferred frame in which the spatial rotations remain intact. Notice that in addition to the first four terms which appear in \cite{Bordin:2018pca}, we have also included a fifth term which has a single derivative and is parity-odd. We in principle have five free parameters but we can fix one using our freedom to normalise $\Sigma^{\mu_{1} \cdots \mu_{S}}$ which leaves us with four: $c, \delta c, m^2$ and $\kappa$ which respectively correspond to two speed-of-sound parameters, the mass, and the \textit{chemical potential}. The free theory for $\Sigma^{i_{1} \cdots i_{S}}$ is then 
\begin{align} \label{SigmaFreeAction}
	S_2=\frac{1}{2 S!}\int d\eta d^3x a(\eta)^{2}\Bigg[&\sigma'^2_{i_1\cdots i_S}-c^2 (\partial_j\sigma_{i_1\cdots i_S})^2-\delta c^2(\partial_j\sigma_{j i_2\cdots i_S})^2 \nonumber \\ &-a(\eta)^2m^2\sigma^2_{i_1\cdots i_S}
	-2Sa(\eta)\kappa\epsilon_{ijk}\sigma_{i l_2\cdots l_S}\partial_j \sigma_{k l_2\cdots l_S}\Bigg]~,
\end{align}
where we have defined $\sigma_{i_{1} \cdots i_{S}} = a^{-S} \Sigma_{i_{1} \cdots i_{S}}$ and have converted to conformal time. Here all scale factors are manifest and indices are raised and lowered with the Kronecker symbol $\delta_{ij}$. We see that the kinetic term for this field is the same as that of a canonical scalar in de Sitter. If we had instead tried to directly construct the most general action with at most two derivatives for $\sigma_{i_{1} \cdots i_{S}}$ that respects rotational invariance and scale invariance, we would also have arrived at \eqref{SigmaFreeAction}.\footnote{A parity-odd term with one spatial derivative and one time derivative is degenerate with the terms in \eqref{SigmaFreeAction} up to a total derivative.} For $\Sigma^{i_{1} \cdots i_{S}}$ to be traceless, the trace has to be taken with respect to the induced metric on constant inflaton slices \cite{Bordin:2018pca}. This implies that in \eqref{SigmaFreeAction} we can take $\sigma_{i_{1} \cdots i_{s}}$ to be traceless with respect to $\delta_{ij}$ up to terms that are quadratic in $\sigma_{i_{1} \cdots i_{S}}$ and at least quadratic in $\pi$. When discussing the free theory for this massive spinning field we therefore take it to satisfy $\delta_{ij} \sigma_{ij l_{3} \cdots l_{S}} = 0$. 

We now convert the action to momentum space and decompose the field in terms of its helicities via
\begin{align}
	\sigma_{i_{1} \cdots i_{S}}(\eta, \bfx) = \sum_{h = -S}^{S}\int_{\mathbf{k}} \sigma_{h}(\eta, k){\rm{e}}^{(h)}_{i_{1} \cdots i_{S}}(\bfk) e^{i \bfk \cdot \bfx}~,
\end{align}
where $\sigma_{h}(\eta, k)$ are the mode functions and the traceless polarisation tensors satisfying
\begin{align}
	\left[{\rm{e}}^{(h)}_{i_{1} \cdots i_{S}}(\bfk) \right]^{\star} &= {\rm{e}}^{(h)}_{i_{1} \cdots i_{S}}(-\bfk)~, \label{PolStar} \\
	{\rm{e}}^{(h)}_{i_{1} \cdots i_{S}}(\bfk){\rm{e}}^{(h')}_{i_{1} \cdots i_{S}}(-\bfk) &= S!\, \delta_{h h'}~, \label{PolNorm}
\end{align}
with the first condition following from the reality of the fields in position space, and the second a normalisation choice. For a given helicity the polarisation tensor is a function of $\hat{\bfk}$ and two polarisation directions $\hat{\textbf{e}}^{\pm}$ which are orthogonal to $\hat{\bfk}$, and satisfy \eqref{PolStar} and \eqref{PolNorm}. More explicitly, we can construct the polarisation directions as
\begin{align} \label{PolDirection}
\hat{\mathbf{e}}^\pm(\hat{\mathbf{k}})=\frac{\hat{\mathbf{n}}-(\hat{\mathbf{n}}\cdot\hat{\mathbf{k}})\hat{\mathbf{k}}\pm i \,\hat{\mathbf{k}}\times\hat{\mathbf{n}}}{\sqrt{2[1-(\hat{\mathbf{n}}\cdot\hat{\mathbf{k}})^2]}}~,
\end{align}
where $\hat{\mathbf{n}}$ is an arbitrary unit vector not parallel to $\hat{\mathbf{k}}$. Modes with $h=0$ are functions of $\hat{\bfk}$ only, modes with $|h| = S$ are functions of $\hat{\textbf{e}}^{\pm}$ only, while intermediate modes are functions of both $\hat{\bfk}$ and $\hat{\textbf{e}}^{\pm}$, with $|h|$ powers of the latter. This structure ensures that modes of different helicity decouple in the quadratic action, while the normalisation choice ensures that each mode has a canonical kinetic term. For example, for $S=1$ we have
\begin{align}
	{\rm{e}}_{i}^{(0)} = i \hat{k}_{i}~, \qquad {\rm{e}}_{i}^{(\pm 1)} = \hat{e}_{i}^{\pm}~,
\end{align}
while for $S=2$ we have 
\begin{align}
	{\rm{e}}_{ij}^{(0)} =\sqrt{3}\left(\hat{k}_{i}\hat{k}_{j} - \frac{1}{3}\delta_{ij} \right)~, \qquad {\rm{e}}_{ij}^{(\pm 1)} =i(\hat{k}_{i} \hat{e}_{j}^{\pm} + \hat{k}_{j} \hat{e}_{i}^{\pm})~, \qquad {\rm{e}}_{ij}^{(\pm 2)} = \sqrt{2}\hat{e}_{i}^{\pm}\hat{e}_{j}^{\pm}~.
\end{align}
Further details on these polarisation structures can be found in e.g. \cite{Lee:2016vti,Goodhew:2021oqg}. Using these properties of the polarisation tensors, \eqref{SigmaFreeAction} becomes a decoupled action for each mode function: 
\begin{align}
	S_{2} = \frac{1}{2}\sum_{h = -S}^{S} \int_{\mathbf{k}} d \eta a^{2}(\eta)\left[\sigma'^2_{h} - c_{h,S}^2 k^2 \sigma^2_{h} - m^2  a(\eta)^{2} \sigma^2_{h} - 2 h\kappa k  a(\eta)\sigma^2_{h} \right]~,
\end{align}
where 
\begin{align}
	c^2_{h,S} = c^2 + \frac{S^2-h^2}{S(2S-1)} \delta c^2~.
\end{align}
In deriving this expression we have used $i  \epsilon_{ijk} k_{j} \hat{e}^{\pm}_{k} = \pm k \hat{e}^{\pm}_{i} $. The fact that the parity-violating term splits the helicities is now clear given the helicity factor $h$ in the final term. In addition to this term, we see that each mode has the same mass but different speed of sound which depends on the two original speed of sound parameters, the spin and the helicity. The equation of motion for each helicity mode is then
\begin{align} \label{CCMeom}
	\left(\eta^2\frac{\partial^2}{\partial \eta^2} - 2\eta \frac{\partial}{\partial \eta} + c_{h,S}^2 k^2\eta^2 + \frac{m^2}{H^2} - \frac{2 h \kappa}{H} k\eta \right) \sigma_{h}(\eta, k) = 0~.
\end{align}
The solution to this equation with Bunch-Davies initial conditions i.e. the solution that has Minkowski-like behaviour in the far past and satisfies the de Sitter Wronskian condition is
\begin{align} \label{CCMmodefunction}
	\sigma_{h}(\eta, k) =  e^{-\pi\tilde{\kappa}/2}  \frac{- H \eta}{\sqrt{2 c_{h,S} k}}  W_{i \tilde{\kappa}, \nu}(2 i c_{h,S} k \eta)~, \quad \nu = \sqrt{\frac{9}{4} - \frac{m^2}{H^2}}~, \quad \tilde{\kappa} = \frac{h \kappa}{c_{h,S} H}~,
\end{align}
where $W_{a,b}(z)$ is the Whittaker $W$-function. This solution is valid for both light ($\text{Im} ~ \nu = 0$) and heavy ($\text{Re} ~ \nu = 0$) fields. In the limit that the chemical potential vanishes we would expect to recover the solution of \cite{Bordin:2018pca} corresponding to the solution of a massive scalar field in de Sitter space. This can be verified using the relation
\begin{align} \label{WtoH}
	W_{0, \nu}(2 i c_{h,S} k \eta) = \sqrt{\frac{\pi}{2}}\sqrt{- c_{h,S} k \eta}e^{i \pi (\nu + 1/2)/2} H_{\nu}^{(1)}(- c_{h,S} k \eta)~,
\end{align}
where $H_{\nu}^{(1)}(z)$ is the Hankel function of the first kind. We will discuss the corresponding power spectra and propagators in Section \ref{PropagatorPropertySection}. Notice that in this CCM scenario, the introduction of chemical potential $\kappa$ is quite natural since it serves as a next-to-leading order correction to the dispersion relation of massive spinning fields in the gradient expansion, and is consistent with spatial rotations and scale invariance. More precisely, it appears as a linear term in momentum in the dispersion relation of a massive field,
\begin{align}
	w^2(\mathbf{k}_p^2,\mathbf{S}\cdot\mathbf{k}_p)=m^2+2\kappa \,\mathbf{S}\cdot\mathbf{k}_p+\left[\left(\delta c^2+\frac{S}{2S-1}\delta c^2\right) \mathbf{k}_p^2-\frac{\delta c^2}{S(2S-1)}(\mathbf{S}\cdot\mathbf{k}_p)^2\right]+\mathcal{O}\left(|\mathbf{k}_p|^3\right)~,
\end{align}
where $\mathbf{k}_p\equiv \mathbf{k}/a(t)$ and $\mathbf{S}$ are the physical momentum and the spin angular momentum of the field mode, respectively. Such a linear correction to the massive field's dispersion relation is not sign-definite, and alters the analytic structure of its equation of motion (\ref{CCMeom}), leading to enhanced particle production when $m\gtrsim \kappa$ \cite{Sou:2021juh}. In the case where $m\lesssim\kappa$, however, modes with a negative linear term may experience a transient tachyonic phase where $w^2<0$ and grow exponentially, i.e. $\sigma\sim e^{-i\omega t}\sim e^{|\omega| t}$. Such a tachyonic growth is eventually halted by the finite mass, leading to $w^2\approx m^2>0$ in the IR limit $\mathbf{k}_p\to 0$. Nevertheless, the exponential growth during the tachyonic period may overproduce particles and threaten perturbativity or even the inflationary background. Therefore, in favour of theoretical control, we will require $m-\kappa\gtrsim -H$ throughout this paper.


\subsection{Cosmological collider physics}

We now turn to the more familiar description of massive spinning fields in de Sitter/inflation where they are representations of the de Sitter group. In this section we primarily follow \cite{Lee:2016vti} and the refer the reader there for further details. Such fields can certainly exist in the absence of the inflaton, in contrast to CCM. A spin-$S$ bosonic field in this CC set-up is described by a symmetric rank-$S$ tensor that satisfies:
\begin{align}
	(\Box - m_{S}^2) \Phi_{\mu_{1} \cdots \mu_{S}} = 0~, \qquad \nabla^{\nu} \Phi_{\nu \mu_{2} \cdots \mu_{S}} = 0~, \qquad \Phi^{\nu}{}_{\nu \mu_{2} \cdots \mu_{S}} = 0~,
\end{align}
where the mass parameter is $m_{S}^2 = m^2 - (S^2-2S-2)H^2$. On-shell we therefore have a transverse and traceless rank-$S$ tensor that satisfies a wave equation. To solve this system we work in a $3+1$ decomposition where the components are of the form $\Phi_{\eta \cdots \eta i_{1} \cdots i_{n}}$ with $ 0 \leq n \leq S$. We further convert to momentum space and decompose each of these components in terms of helicities via 
\begin{align}
	\Phi_{\eta \cdots \eta i_{1} \cdots i_{n}}(\eta, \bfk) = \sum_{h=-n}^{n} \Phi_{n,S}^{h}(\eta, k) {\mathfrak{e}}^{(h)}_{i_{1} \cdots i_{n}}(\hat{\bfk})~.
\end{align}
Each mode function is therefore labelled by three numbers corresponding to the spacetime spin ($S$), spatial spin ($n$), and helicity component of the spatial spin ($h$). We have made a distinction between the polarisation tensors used here (${\mathfrak{e}}$) compared to above in the CCM case (${\rm{e}}$). This is because a different normalisation is employed in \cite{Lee:2016vti} compared to what we used above, and while discussing the CC scenario we will follow the conventions of \cite{Lee:2016vti} to (hopefully) avoid confusion for the reader. The polarisation tensors here are still functions of $\hat{\bfk}$ and two polarisation directions $\hat{\textbf{e}}^{\pm}$, however rather than satisfying \eqref{PolStar} and \eqref{PolNorm}, they are chosen to satisfy
\begin{align} \label{PolStarCC}
	\left[{\mathfrak{e}}^{(h)}_{i_{1} \cdots i_{n}}(\hat{\bfk})\right]^{\star} &= {\mathfrak{e}}^{(-h)}_{i_{1} \cdots i_{n}}(\hat{\bfk}) ~, \\
	{\mathfrak{e}}^{(h)}_{i_{1} \cdots i_{S}}(\bfk)[{\mathfrak{e}}^{(h)}_{i_{1} \cdots i_{S}}(\bfk)]^{\star} &= \frac{(2 S-1)!! (S+ |h|)!}{2^{|h|} [(2 |h|-1)!!]^2 S! (S - |h|)!} {\mathfrak{e}}^{(h)}_{i_{1} \cdots i_{|h|}}(\bfk)[{\mathfrak{e}}^{(h)}_{i_{1} \cdots i_{|h|}}(\bfk)]^{\star}~, \label{PolNormCC}
\end{align}
where the polarisation tensor with lowest index is defined as 
\begin{align}
  {\mathfrak{e}}^{(h)}_{i_{1} \cdots i_{|h|}}(\bfk)=2^{|h|/2}\hat{e}_{i_1}^{h}\cdots\hat{e}_{i_{|h|}}^{h}~.
\end{align}
In both cases (CCM and CC) the numerical factors that multiply the $\hat{\bfk}$ and $\hat{\textbf{e}}^{\pm}$ factors can be made purely real or purely imaginary. The magnitudes of these factors are fixed by \eqref{PolNorm} and \eqref{PolNormCC} in the two cases, while in the CCM case the condition \eqref{PolStar} fixes the factors to be imaginary when there are an odd number of $\hat{\bfk}$'s, and real when there is an even number, while \eqref{PolStarCC} fixes the factors to be always real. This phase difference will ultimately be inconsequential since it can be absorbed into the mode functions which are always only fixed up to a phase.

As an illustration let us spell out the $S=1$ case. If we decompose $\Phi_{\mu}$ into its time and space components $\Phi_{\eta}$ and $\Phi_{i}$, then $(\Box - m_{1}^2) \Phi_{\mu} = 0$ becomes
\begin{align}
	\Phi''_{\eta} - \left(\partial^2_{i} - \frac{m^2}{H^2 \eta^2} + \frac{2}{\eta^2} \right) \Phi_{\eta} = \frac{2}{\eta} \partial_{i} \Phi_{i}~, \\
	\Phi''_{i} - \left(\partial^2_{j} - \frac{m^2}{H^2 \eta^2}  \right) \Phi_{i} = \frac{2}{\eta} \partial_{i} \Phi_{\eta}~,
\end{align}
while the transverse constraint is
\begin{align}
	\Phi'_{\eta}  - \frac{2}{\eta}\Phi_{\eta} = \partial_{i}\Phi_{i}~.
\end{align}
The $\Phi_{\eta}$ field carries only a $h=0$ mode, while the $\Phi_{i}$ components carry both $h=0$ and $h =\pm 1$ modes. We then write
\begin{align}
	\Phi_{\eta} = \Phi_{0,1}^{0}~, \qquad \Phi_{i}^{(0)} = \Phi_{1,1}^{0} {\mathfrak{e}}_{i}^{0}~, \qquad \Phi_{i}^{(\pm 1)} = \Phi_{1,1}^{\pm 1} {\mathfrak{e}}_{i}^{\pm 1}~.
\end{align}
The polarisation vectors are chosen to be 
\begin{align}
	{\mathfrak{e}}_{i}^{(0)} = \hat{k}_{i}~, \qquad {\mathfrak{e}}_{i}^{(\pm 1)} = \hat{e}_{i}^{\pm}~.
\end{align}
The equations of motion then decouple for each mode function and are given by 
\begin{align}
	\Phi_{0,1}^{0}{}^{''} - \frac{2}{\eta} \Phi_{0,1}^{0}{}^{'} + \left(k^2 + \frac{m^2}{H^2 \eta^2} + \frac{2}{\eta^2} \right) \Phi_{0,1}^{0} & = 0~, \\
	\Phi_{1,1}^{0}{}^{''} -\frac{k^2 \eta^2}{k^2 \eta^2 + m^2/H^2} \frac{2}{\eta} \Phi_{1,1}^{0}{}^{'} + \left(k^2 + \frac{m^2}{H^2 \eta^2} \right) \Phi_{1,1}^{0} & = 0~, \\
	\Phi_{1,1}^{\pm 1}{}^{''}  + \left(k^2 + \frac{m^2}{H^2 \eta^2}  \right) \Phi_{1,1}^{\pm 1} & = 0~, 
\end{align}
subject to the transverse constraint
\begin{align}
	\Phi_{1,1}^{0} = -\frac{i}{k} \left(\Phi_{0,1}^{0}{}^{'} - \frac{2}{\eta} \Phi_{0,1}^{0} \right)~.
\end{align}
The equation of motion for the $h = \pm 1$ modes does not contain the Hubble friction term since the Maxwell kinetic term is conformally invariant. The solutions to these equations with Bunch-Davies initial conditions are 
\begin{align}
	\Phi_{0,1}^{0} &= \frac{\sqrt{\pi}}{2}\frac{H k}{ m} e^{i \pi (\nu_{1} + 1/2)/2} (- \eta)^{3/2} H_{\nu_{1}}^{(1)}(- k \eta)~, \\
	\Phi_{1,1}^{0} &= i \frac{\sqrt{\pi}}{4}\frac{H }{ m} e^{i \pi (\nu_{1} + 1/2)/2} (- \eta)^{1/2} [ k \eta ( H_{\nu_{1}+1}^{(1)}(- k \eta)-H_{\nu_{1}-1}^{(1)}(- k \eta)) - H_{\nu_{1}}^{(1)}(- k \eta)]~, \\
	\Phi_{1,1}^{\pm 1} &= \frac{\sqrt{\pi}}{2} e^{i \pi (\nu_{1} + 1/2)/2} (- \eta)^{1/2} H_{\nu_{1}}^{(1)}(- k \eta)~,
\end{align}
where 
\begin{align}
	\nu_{1} = \sqrt{\frac{1}{4} - \frac{m^2}{H^2}}~,
\end{align}
and the normalisation constants have been fixed by demanding that the commutation relations are the usual ones \cite{Lee:2016vti}. We see here a feature that immediately distinguishes this set-up from the CCM one: some of the mode functions in this case are given by a sum of Hankel functions with degenerate order parameters. The dynamics in the two set-ups therefore different.

The story for spin-$S$ is similar. Modes with helicity $h$ can come from all components with $n \geq |h|$, and those with $n = |h|$ satisfy \cite{Lee:2016vti}
\begin{align}
	\Phi_{|h|,S}^{h}{}^{''} - \frac{2(1 - |h|)}{\eta} \Phi_{|h|,S}^{h}{}^{'} + \left( k^2 + \frac{m^2}{H^2 \eta^2} - \frac{(S+|h|-2)(S-|h|+1)}{\eta^2} \right) \Phi_{|h|,S}^{h} =0~.\label{CCModeEoM}
\end{align}
The other mode functions with the same helicity but with $n > |h|$ are then obtained iteratively from the transverse and traceless conditions which fix\footnote{As noticed in \cite{Pimentel:2022fsc}, there is a typo in the corresponding formula in \cite{Lee:2016vti}, (A.70): the coefficient of the $B_{m,n+1}$ terms should be $+1$ rather than $-1$.}
\begin{align}
	\Phi^{h}_{n+1,S} = - \frac{i}{k} \left(\Phi_{n,S}^{h}{}^{'} - \frac{2}{\eta} \Phi_{n,S}^{h} \right) + \sum_{m = |h|}^{n} B_{m,n+1} \Phi_{m,S}^{h}~, \label{ModeFunctionCCDifferentialOperator}
\end{align}
where
\begin{align}
	B_{m,n} = \frac{2^n n!}{m!(n-m)!(2n-1)!!} \frac{\Gamma[\frac{1}{2}(1+m+n)]}{\Gamma[\frac{1}{2}(1+m-n)]}~.
\end{align}
One can also solve the recursion relation above and write the mode functions with higher spatial spin as a linear differential operator $\hat{\mathcal{D}}$ acting on the lowest-spatial-spin one,
\begin{align}
	\Phi^{h}_{n,S}(\eta,k)\equiv \hat{\mathcal{D}}_{h,n}(i\eta,k) \Phi^{h}_{|h|,S}(\eta,k)~, \qquad n > |h|~.
\end{align}
This form of the general mode function will become useful later in Section \ref{CCProprPropySect}. Note that for a given $n$, $B_{m,n}$ is only non-zero if $n$ and $m$ differ by an even number. This is because the terms proportional to $B_{m,n}$ come from subtracting traces from $\Phi_{\eta \cdots \eta i_{1} \cdots i_{n}}$. The factor of $i$ in this expression comes from converting to momentum space, and no additional factors of $i$ appear since here the coefficients of all polarisation tensors are taken to be real. The solution to \eqref{CCModeEoM} with Bunch-Davies initial conditions is 
\begin{align} \label{ModeFunctionCC}
	\Phi_{|h|,S}^{h} = e^{i \pi (\nu_{S} + 1/2)/2} Z_{S}^{h}(- k \eta)^{3/2 - |h|} H_{\nu_{S}}^{(1)}(- k \eta)~,
\end{align}
with
\begin{align}\label{coeff_Zhs}
	\left(Z_{S}^{h} \right)^2 = \frac{\pi}{4} \frac{1}{k} \left( \frac{k}{H} \right)^{2S-2} \frac{[(2 |h| -1)!!]^2 S!(S-|h|)!}{(2S-1)!! (S+ |h|)!} \frac{\Gamma(\frac{1}{2} + |h| + \nu_{S})\Gamma(\frac{1}{2} + |h| - \nu_{S})}{\Gamma(\frac{1}{2} + S + \nu_{S})\Gamma(\frac{1}{2} + S - \nu_{S})}~,
\end{align}
which is again fixed by demanding that we have the usual commutation relations \cite{Lee:2016vti}, and 
\begin{align}
	\nu_{S} = \sqrt{\left(S - \frac{1}{2} \right)^2 - \frac{m^2}{H^2}}~. \label{CCdimlessMass}
\end{align}
As we saw explicitly for $S=1$, the mode functions with $n > |h|$ will be given by a sum of Hankel functions.

The different solutions fall into different classes depending on the mass of the field. The \textit{principle series} corresponds to heavy masses with $\text{Re} ~  \nu_{S} = 0$, i.e. $m^2 \geq H^2 (S - 1/2)^2$. The \textit{complementary series} corresponds to light masses where $\text{Im} ~ \nu_{S} = 0$, however, the Higuchi bound sets a lower bound on the mass such that the theory remains unitary.\footnote{For $S \geq 2$ the necessity of the Higuchi bound can be seen at the level of the equations of motion once the mode functions have been decoupled where it ensures that the mass term contributions do not become tachyonic.} The complementary series is then defined by $S(S-1) < m^2 /H^2 < (S - 1/2)^2$. Finally, we have the discrete series for which $m^2 = H^2[S(S-1)  - T(T-1)]$ for $S,T = 0,1,2, \cdots$, with $T \leq S$. In these cases there is an additional gauge symmetry that reduces the number of propagating degrees of freedom and corresponds to partial masslessness \cite{Deser:2001pe,Deser:2001us}. We will discuss the corresponding power spectra and propagators in the following section.

\subsection{Mass parameter comparison}

\begin{figure}[ht]
\centering
\includegraphics[width=0.8\textwidth]{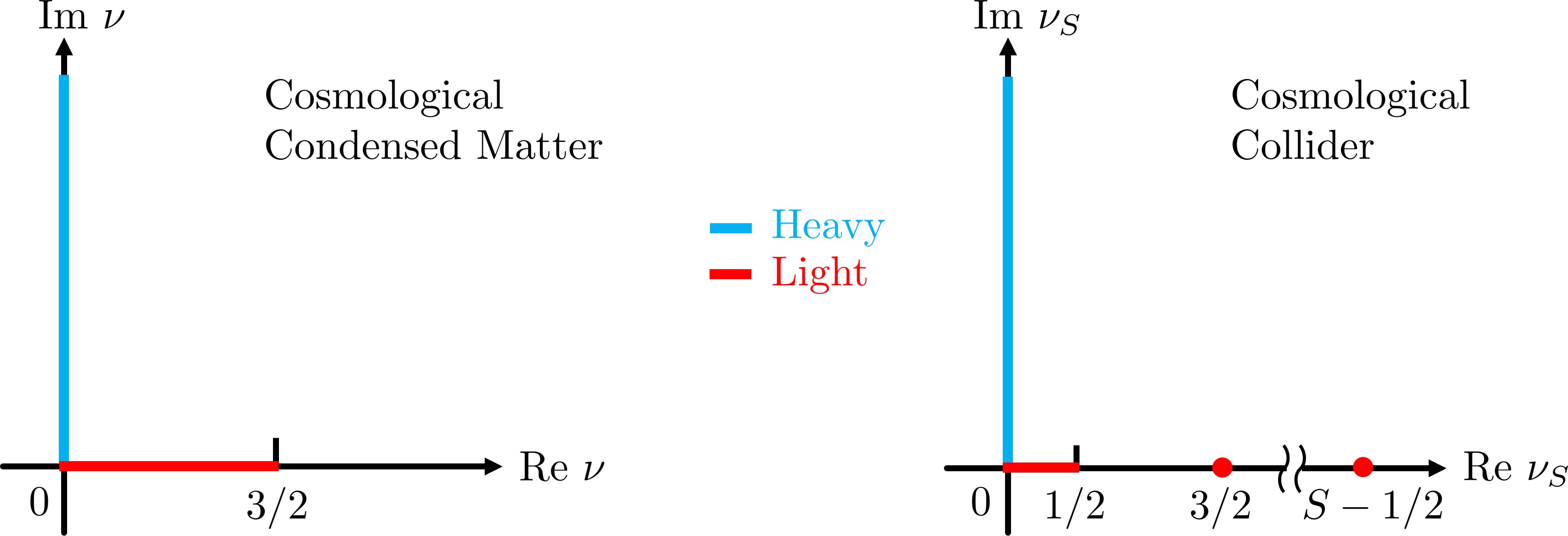}\\
    \caption{The dimensionless mass parameters $\nu/ \nu_S$ for light fields and heavy fields in CCM (left) and CC (right) scenarios.}\label{CCMvsCCmass}
\end{figure}

As emphasised in \cite{Bordin:2018pca}, in the CCM case the masses of the fields have a wider range as they are not constrained by the Higuchi bound. For comparison, in Figure \ref{CCMvsCCmass} we show the distribution of the dimensionless mass parameter on the complex plane for light/heavy fields in both the CCM and the CC scenarios. From \eqref{CCMeom} we see that for light fields in the CCM scenario $\nu$ cannot exceed the massless boundary of $\nu = 3/2$, while for heavy fields we can keep increasing the mass to keep increasing the value of $\text{Im} ~ \nu$. Similarly, in the CC scenario we see from \eqref{CCdimlessMass} that $\text{Im} ~ \nu_S$ continues to increase as we increase the mass while in the principle series. For the complementary series the mass parameter can take values $0 < \nu_S < 1/2$, as dictated by the Higuchi bound. The discrete series, which also lines along the real line in the complex plane of $\nu_S$, corresponds to isolated points and with masses that we will also refer to as ``light". 

\section{General properties of Wick-rotated propagators}\label{PropagatorPropertySection}
We now move on to the propagators that are required to perturbatively compute wavefunction coefficients and cosmological correlators. For each field we have a bulk-boundary propagator for external lines in Feynman diagrams, and a bulk-bulk propagator for internal lines. In this section we will discuss some important properties of the bulk-bulk propagators of massive spinning fields: we will show that after Wick rotation, the time-ordered parts are always purely real regardless of the mass and spin, while for light fields i.e. those in the complementary series, the full bulk-bulk propagator is purely real. To derive these properties we consider the CCM and CC cases separately since the proofs are slightly different in the two cases. For details on the Feynman rules for wavefunction calculations we refer the reader to \cite{COT}. 


\subsection{Cosmological condensed matter physics}\label{CCMPsubsection}
We begin with the power spectrum of massive spinning fields introduced in Section \ref{CCMmodefunctions}. The late-time two-point function is
\begin{align}
	\langle \sigma_{i_{1} \cdots i_{S}}(\eta_{0}, \bfk) \sigma_{j_{1} \cdots j_{S}}(\eta_{0}, -\bfk) \rangle' = \sum_{h = -S}^{S} P^{(h)}_{\sigma}(\eta_{0},k) {\rm{e}}_{i_{1} \cdots i_{S}}^{(h)}(\bfk) {\rm{e}}_{j_{1} \cdots j_{S}}^{(h)}(-\bfk)~,
\end{align}
where 
\begin{align}
	P_{\sigma}^{(h)}(\eta_{0},k) &\equiv \sigma_{h}(\eta_{0},k)\sigma^{\star}_{h}(\eta_{0},k) \\ 
	& =  e^{-\pi \tilde{\kappa}}  \frac{ H^2 \eta_{0}^2}{2 c_{h,S} k}  W_{i \tilde{\kappa}, \nu}(2 i c_{h,S} k \eta_{0})W_{-i \tilde{\kappa}, \nu}(-2 i c_{h,S} k \eta_{0})~.
\end{align}
This expression is valid for both light and heavy fields (i.e. for both purely real and purely imaginary $\nu$), and we have used $[W_{a,b}(z)]^{\star} = W_{a^{\star},b^{\star}}(z^{\star})$ and the symmetry property $W_{a,-b}(z) = W_{a,b}(z)$. The power spectrum of each helicity mode is different due to the $c_{h,S}$ and $\tilde{\kappa}$ dependence. Parity violation is then encoded in the asymmetry between opposite helicities. It can be further shown that the IR expansion of the power spectrum contains enhanced oscillations in $\eta_0$ due to particle production assisted by the chemical potential, which leads to lifted cosmological collider signals (see, e.g. \cite{Tong:2022cdz}). The bulk-boundary propagator of a given helicity mode, which we will denote as $K_{\sigma}^{(h)}(\eta, k)$, should satisfy the equation of motion \eqref{CCMeom} subject to the boundary conditions:
\begin{align}
	\lim_{\eta \rightarrow \eta_{0}}K_{\sigma}^{(h)}(\eta, k) = 1, \qquad \lim_{\eta \rightarrow -\infty(1 - i \epsilon)}K_{\sigma}^{(h)}(\eta, k) = 0~.
\end{align}
The first condition simply requires us to add an appropriate normalisation factor, while the second condition requires the bulk-boundary propagator to be fixed by $\sigma^{\star}_{h}(\eta, k)$ since this is the negative frequency solution that vanishes in the far past and projects onto the Bunch-Davies vacuum. We can therefore write the \textit{indexed} bulk-boundary propagator
\begin{align} \label{BulktoBoundaryCCM}
	K_{i_{1} \cdots i_{S} \,j_{1} \cdots j_{S}}(\eta, \bfk) = \frac{1}{S!}\sum_{h=-S}^{S} K_{\sigma}^{(h)}(\eta, k)  {\rm{e}}^{(h)}_{i_{1} \cdots i_{S}}(\bfk){\rm{e}}^{(h)}_{j_{1} \cdots j_{S}}(-\bfk)~,
\end{align}
as a linear combination of the \textit{helical} bulk-boundary propagator
\begin{align}
	K_{\sigma}^{(h)}(\eta, k) = \frac{\sigma^{\star}_{h}(\eta, k)}{\sigma^{\star}_{h}(\eta_{0}, k)}~.
\end{align}
The factor of ${\rm{e}}^{(h)}_{i_{1} \cdots i_{S}}(\bfk){\rm{e}}^{(h)}_{j_{1} \cdots j_{S}}(-\bfk)$ in \eqref{BulktoBoundaryCCM} ensures that the different helicity modes remain decoupled during propagation, and the factor of $1/S!$ follows from the normalisation \eqref{PolNorm}. The bulk-bulk propagator for each helicity mode, which we denote as the helical bulk-bulk propagator $G_{\sigma}^{(h)}(\eta_{1}, \eta_{2},k)$, satisfies
\begin{align}
	\left(\eta_{1}^2 \frac{\partial^2}{\partial \eta_{1}^2} - 2 \eta_{1}  \frac{\partial}{\partial \eta_{1}} + c_{h,S}^2 k^2 \eta_{1}^2 + \frac{m^2}{H^2} - \frac{2 h \kappa}{H} k \eta_{1} \right) G_{\sigma}^{(h)}(\eta_{1}, \eta_{2},k) = - i H^2 \eta_{1}^2 \eta_{2}^2 \delta(\eta_{1}-\eta_{2})~,\label{PropagatorEoMhelicityBasisCCMPPV}
\end{align}
and is subjected to the boundary conditions:
\begin{align}
	\lim_{\eta_1, \eta_2 \rightarrow \eta_{0}}G_{\sigma}^{(h)}(\eta_1,\eta_2,k) = 0~, \qquad \lim_{\eta_1, \eta_2 \rightarrow -\infty(1 - i \epsilon)}G_{\sigma}^{(h)}(\eta_1, \eta_2, k) = 0~.\label{PropagatorBCs}
\end{align}
The second condition again ensures that we project onto the vacuum in the far past while the first ensures that this propagator takes us between two bulk points rather than from a bulk point to the boundary. The solution to this equation is then
\begin{align}
	G_{\sigma}^{(h)}(\eta_{1}, \eta_{2},k) &=[\sigma_{h}(\eta_{1}, k)\sigma^{\star}_{h}(\eta_{2},k) \theta(\eta_{1}-\eta_{2})  + (\eta_{1} \leftrightarrow \eta_{2})] - \frac{\sigma_{h}(\eta_{0},k)}{\sigma^{\star}_{h}(\eta_{0},k)}\sigma^{\star}_{h}(\eta_{1},k)\sigma^{\star}_{h}(\eta_{2},k)~, \label{bulk-bulkSolution} \\
	& = -2 i P_{\sigma}^{(h)}(\eta_{0},k) K_{\sigma}^{(h)}(\eta_{2}, k) ~ \text{Im} K_{\sigma}^{(h)}(\eta_{1}, k ) \theta(\eta_{1} - \eta_{2}) + (\eta_{1} \leftrightarrow \eta_{2})~.
\end{align}
The manifestly time-ordered parts of this expression correspond to the usual Feynman propagator (time-ordered two-point function), while the $\eta_0$-dependent terms are required to satisfy the future boundary condition. We can again dress the helical propagators with polarisation tensors and write the indexed propagator
\begin{align} \label{BulktoBulkCCM}
	G_{i_{1} \cdots i_{S} \,j_{1} \cdots j_{S}}(\eta_1, \eta_2, \bfk) = \sum_{h=-S}^{S} G_{\sigma}^{(h)}(\eta_1, \eta_2,k)  {\rm{e}}^{(h)}_{i_{1} \cdots i_{S}}(\bfk){\rm{e}}^{(h)}_{j_{1} \cdots j_{S}}(-\bfk)~,
\end{align}
which satisfies 
\begin{align}\label{BulktoBulkCCMallh}
	\Bigg[\left(\eta_{1}^2 \frac{\partial^2}{\partial \eta_{1}^2} - 2 \eta_{1}  \frac{\partial}{\partial \eta_{1}} + c^2 \eta_{1}^2 k^2  +  \frac{m^2}{H^2} \right) \delta_{l(i_{1}} \delta _{|m| i_{2}} -  \frac{ 2 i S \kappa  \eta_{1}}{H} k_{n} \epsilon_{n l (i_{1}} \delta_{i_{2} |m|} \nonumber \\  + \delta c^2 \eta_1^2 \left( k_{l} k_{(i_{1}} \delta_{|m| i_{2}} - \frac{S-1}{2S-1}k_{l} k_{m} \delta_{(i_{1}i_{2}} \right) \Bigg]   G_{i_{3} \cdots i_{S})l m j_{1} \cdots j_{S}}&(\eta_1, \eta_2, \bfk) \nonumber \\   = - i H^2 \eta_{1}^2  \eta_{2}^2  \delta(\eta_{1}-\eta_{2})  \sum_{h=-S}^{S}  &{\rm{e}}^{(h)}_{i_{1} \cdots i_{S}}(\bfk) {\rm{e}}^{(h)}_{j_{1} \cdots j_{S}}(-\bfk)~,
\end{align}
where we symmetrise according to e.g. $S_{(ab)} = (S_{ab} + S_{ba})/2$. We would now like to consider the properties of this indexed bulk-bulk propagator after we Wick rotate both time variables by $90^{\circ}$ in the complex plane. We are therefore interested in the properties of $\tilde{G}_{\sigma}^{(h)}(\chi_1, \chi_2, k)$, c.f. \eqref{WickRotation}.

First consider the LHS of \eqref{BulktoBulkCCMallh} where the differential operator that acts on the bulk-bulk propagator is purely real after Wick rotation: the only term that is odd in $\eta_1$ comes with a factor of $i$ which itself comes from converting to momentum space. It is scale invariance of the quadratic action that ensures that this differential operator is real after Wick rotation. Indeed, time derivatives are forced to appear as $\eta \frac{\partial}{\partial \eta}$, while spatial derivatives yield factors of $i \eta \bfk$. We therefore have a real differential operator acting on the indexed propagator. Moving to the RHS, we first note that the polarisation sum over all helicity modes is purely real which follows from the reality of the fields in position space.\footnote{This can be easily checked using our basis of polarisation tensors by noting that $\hat{\mathbf{e}}^{\pm} (\hat\bfk) = [\hat{\mathbf{e}}^{\mp} (\hat\bfk)]^{*}$, c.f. \eqref{PolDirection}.} After Wick rotation we also do not have the factor of $i$ on the RHS since in this ``Euclidean" picture we are computing $\psi \sim e^{-S}$ rather than $\psi \sim e^{i S}$. The RHS is therefore manifestly real after Wick rotation. \textit{We therefore conclude that the time-ordered parts of the Wick-rotated bulk-bulk propagator, which are the parts that are fixed by \eqref{BulktoBulkCCMallh}, are purely real}. This does not imply that the full bulk-bulk propagator is purely real after Wick rotation as this discussion still allows for the possibility of adding complex contributions that satisfy the homogeneous equation. This also does not imply that the Feynman propagator i.e. the manifestly time-ordered parts in \eqref{bulk-bulkSolution} are purely real after Wick rotation; we may have to add factorised terms that satisfy the homogeneous equation to make this reality property manifest (this is precisely what happens as we will see below).\footnote{Note that in simple massless theories the Feynman propagator is indeed real after rotation, see e.g. \cite{WFCtoCorrelators1}, but for massive mode functions things are more involved.} As long as we are considering fields that are real in position space with scale invariant free theories, this property of the bulk-bulk propagator will hold. Let's see this explicitly by working with \eqref{bulk-bulkSolution}.

\paragraph{Light fields} First consider light fields ($\text{Im} ~ \nu = 0$) where it was shown in \cite{Cabass:2022rhr} that when $\tilde{\kappa} = 0$ the manifestly time-ordered parts are not real, but once we add the factorised term as required by the future boundary condition, the \textit{full} bulk-bulk propagator is real. We can check if this remains true when $\tilde{\kappa} \neq 0$. We first note that for light fields the bulk-bulk propagator is independent of $\eta_0$. Indeed, we have
\begin{align} \label{Canceleta0Whittaker}
	\frac{\sigma_{h}(\eta_{0},k)}{\sigma^{\star}_{h}(\eta_{0},k)} = \frac{W_{i \tilde{\kappa}, \nu}(2 i c_{h,S}k \eta_0)}{W_{-i \tilde{\kappa}, \nu}(-2 i c_{h,S}k \eta_0)}  \xrightarrow{\eta_{0} \rightarrow 0} -e^{(\nu + \frac{1}{2})i \pi} \frac{\Gamma\left(\frac{1}{2}+\nu+ i \tilde{\kappa}\right)}{\Gamma\left(\frac{1}{2}+\nu - i\tilde{\kappa}\right)}~.
\end{align}
The bulk-bulk propagator for a given helicity mode can then be written as
\begin{align}
	G_{\sigma}^{(h)}(\eta_1, \eta_2, k) = \frac{H^2 \eta_1 \eta_2}{2 c_{h,S} k} & \frac{\Gamma\left(\frac{1}{2}+\nu + i \tilde{\kappa}\right)}{\Gamma(1+2 \nu)} \nonumber \\ & \times  \left[M_{-i \tilde{\kappa}, \nu}(-2 i c_{h,S}k \eta_1) W_{-i \tilde{\kappa}, \nu}(-2 i c_{h,S}k \eta_2) \theta(\eta_1 - \eta_2) + (\eta_1 \leftrightarrow \eta_2) \right]~, \label{BulktoBulkMandW}
\end{align}
where we have introduced the Whittaker-$M$ function $M_{a,b}(z)$ which is related to $W_{a,b}(z)$ by
\begin{align} \label{WhittakerWandM}
	\frac{1}{\Gamma(1 + 2 \nu)}M_{i \tilde{\kappa}, \nu}(z) = \frac{e^{\pm(i \tilde{\kappa} - \nu - \frac{1}{2}) i \pi}}{\Gamma\left(\frac{1}{2} +\nu + i \tilde{\kappa}\right)} W_{i \tilde{\kappa}, \nu}(z)+\frac{e^{\mp  \pi \tilde{\kappa}}}{\Gamma\left(\frac{1}{2} +\nu - i \tilde{\kappa}\right)} W_{-i \tilde{\kappa}, \nu}(e^{\pm i \pi}z)~.
\end{align}
In order to arrive at \eqref{BulktoBulkMandW} we have used the rotation that does not cross the branch cut of $W_{a,b}(z)$ which lies on the negative real axis. We have also used
\begin{align}
	M_{i \tilde{\kappa}, \nu}(z e^{\pm i \pi}) = \pm i e^{\pm i \pi \nu}M_{-i \tilde{\kappa}, \nu}(z )~, \label{WhittakerMRotation}
\end{align}
to make sure that the two arguments in \eqref{BulktoBulkMandW} have the same sign. Note that \eqref{Canceleta0Whittaker} follows from \eqref{WhittakerWandM} since for small arguments $W_{a,b}(z)$ dominates over $M_{a,b}(z)$ (for light fields). The function $M_{a,b}(z)$ is another solution to the Whittaker differential equation, and also satisfies $[M_{a,b}(z)]^{\star} =M_{a^{\star},b^{\star}}(z^{\star})$. We can now consider Wick rotating \eqref{BulktoBulkMandW}, and to do so we rotate both time variables clockwise to avoid the branch cuts on the negative real axis (recall that $\eta_1, \eta_2 \leq 0$). We then have
\begin{align}
	\tilde{G}_{\sigma}^{(h)}(\chi_1,\chi_2, k) =  - \frac{H^2 \chi_1 \chi_2}{2 c_{h,S} k} & \frac{\Gamma\left(\frac{1}{2}+\nu + i \tilde{\kappa}\right)}{\Gamma(1+2 \nu)} \nonumber \\  & \times  \left[ M_{-i \tilde{\kappa}, \nu}(2 c_{h,S}k \chi_1) W_{-i \tilde{\kappa}, \nu}(2 c_{h,S}k \chi_2) \theta(\chi_2 - \chi_1) \right]  + (\chi_1 \leftrightarrow \chi_2)~,\label{BulktoBulkMandWRotated}
\end{align}
with $\chi_1, \chi_2 \geq 0 $. The structure of the $\theta$-functions can be understood with a simple flat-space toy example. Consider the bulk-bulk propagator \cite{Arkani-Hamed:2017fdk}
\begin{align}
\nonumber G_{\text{flat}}(\eta_1, \eta_2, k) &= \frac{1}{2 k }\left( e^{i k (\eta_2-\eta_1)}\theta(\eta_1 - \eta_2)+e^{i k (\eta_1-\eta_2)}\theta(\eta_2 - \eta_1) - e^{i k (\eta_1 + \eta_2)} \right) \\
 &= \frac{1}{k} \sinh(-i k \eta_1) e^{i k \eta_2} \theta(\eta_1 - \eta_2) + (\eta_1 \leftrightarrow \eta_2) ~,
\end{align}
where the $\theta$-functions ensure convergence in the far past when $\eta = - \infty(1 - i \epsilon)$. Now consider the clockwise rotation we used above. Since we now integrate from $0$ to $+ \infty$, we need the exponential damping to come from the largest of the two variables. We therefore have  
\begin{align}
\tilde{G}_{\text{flat}}(\chi_1, \chi_2, k) =\frac{1}{k} \sinh(k \chi_1) e^{-k \chi_2}\theta(\chi_2 - \chi_1) + (\chi_1 \leftrightarrow \chi_2)~,
\end{align}
as in \cite{WFCtoCorrelators1}. This example captures the important features for our bulk-bulk propagator given that the Whittaker functions are also exponentially damped ($W$)/growing ($M$) for large argument.

We can now use the properties of the Whittaker functions and the $\Gamma$-functions to conclude that taking the complex conjugate of \eqref{BulktoBulkMandWRotated} can be compensated for by sending $\tilde{\kappa} \rightarrow - \tilde{\kappa}$ which is equivalent to $h \rightarrow - h$. We therefore conclude that the bulk-bulk propagator is helically real i.e.
\begin{align}
	\left[\tilde{G}_{\sigma}^{(h)}(\chi_1, \chi_2, k) \right]^{\star} = \tilde{G}_{\sigma}^{(-h)}(\chi_1, \chi_2, k) \qquad \text{(light fields)}~. \label{ComplexConjugateandHelicityFlip}
\end{align}
This very nice relationship between the bulk-bulk propagator of different helicity modes is not quite enough for us to conclude that the full propagator in \eqref{BulktoBulkCCM} is real; we also need a relationship between the product of polarisation tensors of different helicities. The reality of this full polarisation sum implies that
\begin{align} \label{RealityofPolSum}
	[{\rm{e}}^{(h)}_{i_{1} \cdots i_{S}}(\bfk){\rm{e}}^{(h)}_{j_{1} \cdots j_{S}}(-\bfk)]^{\star} = {\rm{e}}^{(-h)}_{i_{1} \cdots i_{S}}(\bfk){\rm{e}}^{(-h)}_{j_{1} \cdots j_{S}}(-\bfk)~,
\end{align}
since contributions with different $|h|$ have a different number of $\bfk$ factors so cannot be related by complex conjugation, while a single contribution from a single helicity has both real and imaginary components. These two relationships therefore allow us to conclude that the full indexed bulk-bulk propagator \eqref{BulktoBulkCCM}, when the field is light, is real after Wick rotation:
\begin{keyeqn}
	\begin{align}
		\left[ \tilde{G}_{i_{1} \cdots i_{S} \,j_{1} \cdots j_{S}}(\chi_1, \chi_2, \bfk) \right]^{\star} = \tilde{G}_{i_{1} \cdots i_{S} \,j_{1} \cdots j_{S}}(\chi_1, \chi_2, \bfk)~, \qquad \text{(light fields, CCM)}.\label{PropRealityLightCCM}
	\end{align}
\end{keyeqn}

\paragraph{Heavy fields} Let's now consider heavy fields where $\text{Re} ~ \nu = 0$. For convenience let us therefore write $\nu = i \mu$ with $\mu > 0$. The primary difference here compared to the light field case is that the $\eta_0$ dependence in the bulk-bulk propagator no longer cancels out. This is perhaps easiest to see in the $\tilde{\kappa} = 0$ limit where the mode functions are Hankel functions. As we send $\eta_0 \rightarrow 0^{-}$, each Hankel function has two comparable oscillating contributions, in contrast to the light field case where they both have one decaying contribution and one growing one, and this ensures that the ratio is time-dependent. For $\tilde{\kappa} \neq 0$ the story is the same. Furthermore, one can easily check that the Wick-rotated time-ordered parts of $G_{\sigma}^{(h)}$ do not satisfy a relation of the form of \eqref{ComplexConjugateandHelicityFlip}, and therefore the Wick-rotated Feynman propagator once we sum over the helicities is not manifestly real. In contrast to the light field case, the factorised terms we add to satisfy the future boundary condition cannot cancel the imaginary parts of the rotated Feynman propagator since they depend on $\eta_0$ while the Feynman part does not. However, as we have already alluded to, it is possible to add and subtract factorised contributions in such a way that we can make the time-ordered parts of the bulk-bulk propagator manifestly real after Wick rotation. 

To see how this can work let us decompose the full bulk-bulk propagator into two parts which we will refer to as the \textit{connected part} $(C)$ and the \textit{factorised part} $(F)$:
\begin{align}
	G_{i_{1} \cdots i_{S} \,j_{1} \cdots j_{S}}(\eta_1, \eta_2, \bfk) = C_{i_{1} \cdots i_{S} \,j_{1} \cdots j_{S}}(\eta_1, \eta_2, \bfk)+F_{i_{1} \cdots i_{S} \,j_{1} \cdots j_{S}}(\eta_1, \eta_2, \bfk)~,
\end{align}
where for a given helicity mode these parts of the propagator are
\begin{align}
	C_{\sigma}^{(h)}(\eta_{1}, \eta_{2},k) &=[ \sigma_{h}(\eta_{1}, k)\sigma^{\star}_{h}(\eta_{2},k) \theta(\eta_{1}-\eta_{2}) + (\eta_{1} \leftrightarrow \eta_{2})] + \Delta G_{\sigma}^{(h)}(\eta_{1}, \eta_{2},k)~, \label{ConnectedHelicalProp} \\
	F_{\sigma}^{(h)}(\eta_{1}, \eta_{2},k) &= - \frac{\sigma_{h}(\eta_{0},k)}{\sigma^{\star}_{h}(\eta_{0},k)}\sigma^{\star}_{h}(\eta_{1},k)\sigma^{\star}_{h}(\eta_{2},k) - \Delta G_{\sigma}^{(h)}(\eta_{1}, \eta_{2},k)~,
\end{align}
where we have added a new contribution to each, $\Delta G_{\sigma}^{(h)}$, in such a way that the full propagator is unchanged. Diagrammatically, we represent this propagator decomposition as 
\begin{align} \label{DecompositionFigure}
	\begin{gathered}
		\includegraphics[width=0.4\textwidth]{CFdecomp}
	\end{gathered}
\end{align}
This correction term must satisfy the homogeneous equation of motion and so is factorised. We would now like to fix $\Delta G_{\sigma}^{(h)}$ such that $C_{\sigma}^{(h)}$ is manifestly real after Wick rotation. Our argument above suggests that such a $\Delta G_{\sigma}^{(h)}$ exists. There are in principle a number of possible structures that $\Delta G_{\sigma}^{(h)}$ can take, but we would like each term in this decomposition of the bulk-bulk propagator to vanish in the far past such that integrals involving $G$ can be split into integrals involving $C$ and $F$ without losing convergence in the far past. Given that the Feynman propagator vanishes in the far past, thanks to the time ordering, $\Delta G_{\sigma}^{(h)}$ must also vanish. We therefore take it to depend on $\sigma_{h}^{\star}$ rather than $\sigma_{h}$. We would also like it to be symmetric under the exchange of $\eta_1$ and $\eta_2$ such that both $C$ and $F$ are symmetric. We therefore take
\begin{align}
	\Delta G_{\sigma}^{(h)}(\eta_{1}, \eta_{2},k) = \mathcal{A}_h\, \sigma_{h}^{\star}(\eta_1, k) \sigma_{h}^{\star}(\eta_2, k)~,
\end{align}
where $\mathcal{A}_h=\mathcal{A}_h(\kappa, \mu)$ is a $\eta_0$-independent constant, and so is distinct from the terms we need to add to satisfy the future boundary condition. We now want to fix $\mathcal{A}_h$ by demanding the connected propagator is helically real after Wick rotation:
\begin{align}
	\left[\tilde{C}_{\sigma}^{(h)}(\chi_{1}, \chi_{2},k)\right]^{\star}=\tilde{C}_{\sigma}^{(-h)}(\chi_{1}, \chi_{2},k)~.\label{ConnectedReality}
\end{align}
This condition, along with \eqref{RealityofPolSum}, ensures that the connected part of the full bulk-bulk propagator is real after Wick rotation i.e.
\begin{keyeqn}
	\begin{align}
		\left[ \tilde{C}_{i_{1} \cdots i_{S} \,j_{1} \cdots j_{S}}(\chi_1, \chi_2, \bfk) \right]^{\star} = \tilde{C}_{i_{1} \cdots i_{S} \,j_{1} \cdots j_{S}}(\chi_1, \chi_2, \bfk)~, \qquad \text{(heavy fields, CCM)}~.\label{PropRealityHeavyCCM}
	\end{align}
\end{keyeqn}
To summarise, we define the helical connected bulk-bulk propagator to satisfy the following conditions:
\begin{enumerate}
	\item $C_{\sigma}^{(h)}$ satisfies the same equation as the bulk-bulk propagator, i.e. it satisfies \eqref{PropagatorEoMhelicityBasisCCMPPV}.
	\item $C_{\sigma}^{(h)}$ vanishes exponentially fast in the far past under the $i\epsilon$-prescription.
	\item $C_{\sigma}^{(h)}$ is helically real after Wick rotation i.e. it satisfies \eqref{ConnectedReality}.
\end{enumerate}
In order to fix $\mathcal{A}_h$ it is wise to first write the connected propagator in terms of $M_{a,b}(z)$ only since its analytic continuation satisfies a simple relation c.f. \eqref{WhittakerMRotation}, compared to that of $W_{a,b}(z)$ c.f. \eqref{WhittakerWandM}. We can eliminate all copies of $W_{a,b}(z)$ using
\begin{align}
	W_{i \tilde{\kappa}, i \mu}(z) = \frac{\Gamma(-2 i \mu)}{\Gamma\left(\frac{1}{2} - i \tilde{\kappa}- i \mu\right)}M_{i \tilde{\kappa}, i \mu}(z)+\frac{\Gamma(2 i \mu)}{\Gamma\left(\frac{1}{2} - i \tilde{\kappa}+  i \mu\right)}M_{i \tilde{\kappa},- i \mu}(z)~,
\end{align}
and use \eqref{WhittakerMRotation} to make sure that all arguments lie on the positive imaginary axis such that we can rotate each time variable by $90^{\circ}$ clockwise to make all arguments lie on the positive real axis. It is then a simple task to demand \eqref{ConnectedReality} and fix $\mathcal{A}_h$. The most general solution of $\mathcal{A}_h$ is derived in detail in Appendix \ref{GeneralDeltaGhAppendix}. In short, we find
\begin{keyeqn}
	\begin{align}
		\mathcal{A}_h(\kappa, \mu) = \frac{i \pi \sech(\pi \tilde{\kappa})}{\Gamma\left(\frac{1}{2}- i \tilde{\kappa} - i \mu \right)\Gamma\left(\frac{1}{2}- i \tilde{\kappa} + i \mu\right)}~.\label{AhSolution}
	\end{align}
\end{keyeqn}
This now gives us a connected bulk-bulk propagator that is manifestly real after we Wick rotate, which we will use in Section \ref{RandFSection} to prove the reality of total-energy poles, and a factorised bulk-bulk propagator which we will use in Section \ref{ExactTrispectraSection} to compute exact parity-odd trispectra.\footnote{In \cite{Tong:2021wai} and a recent paper \cite{Agui-Salcedo:2023wlq}, a different decomposition of the bulk-bulk propagator is employed where it is split into a retarded propagator and a factorised part. The retarded part enjoys the property that it is purely imaginary (in Lorentzian time), however it does not vanish in the far past. Our connected propagator is not imaginary in Lorentzian time (it is complex), rather it is real in Euclidean time and vanishes in the far past. It would be interesting to understand the results we will derive in this paper using the retarded propagator rather than the connected one. We expect the fact that the contribution to wavefunction coefficients from the retarded propagator is an even function in the exchanged energy, as shown in \cite{Agui-Salcedo:2023wlq}, to play a crucial role in such an analysis. We thank Scott Melville for discussions on these points.} 


\subsection{Cosmological collider physics}\label{CCProprPropySect}

We now turn our attention back to the cosmological collider physics set-up and derive properties of the bulk-bulk propagator. We do not find it necessary to discuss the power spectra and bulk-boundary propagators here since they are not required for our needs and details can be found in \cite{Lee:2016vti}. The full propagator with covariant indices would naturally be written as $G_{\mu_{1} \cdots \mu_{S} \,\nu_{1} \cdots \nu_{S}}(\eta_1, \eta_2, \bfk)$ but as we did when we discussed the mode functions, we will consider a $3+1$ decomposition and consider the properties of propagators with spatial indices:
\begin{align}
	G_{i_{1} \cdots i_{n}  \,j_{1} \cdots j_{n}}(\eta_1, \eta_2, \bfk) \equiv G_{\eta \cdots \eta i_{1} \cdots i_{n} \,\eta \cdots \eta j_{1} \cdots j_{n}}(\eta_1, \eta_2, \bfk)~,
\end{align}
where as before $ 0 \leq n \leq S$. We can further decompose into helicities and write 
\begin{align}
	G_{i_{1} \cdots i_{n}  \,j_{1} \cdots j_{n}}(\eta_1, \eta_2, \bfk) = \sum_{h=-n}^{n} G_{n,S}^{h}(\eta_1, \eta_2, k) {\mathfrak{e}}^{(h)}_{i_{1} \cdots i_{n}}(\bfk) \left[{\mathfrak{e}}^{(h)}_{j_{1} \cdots j_{n}}(\bfk) \right]^{\star}~.
\end{align}
For a given $n$, the mode functions with the same $|h|$ are equivalent since here we do not consider parity-violation. Then, given that the combination
\begin{align}
	{\mathfrak{e}}^{(h)}_{i_{1} \cdots i_{n}}(\bfk) \left[{\mathfrak{e}}^{(h)}_{j_{1} \cdots j_{n}}(\bfk)\right]^{\star} + {\mathfrak{e}}^{(-h)}_{i_{1} \cdots i_{n}}(\bfk) \left[{\mathfrak{e}}^{(-h)}_{j_{1} \cdots j_{n}}(\bfk)\right]^{\star}~,
\end{align}
is real due to \eqref{PolStarCC}, we only need to consider the reality properties of $G_{n,S}^{h}(\eta_1, \eta_2, k)$. As we did above, we will consider light and heavy fields separately. 

\paragraph{Light fields} First consider light fields, i.e. those in the complementary series, and the modes with $n = |h|$ where the solution to the homogeneous equation of motion is given by \eqref{ModeFunctionCC}. Given that this solution involves a single Hankel function, the properties of $G_{|h|,S}^{h}(\eta_1, \eta_2, k)$ will be very similar to what we encountered in the CCM case but with $\tilde{\kappa} = 0$. Indeed, for light fields we therefore already know that this propagator is purely real after Wick rotation. In any case let us show this explicitly, following \cite{Cabass:2022rhr}, since this will allow us to easily see how to extend the proof to the other modes with the same helicity, but with $n > |h|$. We have
\begin{align}
	G_{|h|,S}^{h}(\eta_1, \eta_2, k) = \left(Z_{S}^{|h|} \right)^2 k^{3-2 |h|} &(\eta_1 \eta_2)^{3/2-|h|}H_{\nu_{S}}^{(1)}(- k \eta_1)H_{\nu_{S}}^{(2)}(- k \eta_2) \theta(\eta_1-\eta_2) + (\eta_1 \leftrightarrow \eta_2) \nonumber \\
	& + \left(Z_{S}^{|h|} \right)^2k^{3-2 |h|} (\eta_1 \eta_2)^{3/2-|h|}H_{\nu_{S}}^{(2)}(- k \eta_1)H_{\nu_{S}}^{(2)}(- k \eta_2)~,
\end{align}
where as usual the term on the second line is there to ensure that we satisfy the future boundary condition of the bulk-bulk propagator, and the $\eta_0$ dependence drops out of this term since we are considering light fields. It is simple to see that the second term ensures that we satisfy Dirichlet boundary conditions since for light fields $H^{(1)}_{\nu_{S}}(- k \eta_0) \rightarrow -H^{(2)}_{\nu_{S}}(- k \eta_0)$. We can write this expression more compactly as 
\begin{align}
	G_{|h|,S}^{h}(\eta_1, \eta_2, k) =2  \left(Z_{S}^{|h|} \right)^2 k^{3-2 |h|} &(\eta_1 \eta_2)^{3/2-|h|} J_{\nu_{S}}(- k \eta_1)H_{\nu_{S}}^{(2)}(- k \eta_2)\theta(\eta_1-\eta_2) + (\eta_1 \leftrightarrow \eta_2)~,
\end{align}
where $J_{\nu_S}(z)$ is the Bessel function of the first kind. We can now make use of the following integral representations:
\begin{align}
	H_{\nu_S}^{(2)}(z) = - \frac{e^{\frac{1}{2} \nu_S \pi i}}{\pi i} \int_{- \infty}^{\infty} dt e^{-i z \cosh t - \nu_S t}~, \qquad J_{\nu_S}(z) = \frac{2^{1-\nu_S}z^{\nu_S}}{\sqrt{\pi} \Gamma(\frac{1}{2}+ \nu_S)} \int_{0}^{1} dt (1-t^2)^{\nu_S-\frac{1}{2}} \cos z t~,
\end{align}
which are respectively valid for $- \pi < \text{ph} ~ z <0$ and $\text{Re} ~ \nu_S > -\frac{1}{2}$, to conclude that 
\begin{align}
	J_{\nu_{S}}(-ik\chi_1)H^{(2)}_{\nu_{S}}(- ik\chi_2) = e^{-\frac{1}{2} \nu_{S} \pi i} \times (\text{Real}) \times i \times e^{\frac{1}{2} \nu_{S} \pi i} \times (\text{Real}) = i \times (\text{Real})~,
\end{align}
where we have used $\text{Im} ~ \nu_{S} = 0$. We also have 
\begin{align}
	(-\chi_1 \chi_2)^{3/2-|h|} = i \times (\text{Real})~.
\end{align}
We therefore conclude that the Wick-rotated propagator is purely real i.e. 
\begin{align}
	\left[\tilde{G}_{|h|,S}^{h}(\chi_1, \chi_2, k) \right]^{*} = \tilde{G}_{|h|,S}^{h}(\chi_1, \chi_2, k)~.
\end{align}
Given this result we would immediately expect this property to hold for the other modes too since ultimately they form a single multiplet. Let's see this explicitly by considering modes with the same helicity but with $n > |h|$. The general form of the bulk-bulk propagator is
\begin{align}
	G_{n,S}^{h}(\eta_1,\eta_2,k) = \,\Phi_{n,S}^{h}(\eta_1,k)\Phi_{n,S}^{h \star}(\eta_2,k) & \theta(\eta_1 - \eta_2) + (\eta_1 \leftrightarrow \eta_2) \nonumber \\ & - \frac{\Phi_{n,S}^{h}(\eta_0,k)}{\Phi_{n,S}^{ \star}(\eta_0,k)}\Phi_{n,S}^{h \star}(\eta_1,k) \Phi_{n,S}^{h \star}(\eta_2,k)~, 
\end{align}
where the mode functions $\Phi_{n,S}^{h}$ are related to $\Phi_{|h|,S}^{h}$ by \eqref{ModeFunctionCCDifferentialOperator}. Indeed, we can use this relationship iteratively to write 
\begin{align}
	\Phi_{n,S}^{h}(\eta, k) = \hat{\mathcal{D}}_{h,n}(i \eta, k)\Phi_{|h|,S}^{h}(\eta, k)~, \qquad n > |h|~,
\end{align}
where $\hat{\mathcal{D}}_{h,n}(x, k)$ are real differential operators in $x$ with $k$-dependent coefficients. It follows that $\hat{\mathcal{D}}_{h,n}(i \eta, k)$ are purely real after Wick rotation. Furthermore, they are either purely even in $x$ and therefore purely real when written in terms of $\eta$ (if $n-|h|$ is even) or purely odd in $x$ and therefore purely imaginary when written in terms of $\eta$ (if $n - |h|$ is odd). This final property follows from the fact that the $B_{m,n}$ in \eqref{ModeFunctionCCDifferentialOperator} are real and only non-zero when $n$ and $m$ differ by an even number. We can then write
\begin{align}
	G_{n,S}^{h}(\eta_1,\eta_2,k) = (-1)^{n-|h|} 2 & \left(Z_{S}^{|h|} \right)^2  k^{3 - 2|h|}\hat{\mathcal{D}}_{h,n}(i \eta_1, k)[(-\eta_1)^{3/2-|h|} J_{\nu_{S}}(- k \eta_1)] \nonumber \\ &\times \hat{\mathcal{D}}_{h,n}(i \eta_2, k)[(-\eta_2)^{3/2-|h|} H_{\nu_{S}}^{(2)}(- k \eta_2)] \theta(\eta_1- \eta_2) + (\eta_1 \leftrightarrow \eta_2)~,
\end{align}
where we have used
\begin{align}
	\frac{\hat{\mathcal{D}}_{h,n}(i \eta, k)[(-\eta)^{3/2-|h|} H^{(1)}_{\nu_{S}}(- k \eta)]}{\hat{\mathcal{D}}^{\star}_{h,n}(i \eta, k)[(-\eta)^{3/2-|h|} H^{(2)}_{\nu_{S}}(- k \eta)]} \xrightarrow{\eta \rightarrow 0} (-1)^{n+1-|h|}~. \label{Eta0CancelCC}
\end{align}
We can then use the integral representations above and the fact that $\hat{\mathcal{D}}_{h,n}(i \eta, k)$ is always real after Wick rotation to conclude that this bulk-bulk propagator is real after Wick rotation. We have therefore shown that for light fields
\begin{align}
	\left[\tilde{G}_{i_{1} \cdots i_{n} \,j_{1} \cdots j_{n}}(\chi_1, \chi_2, \bfk) \right]^{\star} = \tilde{G}_{i_{1} \cdots i_{n} \,j_{1} \cdots j_{n}}(\chi_1, \chi_2, \bfk)~, \qquad 
\end{align}
and since this holds for all $0 \leq n \leq S$ we conclude that
\begin{keyeqn}
	\begin{align}
		\left[\tilde{G}_{\mu_{1} \cdots \mu_{S} \,\nu_{1} \cdots \nu_{S}}(\chi_1, \chi_2, \bfk) \right]^{\star} = \tilde{G}_{\mu_{1} \cdots \mu_{S} \,\nu_{1} \cdots \nu_{S}}(\chi_1, \chi_2, \bfk)~,  \qquad \text{(light fields, CC)}~.\label{PropRealityLightCC}
	\end{align}
\end{keyeqn}
It is easy to see that the same reality condition holds for partially-massless fields in the discrete series, since the proof above is valid for arbitrary $\nu_S>0$ as long as the pure gauge modes are excluded.

\paragraph{Heavy fields} Now consider heavy fields i.e. those in the principle series, with $\nu_{S} = i \mu_{S}$. As we have seen in the CCM scenario, the full bulk-bulk propagator will not be real after Wick rotation, so instead we add and subtract factorised terms such that the connected part of the bulk-bulk propagator is real after rotation. As we did above, we work with $G_{i_{1} \cdots i_{n} j_{1} \cdots j_{n}}(\eta_1, \eta_2, \bfk)$ and work helicity-by-helicity. For each mode we define the following decomposition of the bulk-bulk propagator: 
\begin{align}
	G_{n,S}^{h}(\eta_1, \eta_2, k) = C_{n,S}^{h}(\eta_1, \eta_2, k)+F_{n,S}^{h}(\eta_1, \eta_2, k)~,
\end{align}
where 
\begin{align}
	C_{n,S}^{h}(\eta_1, \eta_2, k) &= [\Phi_{n,S}^{h}(\eta_1, k)\Phi_{n,S}^{h \star}(\eta_2, k) \theta(\eta_1 - \eta_2) + (\eta_1 \leftrightarrow \eta_2)] + \Delta G_{n,S}^{h}(\eta_1, \eta_2, k)~, \label{ConnectedBBCC} \\
	F_{n,S}^{h}(\eta_1, \eta_2, k) &= -\frac{\Phi_{n,S}^{h}(\eta_0, k)}{\Phi_{n,S}^{h \star}(\eta_0, k)}\Phi_{n,S}^{h \star}(\eta_1, k)\Phi_{n,S}^{h \star}(\eta_2, k) - \Delta G_{n,S}^{h}(\eta_1, \eta_2, k)~.
\end{align}
As before we take $\Delta G_{n,S}^{h}(\eta_1, \eta_2, k)$ to be factorised, to solve the homogeneous equation of motion, and to vanish in the far past. We therefore write
\begin{align}
	\Delta G_{n,S}^{h}(\eta_1, \eta_2, k) = \mathcal{A}_{h,n} \Phi_{n,S}^{h \star}(\eta_1, k)\Phi_{n,S}^{h \star}(\eta_2, k)~,
\end{align}
where $\mathcal{A}_{h,n} = \mathcal{A}_{h,n}(\mu_S)$ is independent of $\eta_0$. We now want to fix $\mathcal{A}_{h,n}$ such that
\begin{align} \label{CCHeavyReal}
	\left[ \tilde{C}_{n,S}^{h}(\chi_1, \chi_2, k) \right]^{\star} = \tilde{C}_{n,S}^{h}(\chi_1, \chi_2, k)~.
\end{align}
For the $n=|h|$ cases, where the mode function is given by a single Hankel function, we can easily read off the necessary form of $\mathcal{A}_{h,h}$ since it is a special case of what we did in the CCM scenario with $\tilde{\kappa} = 0$. We derived the constraint that $\mathcal{A}_{h,h}$ must satisfy, and the solution, in Appendix \ref{GeneralDeltaGhAppendix}. The result is 
\begin{align}
	\mathcal{A}_{h,|h|}(\mu_S) =i\cosh\pi\mu_S~. 
\end{align}
Note that we could also add any purely real correction to $\mathcal{A}_{h,|h|}$ and still satisfy the necessary condition so this choice is the minimal one required to make $\tilde{C}_{|h|,S}^{h}(\chi_1, \chi_2, k)$ real. With this result in hand, it is simple to deduce what we need to add for modes with $n > |h|$. We have 
\begin{align}
	C_{n,S}^{h}(\eta_1, \eta_2, k) &= \{\hat{\mathcal{D}}_{h,n}(i \eta_1,k)[\Phi_{|h|,S}^{h}(\eta_1, k)]  \hat{\mathcal{D}}^{\star}_{h,n}(i \eta_2,k)[\Phi_{|h|,S}^{h \star}(\eta_2, k)] \theta(\eta_1 - \eta_2) + (\eta_1 \leftrightarrow \eta_2) \} \nonumber \\
	& ~~~~~~~~~~~~~~~~~~~~~~ + \mathcal{A}_{h,n} \hat{\mathcal{D}}^{\star}_{h,n}(i \eta_1,k)[\Phi_{|h|,S}^{h \star}(\eta_1, k)] \hat{\mathcal{D}}^{\star}_{h,n}(i \eta_2,k)[\Phi_{|h|,S}^{h \star}(\eta_2, k)]~,
\end{align}
and by using the fact that $\hat{\mathcal{D}}_{h,n}(i \eta,k)$ is purely real for even $n - |h|$, and purely imaginary for odd $n -|h|$, we can write
\begin{align}
	C_{n,S}^{h}(\eta_1, \eta_2, k) = (-1)^{n-|h|} &\hat{\mathcal{D}}_{h,n}(i \eta_1,k) \hat{\mathcal{D}}_{h,n}(i \eta_2,k)[\Phi_{|h|,S}^{h}(\eta_1, k)  \Phi_{|h|,S}^{h \star}(\eta_2, k) \nonumber \\ & + (-1)^{n-|h|} \mathcal{A}_{h,n} \Phi_{|h|,S}^{h \star}(\eta_1, k) \Phi_{|h|,S}^{h \star}(\eta_2, k)] \theta(\eta_1 - \eta_2) + (\eta_1 \leftrightarrow \eta_2)~.
\end{align}
Given that $\hat{\mathcal{D}}_{h,n}(i \eta,k)$ is always real after Wick rotation, we therefore conclude that the choice
\begin{align}
	\mathcal{A}_{h,n}(\mu_S) = (-1)^{n- |h|}i\cosh\pi\mu_S~, 
\end{align}
ensures that \eqref{CCHeavyReal} is satisfied. Again we can add any purely real correction to $\mathcal{A}_{h,n}(\mu_S)$, but this is the minimal solution that is sufficient for our purposes. We therefore conclude that
\begin{align}
	\left[\tilde{C}_{i_{1} \cdots i_{n}j_{1} \cdots j_{n}}(\chi_1, \chi_2, \bfk) \right]^{\star} = \tilde{C}_{i_{1} \cdots i_{n}j_{1} \cdots j_{n}}(\chi_1, \chi_2, \bfk)~, 
\end{align}
and since this holds for all $0 \leq n \leq S$ we conclude that
\begin{keyeqn}
	\begin{align}
		\left[\tilde{C}_{\mu_{1} \cdots \mu_{S} \,\nu_{1} \cdots \nu_{S}}(\chi_1, \chi_2, \bfk) \right]^{\star} = \tilde{C}_{\mu_{1} \cdots \mu_{S} \,\nu_{1} \cdots \nu_{S}}(\chi_1, \chi_2, \bfk)~,  \qquad \text{(heavy fields, CC)}~.\label{PropRealityHeavyCC}
	\end{align}
\end{keyeqn}

We have therefore shown that for the cosmological collider physics set-up, the full bulk-bulk propagator $\tilde G$ is real for light fields, while for heavy fields this is a property of the connected part $\tilde C$ only. We will use these properties in the next section to prove the reality of total-energy poles in wavefunction coefficients with massless external states, and we use $\tilde F$ in Section \ref{ExactTrispectraSection} to compute exact parity-odd trispectra. 

\section{Reality and factorisation}\label{RandFSection}

In this section we unleash the full power of the reality properties of the indexed propagators for general massive fields that we derived in the previous section, and prove a theorem revealing the universal reality of the total-energy singularities in any tree diagrams with external massless scalars (and massless gravitons) for both the CCM and the CC scenarios. More precisely, we will see that in theories involving light fields of arbitrary mass, spin, couplings and chemical potential, the wavefunction coefficient of external massless scalars is always real. This property also holds for an even number of external conformally coupled scalars, and also external massless gravitons as long as we sum over the two helicities. In more general theories involving heavy fields, despite the fact that the full wavefunction coefficient can be complex, the total-energy singularities remain real. Indeed, these singularities come from the \textit{maximally-connected} parts of the wavefunction coefficients (which we define below) and we prove that these parts are purely real. Based on the universal reality of total-energy singularities, we then prove that all parity-odd correlators are necessarily factorised at tree-level, and are free of any total-energy singularities, under the assumption of IR-convergence.

\subsection{Light fields: the wavefunction reality}

\paragraph{Cosmological condensed matter} We start by considering the most general theory of interacting light fields in the CCM scenario. The particle spectrum $\mathscr{L}$ consists of a set of spinning fields $\sigma_{i_1\cdots i_{S_f}}^{f}$ labelled by their flavour $f\in \mathscr{L}$, and each field $\sigma_{i_1\cdots i_{S_f}}^f$ is equipped with an integer spin $S_f=0,1,2,\cdots$, a dimensionless mass parameter $0<\nu_f\leq 3/2$, a sound speed $c_f$, and a chemical potential $\kappa_f$. Among these fields, we will mainly pay attention to the statistics of a ``visible'' massless scalar field $\phi\equiv \sigma^{f=\phi}$ with $S_\phi=0$ and $\nu_{\phi}=3/2$, while allowing for exchanges of any of the other fields. For instance, in inflationary cosmology, $\phi$ is usually associated with the curvature perturbation $\zeta$ (or equivalently, the Goldstone $\pi$ of broken time translation), while $\sigma_{i_1\cdots i_{S_f}}^f$, $f\neq \phi$ are associated with various isocurvature perturbations. In position space, the most general interaction Lagrangian at a vertex $v$ schematically reads
\begin{align}
	\mathcal{L}_{v}= \lambda_v \,a^{4-k_v-l_v}(\eta)\left[\left(\delta_{ij}\right)^{p_v}\left(\epsilon_{ijk}\right)^{q_v}\left(\partial_\eta\right)^{k_v} \left(\partial_i\right)^{l_v} \prod_{f\in \mathscr{L}}\left\{\sigma_{i_1\cdots i_{S_f}}^f(\eta,\mathbf{x})\right\}^{N_{v,f}}\right]_{\text{contract}}\equiv \lambda_v D_{v} \sigma^{N_v}~,\label{LvCCM}
\end{align}
where $D_{v}$ collectively denotes the scale factors and derivatives at the vertex $v$, and $N_v=\sum_{f\in \mathscr{L}}N_{v,f}$. Here $k_v,l_v\geq 0$ are integers counting the number of derivatives and $p_v,q_v,N_{v,f}\geq 0$ are integers counting the spatial indices and number of fields involved in the interaction vertex $v$.\footnote{As an aside, rotational symmetry dictates that all the spatial indices must be contracted, leading to a constraint $2p_v=3q_v+l_v+\sum_{f\in \mathscr{L}}S_f N_{v,f}$.} The number of scale factors is fixed by diffeomorphism invariance (or its global subgroup of de Sitter scale invariance). Note also that the coupling coefficients are real constants in time i.e.
\begin{align}
	\lambda_v^*=\lambda_v~,\quad \partial_\eta \lambda_v=0~.\label{gvReal}
\end{align}
as a consequence of unitarity and de Sitter scale invariance.

Let us now try to compute the wavefunction coefficient $\psi_n$ for $n$ external massless scalars $\phi$. A general tree diagram that contributes to the wavefunction coefficient $\psi_n$ is obtained via contracting the fields in adjacent vertices to form bulk-bulk propagators $G_{i_1\cdots i_{S_f}\,j_1\cdots j_{S_f}}^f$ and massless bulk-boundary propagators $K_{\phi}$, before finally integrating the interaction time $\eta_v$ at the vertices. The bulk-boundary propagator for the massless scalar is
\begin{align} \label{MasslessBulktoBoundary}
	K_{\phi}(\eta, k) = (1 - i c_\phi k \eta)e^{i c_\phi k \eta},
\end{align}
which is crucially purely real after Wick rotation. Schematically, for a tree diagram with $n$ external lines, $V$ interaction vertices and $I$ internal lines, we have
\begin{align}
	\psi_n=\int_{-\infty(1-i\epsilon)}^{0}\left[\,\prod_{v=1}^V  d\eta_v \,i\lambda_v\, D_{v}\right]\left[\,\prod_{e=1}^n K_e\right] \left[\,\prod_{e'=1}^I G_{e'}\right]~.\label{PsiNBeforeWick}
\end{align}
As dictated by the Feynman rules, we have included a factor of $i$ for each vertex. Now notice that the whole integrand of \eqref{PsiNBeforeWick} is analytic in the second quadrant of complex $\eta$-plane,\footnote{Analyticity in the second quadrant of $\eta$-plane is crucial here, since otherwise singularities passed through when deforming the contour would contribute non-trivially. Such analyticity is satisfied by \textit{local} interactions but can be violated by \textit{non-local} interactions, in which case the contribution from these singularities becomes important \cite{Jazayeri:2023kji}.} which allows us to deform the integration contour by performing a Wick rotation
\begin{align}
	\eta=i e^{i\epsilon}\chi~,\quad \epsilon\to0^+~,\label{WickRotationInSect4}
\end{align}
under which the propagators become
\begin{align}
	K_{\phi}(\eta,k)&=\tilde{K}_{\phi}(\chi,k)~,\\
	G^f_{i_1\cdots i_{S_f}\,j_1\cdots j_{S_f}}(\eta_1,\eta_2,\mathbf{k})&=\tilde{G}^f_{i_1\cdots i_{S_f}\,j_1\cdots j_{S_f}}(\chi_1,\chi_2,\mathbf{k})~.
\end{align}
The original integration contour along the negative real axis is thus deformed to that along the positive imaginary axis, together with an arc at infinity. The $i\epsilon$-prescription and the Bunch-Davies initial condition for the propagators guarantee that as $|\eta|\to \infty$, the integrand of \eqref{PsiNBeforeWick} decays exponentially, leaving a vanishing contribution from the arc at infinity. The only non-vanishing contribution now comes from the integral along the positive imaginary axis where $\chi>0$. In momentum space, the vertex derivative operators transform as
\begin{align}
	\nonumber D_v&= a^{4-k_v-l_v}(\eta)\left[\left(\delta_{ij}\right)^{p_v}\left(\epsilon_{ijk}\right)^{q_v}\left(\partial_\eta\right)^{k_v} \left(i \, k_i\right)^{l_v} \right]_{\text{partially contract}}\\
	\nonumber&=i^{-4+k_v+l_v}~ i^{-k_{v}}~ i^{l_v}\times a^{4-k_v-l_v}(\chi)\left[\left(\delta_{ij}\right)^{p_v}\left(\epsilon_{ijk}\right)^{q_v}\left(\partial_\chi\right)^{k_v} \left(k_i\right)^{l_v} \right]_{\text{partially contract}}\\
	&\equiv \tilde{D}_v~.
\end{align}
Thus the wavefunction coefficient becomes
\begin{align}
	\psi_n=(-1)^{V}\int_{0}^{\infty}\left[\,\prod_{v=1}^V   \, d\chi_v \lambda_v \tilde{D}_{v}\right] \left[\,\prod_{e=1}^n \tilde{K}_e\right] \left[\,\prod_{e'=1}^I \tilde{G}_{e'}\right]~.\label{PsiNWickRotated}
\end{align}
Now we evoke the reality property (\ref{PropRealityLightCCM}) of the Wick-rotated bulk-bulk propagator for light fields, namely
\begin{align}
	\tilde{K}^*_{e}&=\tilde{K}_{e}~,\label{KeReal}\\
	\tilde{G}^*_{e'}&=\tilde{G}_{e'}~,\label{GeImg}
\end{align}
together with the reality of the Wick-rotated derivative operator,
\begin{align}
	\tilde{D}_v^*=\tilde{D}_v~, \label{DvReal}
\end{align}
to see that each individual factor in (\ref{PsiNWickRotated}) is purely real. Therefore, combining (\ref{gvReal}), (\ref{DvReal}), (\ref{KeReal}) and (\ref{GeImg}), we finally arrive at
\begin{align}
	\psi_n^*=\psi_n~.
\end{align}
More precisely speaking, however, what we have managed to prove is only the reality of the \textit{integrand} in the perturbative expression of $\psi_n$. In order to make the final logical leap to the reality of the full-fledged $\psi_n$, we need to also ensure the \textit{convergence} of the integral. Since the propagators and the vertices are well-behaved for any finite Euclidean conformal time $0<\chi<\infty$, we only need to check the convergence at the endpoints. In particular, the UV convergence at $\chi\to\infty$ is guaranteed by the time-ordering and the Bunch-Davies initial condition, as mentioned above. On the other hand, the IR convergence at $\chi\to 0$ is not automatic and is generally model-dependent. For instance, a $\lambda\sigma^4$ self-interaction for a massless $\sigma$ field can bring a logarithmic divergence $\ln (-\eta)$ in the IR limit, which after Wick rotation (\ref{WickRotationInSect4}) brings an imaginary factor of $\ln e^{-i(\pi/2)}=-i\pi/2$. An odd number of such $\lambda\sigma^4$ insertions would lead to a complex $\psi_n$ in general. One can also view this as a consequence of scale invariance spontaneously broken by the IR cutoff \cite{Cabass:2022rhr}. Therefore, we will restrict ourselves to the case of IR-convergent interactions and perfect scale invariance. This establishes the reality of $\psi_n$ for a general tree diagram expressed by (\ref{PsiNWickRotated}). Since the total $\psi_n$ is a sum of all possible diagrams with $n$ external $\phi$ lines, the reality property extends to the full $\psi_n$ at tree-level for the CCM scenario.

\paragraph{Cosmological collider} This proof easily generalises to the CC scenario, where the dimensionless mass of the spinning fields (\ref{CCdimlessMass}) takes $0<\nu_f<1/2$ for the complementary series and $\nu_f=1/2+T,T\in\mathbb{N}$ for the discrete series. The bulk-bulk propagators are labelled with covariant indices, (e.g. $G_{\mu_1\cdots\mu_{S_f},\nu_1\cdots\nu_{S_f}}^f$) and the vertices include contractions of all the covariant indices including that of a time-like unit vector $n_\mu$, which allows us to break de Sitter boosts at the level of the interactions as in \cite{Lee:2016vti}. The Lagrangian vertices are then
\begin{align}
	\nonumber\mathcal{L}_{v}&= \sqrt{-g}\,\lambda_v \left[\left(g^{\mu\nu}\right)^{p_v}\left(\varepsilon_{\mu\nu\rho\sigma}\right)^{q_v}\left(n_\mu\right)^{k_v}\left(\nabla_\nu\right)^{l_v} \prod_{f\in \mathscr{L}}\left\{\Phi_{\mu_1\cdots \mu_{S_f}}^f(\eta,\mathbf{x})\right\}^{N_{v,f}}\right]_{\text{contract}}\\
	\nonumber&=\lambda_v \, a^{4-l_v}(\eta)\,\left[\left(\eta^{\mu\nu}\right)^{p_v}\left(\epsilon_{\mu{\nu}\rho\sigma}\right)^{q_v}\left(\bar n_\mu\right)^{k_v}\left(\partial_\nu\right)^{l_v} \prod_{f\in \mathscr{L}}\left\{\bar{\Phi}_{\mu_1\cdots \mu_{S_f}}^f(\eta,\mathbf{x})\right\}^{N_{v,f}}\right]_{\text{contract}}+\mathcal{O}(\partial^{l_v-1})\\
	&\equiv \lambda_v D_{v} \bar{\Phi}^{N_v}+\mathcal{O}(\partial^{l_v-1})~.
\end{align}
Here, $\varepsilon_{\mu\nu\rho\sigma}=\sqrt{-g}\epsilon_{\mu\nu\rho\sigma}$ is the covariant Levi-Civita tensor density. For simplicity, we have introduced rescaled quantities that are related to the original covariant ones by $\bar{n}_\mu=a^{-1} n_\mu$ and $\bar{\Phi}_{\mu_1\cdots\mu_{S_f}}^f=a^{-S_f}\Phi_{\mu_1\cdots\mu_{S_f}}^f$. Note that these rescaled quantities are contracted with the flat metric tensor $\eta^{\mu\nu}$, and $\mathcal{O}(\partial^{l_v-1})$ are terms with fewer derivatives that take an analogous form to $D_v$.\footnote{Expanding the covariant derivative in terms of ordinary derivatives generates new terms proportional to de Sitter connections and curvature tensors, which enjoy simple forms due to the conformal flatness of de Sitter. By spatial diffeomorphism invariance, these terms must also take the form of $\lambda_v D_{v} \bar{\Phi}^{N_v}$ with a different $D_v$ with fewer derivatives.}  Henceforth, it suffices to consider the leading operator $\lambda_v D_{v} \bar{\Phi}^{N_v}$ as a general parametrisation of the interactions. In other words, we have expanded all the covariant interaction operators and re-organised the expansion in powers of ordinary derivatives on fields.

The computation of $\psi_n$ at tree-level is completely analogous to (\ref{PsiNBeforeWick}), with $D_v$ and $G_{e'}$ replaced by their covariant cousins. We have shown in (\ref{PropRealityLightCC}) that the indexed propagator of $\Phi^f$ satisfies the reality condition
\begin{align}
	\left[\tilde{G}_{\mu_1\cdots\mu_{S_f}\,\nu_1\cdots\nu_{S_f}}^f(\chi_1,\chi_2,\mathbf{k})\right]^*=\tilde{G}_{\mu_1\cdots\mu_{S_f}\,\nu_1\cdots\nu_{S_f}}^f(\chi_1,\chi_2,\mathbf{k})~.
\end{align}
Thus the propagator of the rescaled field $\bar{\Phi}^f$ also satisfies the reality condition after Wick rotation
\begin{align}
	&\left[a^{-S_f}(i\chi_1)a^{-S_f}(i\chi_2)\,\tilde{G}_{\mu_1\cdots\mu_{S_f}\,\nu_1\cdots\nu_{S_f}}^f(\chi_1,\chi_2,\mathbf{k})\right]^* =a^{-S_f}(i\chi_1)a^{-S_f}(i\chi_2)\,\tilde{G}_{\mu_1\cdots\mu_{S_f}\,\nu_1\cdots\nu_{S_f}}^f(\chi_1,\chi_2,\mathbf{k})~.
\end{align}
In addition, the vertex derivative operators transforms as
\begin{align}
	\nonumber D_v&=a^{4-l_v}(\eta)\left[\left(\eta^{\mu\nu}\right)^{p_v}\left(\epsilon_{\mu\nu\rho\sigma}\right)^{q_v}\left(\bar{n}_\mu\right)^{k_v}\left(\partial_\eta,i \,k_i\right)^{l_v} \right]_{\text{partially contract}}\\
	\nonumber&=i^{-4+l_v}~ i^{-l_v}\times a^{4-l_v}(\chi)\left[\left(\eta^{\mu\nu}\right)^{p_v}\left(\epsilon_{\mu\nu\rho\sigma}\right)^{q_v}\left(\bar{n}_\mu\right)^{k_v}\left(\partial_\chi,-k_i\right)^{l_v}  \right]_{\text{partially contract}}\\
	&\equiv \tilde{D}_v~,
\end{align}
with the same reality property as in (\ref{DvReal}) i.e.
\begin{align}
	\tilde{D}_v^*=\tilde{D}_v~.
\end{align}
Thus all the ingredients in the perturbative computation of $\psi_n$ transform identically as in the CCM scenario, leading to the same conclusion:

\begin{keythrm}
	\begin{theorem}
		{\rm\textbf{($\psi_n$-reality)}} The tree-level wavefunction coefficient of massless scalar fields is purely real, i.e. $\Im\psi_n=0$, in theories containing an arbitrary number of fields of any light mass, spin, coupling, sound speed and chemical potential, under the assumption of locality, unitarity, scale invariance, IR convergence and a Bunch-Davies vacuum. \label{psiNRealityTheorem}
	\end{theorem}
\end{keythrm}

\paragraph{Discussion} Before we move on to the inclusion of heavy fields, let us make a few remarks:
\begin{enumerate}
	\item[$\bullet$] Although we have chosen a specific massless scalar field $f=\phi$ as the visible sector, and focused on its wavefunction coefficient $\psi_n$, the same proof straightforwardly generalises to multiple massless scalar fields with different flavours,
	\begin{align}
		\Im \psi_{f_1\cdots f_n}=0~,~ \text{with}\quad\nu_{f_1}=\cdots=\nu_{f_n}=3/2~,
	\end{align}
	since their bulk-boundary propagators are all real after Wick rotation, and we did not use any Bose symmetry properties in the proof.
	\item[$\bullet$] Apart from massless scalar external lines, the particle spectrum may include massless spinning fields such as the graviton. Since the external polarisation tensors are complex, the \textit{helical} wavefunction coefficients are in general complex. Thus it is more convenient to work with the \textit{indexed} wavefunction coefficients where the helicities are added together. The indexed bulk-boundary propagator of the massless graviton is 
	\begin{align}
		K_{i_1 i_2 \,j_1 j_2}^\gamma(\eta,\mathbf{k})= \frac{1}{2}\sum_{h=\pm 2} K_{\gamma}(\eta,k) \mathrm{e}^{(h)}_{i_1 i_2}(\mathbf{k}) \mathrm{e}^{(h)}_{j_1 j_2}(-\mathbf{k})~,
	\end{align}
	where 
	\begin{align}
		K_{\gamma}(\eta,k) = (1 - i k \eta)e^{i k \eta}~.
	\end{align}
	This indexed bulk-boundary propagator is purely real after Wick-rotating $\eta$. Thus after going through the same argument as above, we can further extend the $\psi_n$-reality theorem to the indexed wavefunction coefficients with $m$ massless scalars and $n-m$ massless gravitons:
	\begin{align}
		\Im \psi_{m;\underbrace{\scriptstyle{(i_1 i_2)\cdots (j_1 j_2)}}_{n-m}}=0~,\quad 0\leq m\leq n~.
	\end{align}
	
	\item[$\bullet$] We can also consider conformally-coupled scalars $\varphi$ on the external lines where the bulk-boundary propagators are
	\begin{align}
		K_{\varphi}(\eta, k) =  \frac{\eta}{\eta_0}e^{i k (\eta-\eta_0)}~,
	\end{align}
	where the $\eta_0$-dependence ensures that we satisfy the future boundary condition. Clearly this propagator is not real after Wick rotation, instead it is purely imaginary. Our reality theorem then extends to these fields if we have an even number of them on external lines.
    
	\item[$\bullet$] The $\psi_n$-reality automatically implies the reality of the total-energy $k_T$-singularities within. As we shall see in the next subsection, after including heavy fields, the wavefunction coefficient can become complex, but the $k_T$-reality remains true.
\end{enumerate}

\subsection{Adding heavy fields: the total-energy reality}\label{HeavyFieldkTRealitySubSect}

Now let us move on to the case with heavy fields included. We again label the heavy fields by their flavor $f\in \mathscr{H}$, with a dimensionless mass $\mu_f=-i\nu_f>0$. In the CCM scenario, the most general interaction vertex straightforwardly generalises to
\begin{align}
	\mathcal{L}_{v}= \lambda_v \,a^{4-k_v-l_v}(\eta)\left[\left(\delta_{ij}\right)^{p_v}\left(\epsilon_{ijk}\right)^{q_v}\left(\partial_\eta\right)^{k_v} \left(\partial_i\right)^{l_v} \prod_{f\in \mathscr{L}\cup \mathscr{H}}\left\{\sigma_{i_1\cdots i_{S_f}}^f(\eta,\mathbf{x})\right\}^{N_{v,f}}\right]_{\text{contract}}\equiv \lambda_v D_{v} \sigma^{N_v}~.
\end{align}
In contrast to light fields, the Wick-rotated bulk-bulk propagator of heavy fields does not enjoy the reality property by itself. However, as we showed in Section \ref{PropagatorPropertySection}, we can always achieve reality for the connected part of the propagator by adding appropriate solutions of the homogeneous equation of motion, regardless of the mass of the propagating field. This leads us to a decomposition of the indexed bulk-bulk propagator of any mass,
\begin{align}
	G_{i_1\cdots i_{S_f}\,j_1\cdots j_{S_f}}^f(\eta_1,\eta_2,\mathbf{k})=C_{i_1\cdots i_{S_f}\,j_1\cdots j_{S_f}}^f(\eta_1,\eta_2,\mathbf{k})+F_{i_1\cdots i_{S_f}\,j_1\cdots j_{S_f}}^f(\eta_1,\eta_2,\mathbf{k})~,\label{CFDecomp}
\end{align}
and as shown in (\ref{PropRealityHeavyCCM}), after Wick rotation, the connected part
\begin{align}
	C^f_{i_1\cdots i_{S_f} \,j_1\cdots j_{S_f}}(\eta_1,\eta_2,\mathbf{k})=\tilde{C}^f_{i_1\cdots i_{S_f} \,j_1\cdots j_{S_f}}(\chi_1,\chi_2,\mathbf{k})
\end{align}
enjoys the reality property
\begin{align}
	\left[\tilde{C}^f_{i_1\cdots i_{S_f} \,j_1\cdots j_{S_f}}(\chi_1,\chi_2,\mathbf{k})\right]^*=\tilde{C}^f_{i_1\cdots i_{S_f} \,j_1\cdots j_{S_f}}(\chi_1,\chi_2,\mathbf{k})~.
\end{align}
Now the key insight is that the time-ordering $\theta$-functions only appear in the connected part $C$, and the factorised part $F$ is a sum of products of functions of the vertex times. Therefore, in a general tree diagram, one can extract the maximally-connected contribution by isolating the all-$C$ piece after the decomposition (\ref{CFDecomp}),
\begin{align}
	\nonumber\psi_n&=\int_{-\infty(1-i\epsilon)}^{0}\left[\,\prod_{v=0}^V  d\eta_v \,i\lambda_v\, D_{v}\right] \left[\,\prod_{e=1}^n K_e\right] \left[\,\prod_{e'=1}^I G_{e'}\right]\\
	\nonumber&=\int_{-\infty(1-i\epsilon)}^{0}\left[\,\prod_{v=0}^V  d\eta_v \,i\lambda_v\, D_{v}\right] \left[\,\prod_{e=1}^n K_e\right] \left[\,\prod_{e'=1}^I (C_{e'}+F_{e'})\right] \\
	\nonumber&=\int_{-\infty(1-i\epsilon)}^{0}\left[\,\prod_{v=0}^V  d\eta_v \,i\lambda_v\, D_{v}\right] \left[\,\prod_{e=1}^n K_e\right] \left[\,\prod_{e'=1}^I C_{e'}\right]+\text{factorised}\\
	&\equiv \psi_n^C(k_T,\cdots)+\text{factorised}~,
\end{align}
where we have explicitly spelled out the $k_T$ dependence in $\psi_n^C$, with $(\cdots)$ denoting other kinematic variables. For $n=4$ (where we only have a single bulk-bulk propagator) the factorised part here is completely factorised in the sense that there are no $\theta$-functions, while for higher-point coefficients the factorised part can still contain $\theta$-functions but crucially fewer than those contained in the all-$C$ (maximally-connected) piece. The total-energy singularities only arise from this maximally-connected part, whereas the factorised piece can only have functional dependence on $k_T$ that is analytic at $k_T\to 0$. In particular, all the total-energy poles are contained in $\psi_n^C$,
\begin{align}
	\underset{k_T\to 0}\Res\Big(k_T^m \,\psi_n\Big)=\underset{k_T\to 0}\Res\Big(k_T^m \,\psi_n^C\Big)~, \quad m,n\in\mathbb{N}.
\end{align}
Now given that the connected part of the propagator $C$ enjoys the reality property for both light and heavy fields regardless of their mass, spin, sound speed and chemical potential, we can go through the same proof in the previous subsection, and conclude that the maximally-connected piece must be real:
\begin{align}
	\Im\psi_n^C(k_T,\cdots)=0~,\label{ImPsiCIs0}
\end{align}
which immediately implies the reality of all the total-energy poles:
\begin{align}
	\Im \underset{k_T\to 0}\Res\Big(k_T^m \,\psi_n\Big)=0~, \quad m,n\in\mathbb{N}~.
\end{align}
The proof in the CC scenario is analogous to that above after doing the same decomposition and utilizing the reality property of the covariant indexed connected propagator of $\bar{\Phi}^f$ based on (\ref{PropRealityHeavyCC}),
\begin{align}
	\left[a^{-S_f}(i\chi_1)a^{-S_f}(i\chi_2)\,\tilde{C}_{\mu_1\cdots\mu_{S_f} \,\nu_1\cdots\nu_{S_f}}^f(\chi_1,\chi_2,\mathbf{k})\right]^*=a^{-S_f}(i\chi_1)a^{-S_f}(i\chi_2)\,\tilde{C}_{\mu_1\cdots\mu_{S_f} \,\nu_1\cdots\nu_{S_f}}^f(\chi_1,\chi_2,\mathbf{k})~.
\end{align}
Therefore, we conclude with a reality theorem on the $k_T$-poles of tree-level wavefunction coefficients:

\begin{keythrm}
	\begin{theorem}
		{\rm\textbf{($k_T$-reality)}} The maximally-connected piece of a tree-level wavefunction coefficient for massless scalar fields, along with all the total-energy poles therein, is purely real, i.e. $\Im \psi_n^C(k_T,\cdots)=\Im \underset{k_T\to 0}\Res\Big(k_T^m \,\psi_n\Big)=0~,~ m,n\in\mathbb{N}$, in theories containing an arbitrary number of fields of any mass, spin, non-linear couplings, sound speed and chemical potential, under the assumption of locality, unitarity, scale invariance, IR convergence and a Bunch-Davies vacuum.\label{kTRealityTheorem}
	\end{theorem}
\end{keythrm}

\paragraph{Discussion} Notice the $k_T$-reality is concretely established only for $k_T>0$. For $k_T$-poles inside $\psi_n$, the reality of their residues automatically follows from the reality along the positive real axis. However, in a general tree diagram, the $k_T\to 0$ limit may possess singularities other than just poles. For instance, a 4-point exchange diagram with a massive field could contain a logarithmic singularity \cite{CosmoBootstrap1,Goodhew:2021oqg},
\begin{align}
	\lim_{k_T\to 0} \psi_4\sim c_p\times k_T^p \ln k_T~,\quad p\in \mathbb{N}~,
\end{align}
which comes with a branch cut with the branching point $k_T=0$. More generally, one may expect other types of singularities to occur at $k_T=0$, but none of them can lie along the positive real axis which is physically accessible. Then what we can say is that if such singularities are part of a function analytic along a strip $k_T>\epsilon>0$, then the coefficient of such a function must be real. To illustrate the idea, suppose we have a branch-cut singularity and an essential singularity at $k_T=0$,
\begin{align}
	\lim_{k_T\to 0^+} \psi_n\sim c_\alpha \times k_T^\alpha \ln k_T+c_{\beta\gamma}\times \exp\left({-\frac{\gamma}{k_T^{\beta}}}\right)+\cdots~,
\end{align}
then the $k_T$-reality states that $\alpha,\beta,\gamma\in\mathbb{R}$ and $\Im c_\alpha=\Im c_{\beta\gamma}=0$. Notice that the full analytic structure at $k_T\to 0$ should be completely fixed by the perturbative structure at tree-level and the analytic property of the mode functions, and there could be a constraint on the type of allowed total-energy singularities. Hence some of the singularities (for instance, an essential singularity) may not exist at least at tree-level. However, the study of the analytic structure of singularities in the cosmological wavefunction is still in its infancy, and we leave a more detailed analysis to future work.

\begin{figure}[ht]
	\centering
	\includegraphics[width=0.4\textwidth]{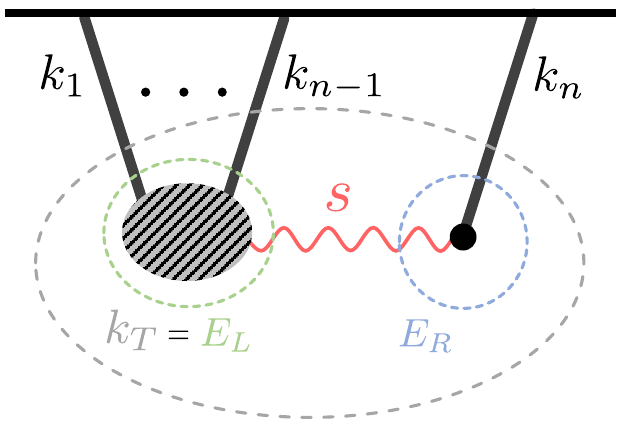}\\
	\caption{In diagrams with a linear-mixing vertex on the external line, the partial-energy singularities (in \textcolor{green3}{light green}) and the total-energy singularities (in \textcolor{gray}{grey}) become indistinguishable because momentum conservation $\mathbf{s}=\mathbf{k}_n$ enforces $s=k_n$ and thus $E_L=k_T$.}\label{ELvskT}
\end{figure}

In the above $k_T$-reality theorem \eqref{kTRealityTheorem} we specified that the couplings should be non-linear. This is because the notion of total-energy poles is more subtle in a theory with linear mixings between different fields. Indeed, with linear mixings momentum conservation can turn a partial-energy pole into a total-energy one. This is illustrated in Figure \ref{ELvskT}. Due to momentum conservation at the linear-mixing vertex, we have $s=k_n$ and therefore $E_L=\sum_{\sf a=1}^{n-1} k_{\sf a}+s=k_T$ i.e. the energy flowing into the left-hand sub-diagram is indistinguishable from the energy flowing into the whole diagram, thus the $E_L$-pole is camouflaged into a $k_T$-pole. Since partial-energy poles are not contained in the maximally connected piece, these fake total-energy poles can be complex in general. We therefore stress that it is only the maximally connected piece and the genuine total-energy poles therein that are purely real by our reality theorem. 

The reality of the $k_T$-singularities (or equivalently, the maximally-connected component $\psi_n^C$) can be physically understood in a few different ways:
\begin{enumerate}
	\item[$\bullet$]  First, from the large-mass EFT perspective, one can choose to integrate out the heavy degrees of freedom in the theory and be left with an EFT of light fields only. This procedure is equivalent to performing a large-mass expansion of the heavy field propagators in a given Feynman diagram. After such an expansion, the heavy field propagators are contracted to contact interactions with a tower of derivative couplings that respect scale invariance. Then by the $\psi_n$-reality theorem (\ref{psiNRealityTheorem}), each contracted diagram is purely real since they only involve light fields. On the other hand, the large-mass expansion preserves part of the maximally-connected wavefunction $\psi_n^C$. This preserved part is free of any partial-energy singularities involving momenta of the expanded heavy field propagators, but includes part of the original total-energy singularities. Hence the reality of these preserved total-energy singularities can be derived from the $\psi_n$-reality after the large-mass expansion, and is consistent with the full $k_T$-reality. 
	
	\item[$\bullet$] Second, from the amplitude perspective, the $k_T$-singularities are generated by the time integral of the connected diagram in the past infinity where $\eta\to-\infty$. Namely, we have
	\begin{align}
		\lim_{k_T\to 0}\psi_n = \lim_{k_T\to 0}\psi_n^C\sim\int_{-\infty} d\eta \,\eta^p e^{i k_T\eta}\sim \frac{\mathcal{A}_n}{k_T^{p+1}}~.
	\end{align}
	In the infinite past, the physical momenta of different modes are much larger than their mass scale as well as the Hubble scale, thus all the fields are effectively massless and interacting in a flat spacetime. The limit $k_T\to 0$ therefore probes the energy-conserving scattering processes in the analytically continued sense \cite{Arkani-Hamed:2017fdk}. In a de Sitter-invariant theory, the residue of the leading $k_T$-pole is expected to recover the on-shell (Lorentz-invariant) scattering amplitude of massless particles $\mathcal{A}_n$ in flat spacetime \cite{Arkani-Hamed:2017fdk,Arkani-Hamed:2018bjr,Benincasa:2019vqr}. However, it is known that a Lorentz-invariant tree amplitude of scalars in flat spacetime is manifestly real unless one of the internal lines hit the mass shell and become disconnected.\footnote{To see this fact, one simply counts the factors of $i$ in a tree diagram: $i^V$ from vertices, $(-i)^I$ from off-shell internal lines, $i^{2n}$ from $2n$ vertex derivatives, and an overall $i$ from convention. This leads to $i^{I-V+1}=1$ for a tree topology with $0=I-V+1$. The imaginary part can only come from $\Im (p^2+m^2-i\epsilon)^{-1}=\pi \delta(p^2+m^2)$, where the diagram is factorised on-shell.} This implies that the leading total-energy pole is real. Here we have also shown that sub-leading ones are also real, which can be argued for given the Manifestly Local Test (MLT) of \cite{MLT} which states that wavefunction coefficients of massless scalars satisfy
	\begin{align}  \label{MLT}
		\frac{\partial \psi_{n} }{\partial k_{a}} \Big|_{k_{a}=0}=0,
	\end{align}
	where the derivative with respect to an external energy is taken while holding all other variables fixed. This equation should be satisfied by both contact and exchange diagrams, and is oblivious to the type of state that is being exchanged. The MLT follows from the simple observation that the bulk-boundary propagator of a massless scalar in de Sitter does not contain a term linear in $k$:
	\begin{align}
		K_{\phi}(\eta, k) = (1- i k \eta)e^{i k \eta} = 1 + \mathcal{O}(k^2).
	\end{align}
	This constraint relates sub-leading total-energy poles to leading ones, and given that it is a real constraint, it implies that sub-leading poles are real once the leading ones are. It can be the case that there are sub-leading total-energy poles that are not tied to the leading ones by this constraint. However, in those cases we expect that the leading pole of such terms also has an amplitude interpretation, coming from an interaction with fewer derivatives, since it has a lower-order pole, and then the argument can be run again.\footnote{We thank Austin Joyce for discussions on these points.}
\end{enumerate}

We end this subsection by pointing out the $k_{T}$-reality also applies to external massless gravitons and an even number of conformally coupled scalars, with the argument mirroring that of light fields which we gave above.

\subsection{Factorising parity-odd correlators}
The universal reality of $k_T$-singularities is not only of theoretical interest, but also provides a powerful tool for the computation of phenomenologically interesting parity-odd correlators. We will show in this subsection that all parity-odd correlators of massless fields must be factorised at tree-level and cannot contain $k_T$-singularities as a consequence of the reality theorems we have just derived. The reason why parity is relevant is simple: for any boundary Hermitian operator $\phi^\dagger(\mathbf{x})=\phi(\mathbf{x})$, its Hermitian conjugate in momentum space is equivalent to a spatial reversal,
\begin{align}
	\phi^\dagger(\mathbf{k})=\left(\int d^3x e^{-i\mathbf{k}\cdot\mathbf{x}} \phi(\mathbf{x})\right)^\dagger=\int d^3x e^{i\mathbf{k}\cdot\mathbf{x}}\phi(\mathbf{x})=\phi(-\mathbf{k})~.
\end{align}
If in addition, $\phi(\mathbf{x})$ is a parity-even scalar (such as the CMB temperature fluctuations), a spatial reversal is equivalent to a parity transformation. Therefore, $n$-point correlation functions in momentum space can always be decomposed into a parity-even part and a parity-odd part:
\begin{align}
	\left\langle\phi(\mathbf{k}_{1}) \cdots \phi(\mathbf{k}_{n})\right\rangle=\left\langle\phi(\mathbf{k}_{1}) \cdots \phi(\mathbf{k}_{n})\right\rangle^{\mathrm{PE}} + \left\langle\phi(\mathbf{k}_{1}) \cdots \phi(\mathbf{k}_{n})\right\rangle^{\mathrm{PO}}~,
\end{align}
with
\begin{align}
	\left\langle\phi(\mathbf{k}_{1}) \cdots \phi(\mathbf{k}_{n})\right\rangle^{\mathrm{PE}}&=\frac{1}{2}\left[\left\langle\phi(\mathbf{k}_{1}) \cdots \phi(\mathbf{k}_{n})\right\rangle+\left\langle\phi(-\mathbf{k}_{1}) \cdots \phi(-\mathbf{k}_{n})\right\rangle\right]~,\\
	\left\langle\phi(\mathbf{k}_{1}) \cdots \phi(\mathbf{k}_{n})\right\rangle^{\mathrm{PO}}&=\frac{1}{2}\left[\left\langle\phi(\mathbf{k}_{1}) \cdots \phi(\mathbf{k}_{n})\right\rangle-\left\langle\phi(-\mathbf{k}_{1}) \cdots \phi(-\mathbf{k}_{n})\right\rangle\right]~.
\end{align}
The Hermiticity of $\phi(\mathbf{x})$ implies the parity-even part is always real while the parity-odd part is always imaginary,
\begin{align}
	\left\langle\phi(\mathbf{k}_{1}) \cdots \phi(\mathbf{k}_{n})\right\rangle^{\mathrm{PE}}&=\Re\left\langle\phi(\mathbf{k}_{1}) \cdots \phi(\mathbf{k}_{n})\right\rangle^{\mathrm{PE}}~,\\
	\left\langle\phi(\mathbf{k}_{1}) \cdots \phi(\mathbf{k}_{n})\right\rangle^{\mathrm{PO}}&=i\, \Im\left\langle\phi(\mathbf{k}_{1}) \cdots \phi(\mathbf{k}_{n})\right\rangle^{\mathrm{PO}}~. \label{POcorrelator}
\end{align}
These boundary correlators are computed by a functional integral of the modulus square of the wavefunction,
\begin{align}
	\left\langle\phi(\mathbf{k}_{1}) \cdots \phi(\mathbf{k}_{n})\right\rangle=\int \mathcal{D}\sigma \, \Big|\Psi[\sigma,\eta_0]\Big|^2 \phi(\mathbf{k}_{1}) \cdots \phi(\mathbf{k}_{n})~,
\end{align}
where $\sigma=\{\sigma_{i_1,\cdots i_{S_f}}^f | f\in\mathcal{L}\cup\mathcal{H}\}|_{\eta=\eta_0}$ collectively denotes all the bulk fields evaluated at the future boundary. The wavefunction exponent is organised as a sum over field products\footnote{We will adopt the notation that $\psi_{n-m;f_1\cdots f_m}$ represents the wavefunction coefficient before the term with at least $n-m$ factors of $\phi$ and $m$ factors of other fields (including $\phi$) labelled by the flavor indices. Note that we also abbreviate $\psi_n=\psi_{n;}$ and $\psi_{f_1\cdots f_n}=\psi_{0;f_1\cdots f_n}$. The same notation will be used for the density matrix diagonals below.}
\begin{align} 
	\nonumber\Psi[\sigma,\eta_0] &= \exp\left[\,\sum_{n=2}^{\infty}\frac{1}{n!} \int_{\mathbf{k}_{1} \cdots \mathbf{k}_{n}} \psi_{f_1\cdots f_n}(\{ k \}, \{ \mathbf{k} \}) (2 \pi)^3 \delta^3 \bigg(\sum_{{\sf a}=1}^n \mathbf{k}_a \bigg)\sigma^{f_1}(\mathbf{k}_{1}) \cdots \sigma^{f_n}(\mathbf{k}_{n}) \right]\\
	&=\exp\left[\,\sum_{n=2}^{\infty}\frac{1}{n!} \int_{\mathbf{k}_{1} \cdots \mathbf{k}_{n}} \psi_{n}(\{ k \}, \{ \mathbf{k} \}) (2 \pi)^3 \delta^3 \bigg(\sum_{{\sf a}=1}^n \mathbf{k}_a \bigg)\phi(\mathbf{k}_{1}) \cdots \phi(\mathbf{k}_{n})+(\cdots) \right]~,
\end{align}
where $(\cdots)$ in the second row denotes terms including fields other than the massless scalar $\phi$. Due to the modulus square, the phase information of the wavefunction $\Psi[\sigma,\eta_0]$ is washed out in correlators, which solely depend on the combination
\begin{align} \label{WavefunctionRho}
	\rho_{f_1\cdots f_n}(\{ k \}, \{ \mathbf{k} \}) \equiv \psi_{f_1\cdots f_n}(\{ k \}, \{ \mathbf{k} \})+\psi_{f_1\cdots f_n}^*(\{ k \}, \{- \mathbf{k} \}),
\end{align}
in the probability distribution functional
\begin{align}
	\Big|\Psi[\sigma,\eta_0]\Big|^2=\exp\left[\,\sum_{n=2}^{\infty}\frac{1}{n!} \int_{\mathbf{k}_{1} \cdots \mathbf{k}_{n}} \rho_{f_1\cdots f_n}(\{ k \}, \{ \mathbf{k} \}) (2 \pi)^3 \delta^3 \bigg(\sum_{{\sf a}=1}^n \mathbf{k}_a \bigg)\sigma^{f_1}(\mathbf{k}_{1}) \cdots \sigma^{f_n}(\mathbf{k}_{n}) \right]~.
\end{align}
The final correlator receives contributions from various partitions of all possible diagrams at tree-level,
\begin{align}
	\nonumber\left\langle\phi(\mathbf{k}_{1}) \cdots \phi(\mathbf{k}_{n})\right\rangle' =\frac{1}{\rho_2^n}\Bigg(\rho_n&+\sum_{m;f} q_{nm}\,\rho_{n-m;f}\,\frac{1}{\rho_{f}}\,\rho_{m;f}\\
	&+\sum_{m,l;f_1,f_2} q_{nml}\,\rho_{n-m-l;f_1}\,\frac{1}{\rho_{f_1}}\,\rho_{l;f_1,f_2}\,\frac{1}{\rho_{f_2}}\rho_{m;f_2}+\cdots\Bigg)~,\label{rhoToCorrelators}
\end{align}
where $q_{nm},q_{nml},\cdots$ are real rational numbers counting the combinatorics of partitions, and momentum conservation is implicit in the individual $\rho$'s. Notice that except the first term, all the other contributions in (\ref{rhoToCorrelators}) are \textit{factorised} in the sense that they do not contain any $k_T$-singularities. The only non-factorisable contribution that contains $k_T$-singularities comes from the maximally-connected part of the first term,
\begin{align}
	\rho_n=\rho_n^C+\text{factorised}~,
\end{align}
with
\begin{align}
	\rho_n^C(\{ k \}, \{ \mathbf{k} \}) =\psi_n^C(\{ k \}, \{ \mathbf{k} \})+\psi_n^{C*}(\{ k \}, \{- \mathbf{k} \})~.
\end{align}
The $k_T$-reality theorem (\ref{kTRealityTheorem}) tells us that the maximally-connected massless wavefunction is always real, i.e. $\psi_n^{C*}=\psi_n^C$. This implies
\begin{align}
	\rho_n^C(\{ k \}, \{ \mathbf{k} \}) =\psi_n^C(\{ k \}, \{ \mathbf{k} \})+\psi_n^{C}(\{ k \}, \{- \mathbf{k} \})~,
\end{align}
i.e. the maximally-connected density matrix is parity-even. We can then check what the consequence of this is for parity-odd $n$-point correlators. We have
\begin{align}
	\nonumber\left\langle\phi(\mathbf{k}_{1}) \cdots \phi(\mathbf{k}_{n})\right\rangle^{\prime\mathrm{PO}}&\equiv\frac{\left\langle\phi(\mathbf{k}_{1}) \cdots \phi(\mathbf{k}_{n})\right\rangle'-\left\langle\phi(-\mathbf{k}_{1}) \cdots \phi(-\mathbf{k}_{n})\right\rangle'}{2}\\
	\nonumber&=\frac{1}{\rho_2^n}\frac{\rho_n^C(\{ k \}, \{ \mathbf{k} \})-\rho_n^C(\{ k \}, \{- \mathbf{k} \})}{2}+\text{factorised}\\
	&=0+\text{factorised}~.
\end{align}
We therefore see that parity-odd correlators of massless scalars are factorised and therefore cannot have genuine total-energy singularities,
\begin{align}
	\lim_{k_T\to 0^+}\frac{d^m}{d k_T^m}\left\langle\phi(\mathbf{k}_{1}) \cdots \phi(\mathbf{k}_{n})\right\rangle^{\prime\mathrm{PO}}=\text{finite}~,\quad m\in \mathbb{N}~.
\end{align}
This also ensures that parity-odd correlators admit a well-defined Taylor expansion around $k_T=0$ (this holds where there are no linear-mixings i.e. when partial-energy singularities can be distinguished from total-energy ones). The proof straightforwardly generalises to the CC scenario, and provides an understanding of why the final parity-odd trispectrum of \cite{Jazayeri:2023kji}, computed in a non-local EFT with the massive spinning field integrated out, is manifestly factorised (we will discuss this further in Section \ref{Example1}). Thus we conclude with the following theorem,
\begin{keythrm}
	\begin{theorem}
		{\rm\textbf{(Parity-odd factorisation)}} The parity-odd part of any tree-level correlator of massless scalar fields is factorised and, in the absence of linear mixings, admits a Taylor expansion around $k_T=0$, in theories containing an arbitrary number of fields of any mass, spin, coupling, sound speed and chemical potential, under the assumptions of locality, unitarity, scale invariance, IR convergence and a Bunch-Davies vacuum.\label{POFactorizationTheorem}
	\end{theorem}
\end{keythrm}

\paragraph{Discussion} Note that despite the subtlety of linear mixings in stating the $k_T$-reality theorem \eqref{kTRealityTheorem}, we have included this possibility in the parity-odd factorisation theorem here. This is because the proof of parity factorisation relies solely on the reality of the maximally-connect wavefunction $\psi_n^C$ but not the $k_T$-poles. The former is always true regardless of linear mixings. Hence the factorisation property follows straightforwardly. The statement about the smoothness near $k_T=0$, however, does depend on the absence of linear mixings. To better illustrate this point, we present an explicit example of a parity-odd trispectrum in a scalar theory with a linear-mixing in Section \ref{Example4}.

Interestingly, although the parity-odd correlator factorises for the exchange of both light fields and heavy fields, they factorise following different routes. For light fields, the whole wavefunction coefficient $\psi_n$ is itself real, leading to
\begin{align}
	\nonumber\rho_n^{\mathrm{PO}}&=\frac{1}{2}\left[\rho_n(\{ k \}, \{ \mathbf{k} \})-\rho_n(\{ k \}, \{ -\mathbf{k} \})\right]\\
	\nonumber&=\frac{1}{2}\left[\psi_n(\{ k \}, \{ \mathbf{k} \})+\psi_n(\{ k \}, \{ -\mathbf{k} \})-\psi_n(\{ k \}, \{ -\mathbf{k} \})-\psi_n(\{ k \}, \{ \mathbf{k} \})\right]\\
	&=0~.
\end{align}
Thus the parity-odd $n$-point correlator sourced by light fields solely receives contributions from lower-point wavefunction coefficients $\psi_{f_1\cdots f_m}$ with $m\leq n-1$. In contrast, when heavy fields are involved, $\psi_n$ is no longer real by itself, and one has to perform the C-F decomposition to isolate the factorised parts of $\psi_n$, which will also contribute to the final parity-odd $n$-point correlator. Alternatively, one can say that for light fields, $F=0$ and there is nothing to isolate away from $\psi_n$. However, such a computational distinction is an artefact of the wavefunction formalism. Namely, the Dirichlet boundary condition at $\eta=\eta_0$ is introduced as an intermediate tool to organise the perturbative expansion. In the absence of IR divergences, the final correlator does not depend on $\eta_0$, and the $\eta_0$ dependence in each contributing piece must cancel out. Yet the Dirichlet boundary condition does not respect the continuity of the mass parameter at $\nu=i\mu=0$, since the IR behaviour of light and heavy fields are different: light fields split into two scaling modes with one dominating over the other in the limit $\eta_0\to 0$:
\begin{align}
	\sigma_h(\eta_0,k)\sim A_h(k,\nu,\tilde{\kappa})(-k\eta_0)^{\frac{3}{2}-\nu}+B_h(k,\nu,\tilde{\kappa})(-k\eta_0)^{\frac{3}{2}+\nu}~.
\end{align}
Heavy fields split into two oscillatory modes with the same damping power, and are equally important as $\eta_0\to 0$:
\begin{align}
	\sigma_h(\eta_0,k)\sim A_h(k,-i\mu,\tilde{\kappa})(-k\eta_0)^{\frac{3}{2}-i\mu}+B_h(k,i\mu,\tilde{\kappa})(-k\eta_0)^{\frac{3}{2}+i\mu}~.
\end{align}
Thus implementing a Dirichlet boundary condition at $\eta_0$ is sensitive to the mass of the bulk fields, which should be artificial. This is because there is nothing physically problematic\footnote{In contrast, the Higuchi bound at $\nu_H=1/2$ \textit{is} problematic in de Sitter-invariant theories due to the loss of unitarity beyond $\nu_H$.} at $\nu=i\mu=0$, and all the physical observables such as the boundary correlator should be continuous across this point. As we shall see in Section \ref{ExactTrispectraSection}, this is indeed the case for parity-odd 4-point correlators. Namely, the correlator for exchanging a heavy field can be directly obtained via the analytic continuation $\nu\to i\mu$ in the final result of light field exchange. To complement this discussion and the proofs we have outlined in this section, in Appendix \ref{InIn} we show how to understand our results in the in-in/Schwinger-Keldysh formalism where the subtlety of the role of $\eta_0$ does not appear.


\section{Exact parity-odd trispectra} \label{ExactTrispectraSection}

In this section we present three examples of parity-odd trispectra, which we are able to compute exactly given our theorem that parity-odd correlators are factorised. Indeed, to arrive at these exact shapes we only need to compute time integrals associated with cubic diagrams, without having to worry about the complicated nested integrals that one usually encounters when computing trispectra. Given these examples, it would be straightforward to extend our methods to more general examples corresponding to other interactions, other spins, etc. 

Before diving into technical details, we first outline the overall algorithm for computing such parity-odd trispectra. Based on the C-F decomposition that isolates the connected propagator $C$ which satisfies helical-reality,
\begin{align}
	\begin{gathered}
		\includegraphics[width=0.4\textwidth]{CFdecomp}
	\end{gathered}~,
\end{align}
we can compute the $s$-channel parity-odd trispectrum as a sum of three terms,
\begin{align}
	\begin{gathered}
		\includegraphics[width=0.8\textwidth]{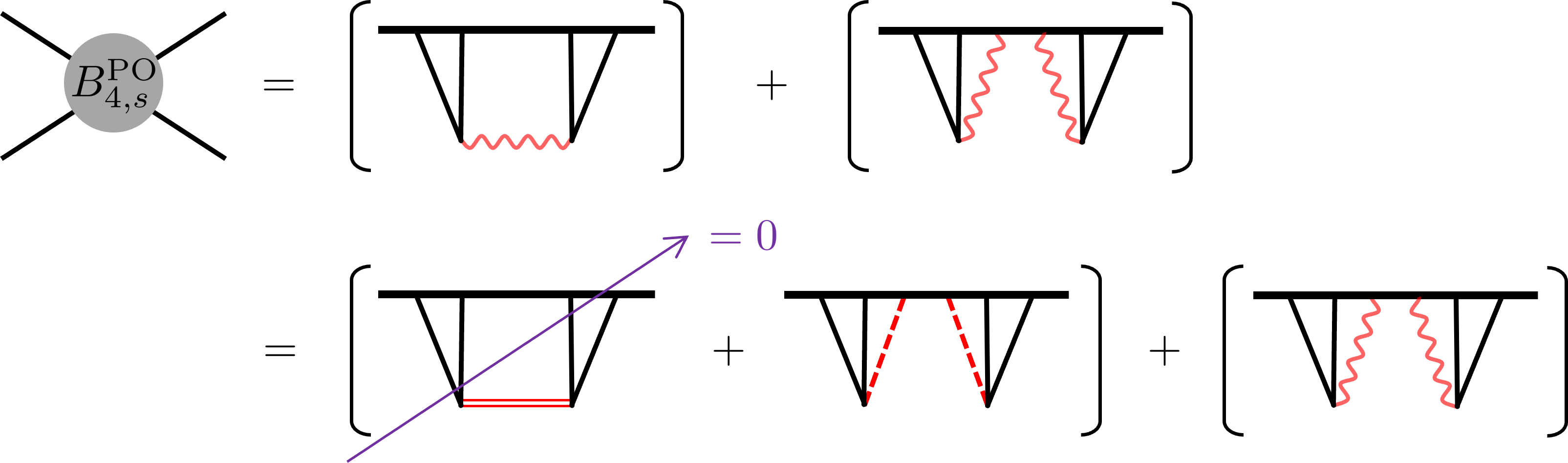}
	\end{gathered}~.
\end{align}
The first term on the second line vanishes due to the reality of the maximally-connected wavefunction $\psi_4^C$. We therefore only need to compute the second and third terms which are products of exactly computable cubic time integrals:
\begin{enumerate}
	\item[$\bullet$] For the exchange of light fields we have $F=0$, and so the second term on the second line above vanishes, leaving only the third term from the product of cubic wavefunction coefficients:
	\begin{align}
		\begin{gathered}
			\includegraphics[width=0.38\textwidth]{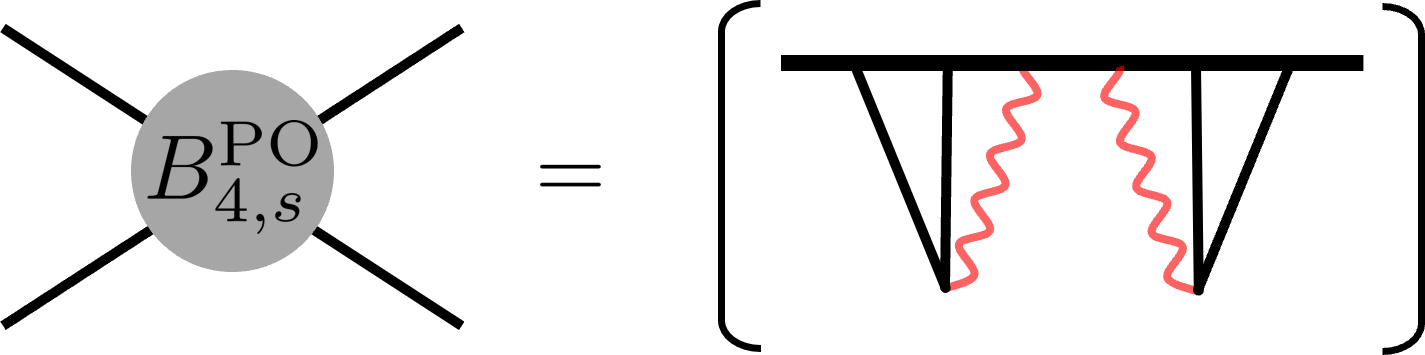}
		\end{gathered}~,\quad\text{(light fields).}\label{BPOLightFormula}
	\end{align}
	
	\item[$\bullet$] For the exchange of heavy fields, the second term is non-zero and so we need to compute both factorised diagrams:
	\begin{align}
		\begin{gathered}
			\includegraphics[width=0.65\textwidth]{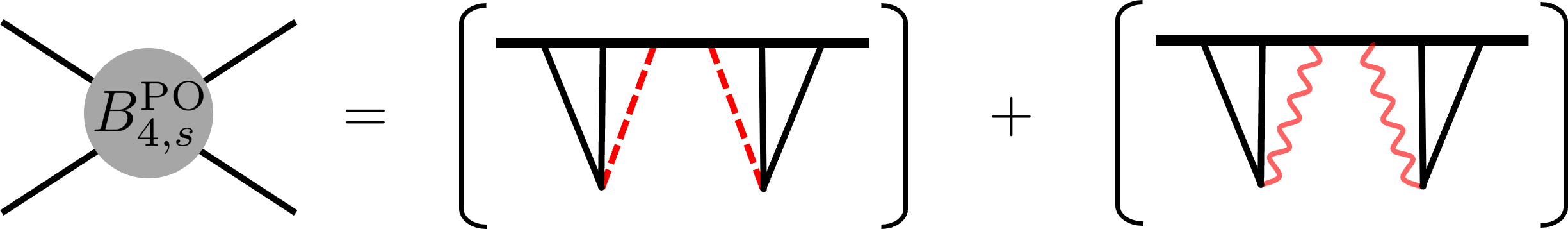}
		\end{gathered}~,\quad\text{(heavy fields).}\label{BPOHeavyFormula}
	\end{align}
 Both diagrams depend on $\eta_0$, and there is an intricate cancellation between the two which ensures that the final result does not depend on this late-time cut-off.
\end{enumerate}
Again we stress that the final results for heavy fields and light fields are related to each other by analytical continuation (as we will see below). 

\subsection{Example 1: spin-1 exchange in CC with chemical potential} \label{Example1}
One particularly interesting example is the exchange of a massive spin-1 field with a chemical potential. In this subsection we consider this case within the CC scenario (although things are essentially the same in the CCM as we will explain). The chemical potential in this case is naturally introduced via a dimension-5 Chern-Simons coupling $\phi F \tilde{F}$ that respects the shift symmetry $\phi \rightarrow \phi+c$. In a recent work \cite{Jazayeri:2023kji}, it has been demonstrated that with the assistance of the chemical potential and a reduced Goldstone sound speed, a large parity-odd trispectrum can be generated. In \cite{Jazayeri:2023kji}, the massive vector field is integrated out, resulting in a non-local EFT description which is organized as a time-derivative expansion. From the perspective of the single-field effective theory, the parity-odd trispectrum emerges from the (spatially) non-local feature of the self-interactions of the inflaton and this non-locality allows the evasion of the no-go theorems of \cite{Cabass:2022rhr,Liu:2019fag}. Utilizing the non-local EFT, one can derive a simple analytical expression that serves as a good approximation to the parity-odd trispectrum and matches well with numerics in certain parameter regimes. In contrast, in this work, we will set out to obtain the \textit{exact} result of the parity-odd trispectrum using our factorisation theorem.\footnote{Notice that for a unit sound speed $c_s=1$, the complete trispectrum has been solved in \cite{Qin:2022fbv}. However, the case with a non-unit sound speed (more specifically, $c_s<1$) has not yet been fully understood in the whole kinematic domain. The main difficulty lies in the analytic continuation beyond the spurious collinear singularities \cite{Jazayeri:2022kjy}. Our work serves as a first complete result in the parity-odd sector for non-unit sound speeds.}

The action of a massive vector field with a chemical potential is
\begin{align} \label{MassiveVector}
	S = \int d^4x \sqrt{-g} \left[ -\frac{1}{4} F_{\mu\nu}^2 -  \frac{m^2}{2} \Phi_\mu^2 + \frac{\phi}{4 \Lambda_c } F_{\mu\nu}\tilde{F}^{\mu\nu} \right]~,
\end{align}
where $F_{\mu\nu} = \nabla_{\mu} \Phi_{\nu} - \nabla_{\nu} \Phi_{\mu}$, $\tilde{F}^{\mu\nu} = \varepsilon^{\mu\nu\alpha\beta}F_{\alpha\beta}$ and $\varepsilon^{\mu\nu\alpha\beta} = \frac{\epsilon^{\mu\nu\alpha\beta}}{\sqrt{-g}}$ is the contravariant Levi-Civita tensor density. The mass term breaks the $U(1)$ gauge symmetry of the spin-$1$ field. As we will explain in more detail below, all of our results also apply in the massless limit corresponding to axion-$U(1)$ gauge field inflation, see e.g. \cite{Turner:1987bw,Garretson:1992vt,Anber:2009ua,Barnaby:2011vw,Barnaby:2010vf,Barnaby:2011qe,Pajer:2013fsa,Domcke:2019lxq,Sorbo:2011rz,Maleknejad:2012fw,Niu:2022fki,Caravano:2021bfn,Caravano:2022epk}, however throughout this subsection we will be more general and keep the mass term.

The inflaton background can be expanded as $\phi=\text{const}+\dot{\phi}_0 t$, where higher order terms are suppressed by slow-roll parameters. The constant term has no dynamical effects due to the shift symmetry of this dimension-5 operator, and the chemical potential $\kappa$ is equal to $\dot{\phi}_0/\Lambda_c$. The equation of motion of this spin-1 field is then
\begin{align}
	\Box \Phi^\nu - \nabla_\mu \nabla^\nu \Phi^\mu - m^2 \Phi^\nu - 2 \kappa\, \varepsilon^{0\nu\alpha\beta} \nabla_{\alpha}\Phi_{\beta} = 0~.
\end{align}
The second term can be eliminate using
\begin{align}
	\nabla_\mu \nabla^\nu \Phi^\mu = \nabla^{\nu} \nabla_{\mu} \Phi^{\mu} + 3 H^2 \Phi^{\nu}~,
\end{align}
and taking the divergence of both sides yields the transverse constraint $\nabla_{\nu}\Phi^{\nu} = 0$. The final equation of motion is then 
\begin{align}
	\left[\Box - (m^2 + 3H^2)\right]\Phi^\nu = 2\kappa\varepsilon^{0\nu\alpha\beta}\nabla_\alpha \Phi_\beta~.
\end{align}
We now convert to momentum space and decompose into the different helicities. We write
\begin{align}
	\Phi_{\mu}(\eta, \bfx) = \sum_{h =-1}^{1} \int_{\mathbf{k}}  \Phi^{h}_{\mu}(\eta, \bfk) e^{i \mathbf{k} \cdot \mathbf{x}}~,
\end{align}
with 
\begin{align}
	& \Phi_{\eta}(\eta, \bfk) = \Phi_{0,1}^{0}(\eta, k)~, \\ 
	& \Phi^{0}_{i}(\eta, \bfk) = \Phi_{1,1}^{0}(\eta, k){\mathfrak{e}}^{0}_{i}(\bfk)~, \qquad \Phi^{\pm 1}_{i}(\eta, \bfk) = \Phi_{1,1}^{\pm 1}(\eta, k){\mathfrak{e}}^{\pm 1}_{i}(\bfk)~.
\end{align}
The equations of motion then decouple for each mode, and only the transverse mode will be affected by the addition of the chemical potential, while the temporal and longitudinal mode remain the same as in Section \ref{sec:massivefields}. Since we ultimately care about the parity-odd contributions, which cannot come from the exchange of $h=0$ modes, let us only focus on the transverse modes which are subject to
\begin{align}
	\Phi_{1,1}^{\pm 1}{}''+(k^2\pm 2 a \kappa k+a^2m^2)\Phi_{1,1}^{\pm 1}=0~,
\end{align}
and the solution to this equation with Bunch-Davies vacuum conditions is given by the Whittaker-$W$ function:
\begin{align}
	\Phi_{1,1}^{h}(\eta,k)=\frac{e^{- \pi \tilde{\kappa}/2}}{\sqrt{2k}}W_{i\tilde{\kappa},\nu}(2ik\eta)~,
\end{align}
with $\tilde{\kappa}\equiv h\kappa/{H}$. This is familiar from our discussion of the cosmological condensed matter scenario but now the mass parameter is different, as is the scaling dimension of the field. Here we have
\begin{align}
\nu=\sqrt{1/4-m^2/H^2}~.
\end{align}
Given that only helicity states with $\pm 1$ are relevant here (and they have the same speed), we have set the speed of sound of the internal massive field to unity, and incorporated a dependence on the speed of sound of the external Goldstone boson $c_s$. This convention has been adopted in \cite{Jazayeri:2023kji} and will make the comparison between the two sets of results more transparent.

We now need to choose interaction vertices of the form $\pi \pi \Phi$, and in order to make use of our factorisation theorem the interactions need to be IR-finite. EFToI operators that are quadratic or cubic in building blocks can both yield the desired interactions where here we define a building block as an object that starts at linear order in fluctuations. By using only these operators the tadpole cancellation is guaranteed. For operators that are quadratic in building blocks the presence of $\pi \pi \Phi$ couplings also induces $\pi \Phi$ couplings. Such couplings have two primary effects: they can contribute to the bispectrum of curvature perturbations through single-exchange diagrams, and yield new trispectrum diagrams which perturbatively capture the corrections to the linear theory. See \cite{Jazayeri:2022kjy,Pimentel:2022fsc} for recent works bootstrapping such single-exchange contributions to the bispectrum. Furthermore, for such interactions convergence of the time integrals we must compute is more subtle, and is not guaranteed, so we present a separate example that considers linear mixings in Section \ref{Example4}. In our quest to write down exact shapes, we will therefore concentrate on EFToI operators for which the leading vertices are $\pi \pi \Phi$ i.e. those that do not necessarily come with $\pi \Phi$ couplings by symmetry. This essentially tells us that $\pi$ must appear as $\pi'$ or $\partial_{i} \partial_{j} \pi$ (which come from the EFToI operators $\delta g^{00}$ and $\delta K_{\mu\nu}$), and we can add extra derivatives to these objects. By requiring that $\pi$ always appears in this way we ensure convergence of all time integrals at the conformal boundary. Indeed, the two spatial derivatives come with two factors of $\eta$ by scale invariance while $\pi'$ yields one power of $\eta$ by scale invariance but then we have 
\begin{align}
	K_{\pi}'(k, \eta) = k^2 \eta e^{i k \eta}~,
\end{align}
which yields an additional factor. The net contribution from the two factors of $\pi$ in each vertex is then at least four powers of $\eta$ which cancels the four inverse powers coming from the integration measure. Adding additional derivatives can only improve convergence thanks to the additional powers of $\eta$ which are dictated by scale invariance. We can also restrict to at most one time derivative on each of the external bulk-boundary propagators given that higher order ones can be eliminated by the scalar field's equation of motion. The lowest dimension operators which satisfies these properties are dimension-7 \cite{Cabass:2022rhr}, and for concreteness we will use\footnote{There is a dimension-6 operator, $\pi' \partial_i \pi' \Phi_{i}$, that satisfies our requirements but since only the $h=\pm 1$ modes contribute we can take $\Phi_{i}$ to be transverse and then this operator is a total spatial derivative. The resulting correlator will then vanish once we impose momentum conservation. The other dimension-7 operator that we could use is $\partial^2 \pi \partial_i \pi' \Phi_{i}$. The correlator arising from this vertex will only differ from the one we are going to compute in the kinematic factors since the time integrals will be the same.}
\begin{align}\label{ccspin1int}
	S_{\text{int}}=\int d^3 x d\eta \left(\frac{a^{-1}}{\Lambda^3}\partial_j\pi'_c\partial_i\partial_j\pi_c \Phi_i\right)~,
\end{align}
which originates from the EFT operator $\nabla_\mu \delta g^{00} \delta K^{\mu}{}_{\nu}\Phi^{\nu}$. The number of scale factors can be understood from scale invariance given that under a scale transformation the vector transforms in the same way as $a(\eta) \pi_c$ (since it is spin-$1$). Here $\pi_c$ is the canonically normalized Goldstone boson $\pi_c=c_s^{-3/2} f^2_\pi\,\pi$ with 
$f^4_\pi=H^4/(2\pi\Delta_\zeta)^2$, and $\Delta_\zeta^2\approx 2\times10^{-9}$ is the observed  dimensionless power spectrum. In the following we compute the parity-odd trispectrum due to the exchange of this massive vector due to \eqref{ccspin1int}. We consider light and heavy fields separately. 

\paragraph{Light mass case.} Let us first consider the light mass case where $m<H / 2$. We remind the reader that the $s$-channel contribution to the trispectrum is, c.f. \eqref{rhoToCorrelators},
\begin{align}
	\langle \pi_c^4\rangle'_s=\prod_{a=1}^{4}P_{\pi_c}(k_a)\left(\rho_4(\{k\};s;\{\mathbf{k}\})+\sum_{h=-1}^{1}P_h(s)\rho^{(h)}_{3}(\mathbf{k}_1,\mathbf{k}_2,-\mathbf{s})\rho^{(h)}_{3}(\mathbf{k}_3,\mathbf{k}_4,\mathbf{s})\right)~.
\end{align}
According to the proof in Section \ref{RandFSection}, the full $\psi_4$ is real and $\rho_4$ does not contribute to the parity-odd trispectrum (see \eqref{BPOLightFormula}). Here we therefore directly compute the factorised contributions i.e. the cubic wavefunction coefficients. We have
\begin{align}
	\nonumber\psi_{3}^{(h)}(\mathbf{k}_1,\mathbf{k}_2,\mathbf{-s})  =&\quad\begin{gathered}
	\includegraphics[width=0.1\textwidth]{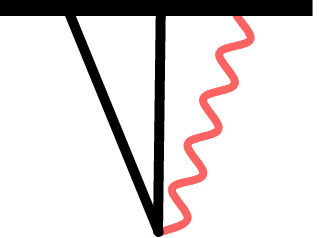}
	\end{gathered}\\
	\nonumber=&- \frac{H}{\Lambda^3} \left(\mathbf{k}_1\cdot\mathbf{k}_2\right)\left(\mathbf{k}_1\cdot \bm{\mathfrak{e}}^{(h)}(\hat{\mathbf{s}})\right)^*\times \int d\eta\,\eta K_{\pi_c}(k_1,\eta)\partial_{\eta}K_{\pi_c}(k_2,\eta)K_{h}(s,\eta) \\
 &-\frac{H}{\Lambda^3} \left(\mathbf{k}_1\cdot\mathbf{k}_2\right)\left(\mathbf{k}_2\cdot \bm{\mathfrak{e}}^{(h)}(\hat{\mathbf{s}})\right)^*\times \int d\eta\,\eta K_{\pi_c}(k_2,\eta)\partial_{\eta}K_{\pi_c}(k_1,\eta)K_{h}(s,\eta) ~,
\end{align}
with
\begin{align}
	K_{\pi_c}=(1-i c_s k \eta)e^{i c_s k \eta},\qquad K_{h}=\frac{W_{-i\tilde{\kappa},\nu}(-2ik\eta)}{W_{-i\tilde{\kappa},\nu}(-2ik\eta_0)}~.
\end{align}
Here the superscript denotes the helicity of the external vector field, and we have used $[\bm{\mathfrak{e}}^{(h)}(\hat{\mathbf{s}})]^*=\bm{\mathfrak{e}}^{(-h)}(\hat{\mathbf{s}}) =\bm{\mathfrak{e}}^{(h)}(-\hat{\mathbf{s}})$ for $h=\pm1$. The dynamical integral can be evaluated exactly using the Laplace transformation of the Whittaker function,
\begin{align}
	\mathcal{I}^h_n(a,b,\nu)&\equiv a^{n+1}\int_{0}^{\infty}x^n W_{-i\tilde\kappa,\nu}(2 a x)e^{-b x}dx\nonumber\\
	&=2^{-1-n}\Gamma\left(\frac{3}{2}+n-\nu\right)\Gamma\left(\frac{3}{2}+n+\nu\right){}_2\tilde{\rm{F}}_1\Bigg[\begin{array}{c} \frac{3}{2}+n-\nu, \frac{3}{2}+n+\nu\\[2pt] 2+n+i\tilde\kappa \end{array}\Bigg|\,\frac{1}{2}-\frac{b}{2a}\Bigg]~,
\end{align}
where ${}_2\tilde{\rm{F}}_1$ is the regularized hypergeometric function:
\begin{align}
	{}_2\tilde{\rm{F}}_1\Bigg[\begin{array}{c} a,b\\[2pt] c \end{array}\Bigg|\,z\Bigg]\equiv
	{}_2{\rm{F}}_1\Bigg[\begin{array}{c} a,b\\[2pt] c \end{array}\Bigg|\,z\Bigg]/\Gamma(c)~.
\end{align} 
This integral enjoys the helical reality property we have seen many times above,
\begin{align}
	\mathcal{I}^{h*}_n(a,b,\nu)=\mathcal{I}^{-h}_n(a,b,\nu)~.
\end{align}
The cubic wavefunction coefficient is therefore given by
\begin{align}
	\nonumber\psi_{3}^{(h)}(\mathbf{k}_1,\mathbf{k}_2,\mathbf{-s}&)=-\frac{ic_s^2}{s^3W_{-i\tilde{\kappa},\nu}(-2is\eta_0)}\frac{H}{\Lambda^3}\left(\mathbf{k}_1\cdot\mathbf{k}_2\right)\left(\mathbf{k}_1\cdot \bm{\mathfrak{e}}^{(h)}(\hat{\mathbf{s}})\right)^*k_2^2\left(1-k_1\frac{\partial}{\partial k_{1}}\right)\mathcal{I}^h_2(s,c_sk_{12},\nu) \\
     &-\frac{ic_s^2}{s^3W_{-i\tilde{\kappa},\nu}(-2is\eta_0)}\frac{H}{\Lambda^3}\left(\mathbf{k}_1\cdot\mathbf{k}_2\right)\left(\mathbf{k}_2\cdot \bm{\mathfrak{e}}^{(h)}(\hat{\mathbf{s}})\right)^*k_1^2\left(1-k_2\frac{\partial}{\partial k_{2}}\right)\mathcal{I}^h_2(s,c_sk_{12},\nu) ~,\label{spin1_cubiccoeff}
\end{align}
and the cubic density matrix coefficient reads
\begin{align}
	\rho^{(h)}_{3}\left(\mathbf{k}_1,\mathbf{k}_2,-\mathbf{s}\right)&\nonumber=\psi_{3}\left(\mathbf{k}_1,\mathbf{k}_2,-\mathbf{s}\right)+\psi^*_{3}\left(-\mathbf{k}_1,-\mathbf{k}_2,\mathbf{s}\right)\\
    \nonumber=&-\frac{2 ic_s^2H}{\Lambda^3}\left(\mathbf{k}_1\cdot\mathbf{k}_2\right)\left(\mathbf{k}_1\cdot \bm{\mathfrak{e}}^{(h)}(\hat{\mathbf{s}})\right)^*\frac{k_2^2}{s^3}\left(1-k_1\frac{\partial}{\partial k_{1}}\right)\Re\left[\frac{\mathcal{I}^h_2(s,c_sk_{12},\nu)}{W_{-i\tilde{\kappa},\nu}(-2is\eta_0)}\right] \\
    &-\frac{2 ic_s^2H}{\Lambda^3}\left(\mathbf{k}_1\cdot\mathbf{k}_2\right)\left(\mathbf{k}_2\cdot \bm{\mathfrak{e}}^{(h)}(\hat{\mathbf{s}})\right)^*\frac{k_1^2}{s^3}\left(1-k_2\frac{\partial}{\partial k_{2}}\right)\Re\left[\frac{\mathcal{I}^h_2(s,c_sk_{12},\nu)}{W_{-i\tilde{\kappa},\nu}(-2is\eta_0)}\right].
\end{align}
The product of polarisation factors is given by
\begin{align}
	\left(\mathbf{k}_1\cdot \bm{\mathfrak{e}}^{(h)}(\mathbf{\hat{s}})\right)^*\left(\mathbf{k}_3\cdot \bm{\mathfrak{e}}^{(h)}(\mathbf{\hat{s}})\right)
	=\left\{\begin{gathered}
		\frac{1}{2}\left[\mathbf{k}_1\cdot\mathbf{k}_3-(\mathbf{k}_1\cdot\mathbf{\hat{s}})(\mathbf{k}_3\cdot\mathbf{\hat{s}})+ih \mathbf{\hat{s}}\cdot(\mathbf{k}_1\times\mathbf{k}_3)\right],\quad h=\pm1\\
		(\mathbf{k}_1\cdot\mathbf{\hat{s}})(\mathbf{k}_3\cdot\mathbf{\hat{s}})\qquad\qquad\qquad\qquad\qquad\quad\quad,\quad h=0
	\end{gathered}\right.~,
\end{align}
and by putting everything together, projecting onto the parity-odd part of the full trispectrum c.f. \eqref{POcorrelator}, and transferring $\pi_c$ to the curvature perturbations using the relation $\zeta=-H \pi=-(2\pi \Delta_\zeta c_s^{3/2}/H)\pi_c$, we arrive at 
\begin{keyeqn}
	\begin{align}\label{Spin1_light_chemical}
		\nonumber B_4^{\zeta,\mathrm{PO}}
		=&\,i\left(\frac{H}{\Lambda}\right)^6 \frac{\pi^4\Delta_\zeta^4}{2c_s^{2}}\frac{\left(\mathbf{k}_1\cdot\mathbf{k}_2\right)\left(\mathbf{k}_3\cdot\mathbf{k}_4\right)}{k_1 k_2 k_3k_4}\frac{\mathbf{s}\cdot\left(\mathbf{k}_1\times\mathbf{k}_3\right)}{k^2_1k^2_3s^8}\left(1-k_1\frac{\partial}{\partial k_{1}}\right)\left(1-k_3\frac{\partial}{\partial k_{3}}\right)\\
		&\times\Bigg\{\frac{\pi i}{\Gamma\left(\frac{1}{2}-i\tilde{\kappa}+\nu\right)\Gamma\left(\frac{1}{2}-i\tilde{\kappa}-\nu\right)}\mathcal{I}^{(+1)}_2(s,c_s k_{12},\nu)\mathcal{I}^{(+1)}_2(s,c_s k_{34},\nu)\nonumber\\
		&~+e^{\pi\tilde{\kappa}}\Re\left[\mathcal{I}^{(+1)}_2(s,c_s k_{12},\nu)\mathcal{I}^{(-1)}_2(s,c_s k_{34},\nu)\right]-\big(\tilde{\kappa}\rightarrow -\tilde{\kappa}\big)\Bigg\}+\text{3 perms} \nonumber \\
   & ~ + (t \text{-channel}) + (u \text{-channel}),
	\end{align}
\end{keyeqn}
where we have implicitly assumed $h=1$ in the  $\tilde{\kappa}$. The $+\text{3 perms}$ yields the correct $s$-channel trispectrum and then we add the $t$ and $u$ channel permutations to give us the full symmetric trispectrum. To cancel the $\eta_0$ factors we used the relation \eqref{Canceleta0Whittaker}, and we see that this result is purely imaginary as it should be. Scale invariance tells us that $B_{4} \sim k^{-9}$, and we can check that this is indeed the case given that $\mathcal{I}_{n}^{h} \sim k^{0}$. This final result therefore is indeed of the form \eqref{B4Schematic}, i.e. a sum of terms containing kinematic factors multiplied by a product of hypergeometric functions which come from time evolution. The factor of $\mathbf{s}\cdot\left(\mathbf{k}_1\times\mathbf{k}_3\right) = \mathbf{k}_2\cdot\left(\mathbf{k}_1\times\mathbf{k}_3\right)$ will appear as a factor in all $s$-channel, parity-odd trispectra.  

\begin{figure}[ht]
	\centering
	\includegraphics[width=\textwidth]{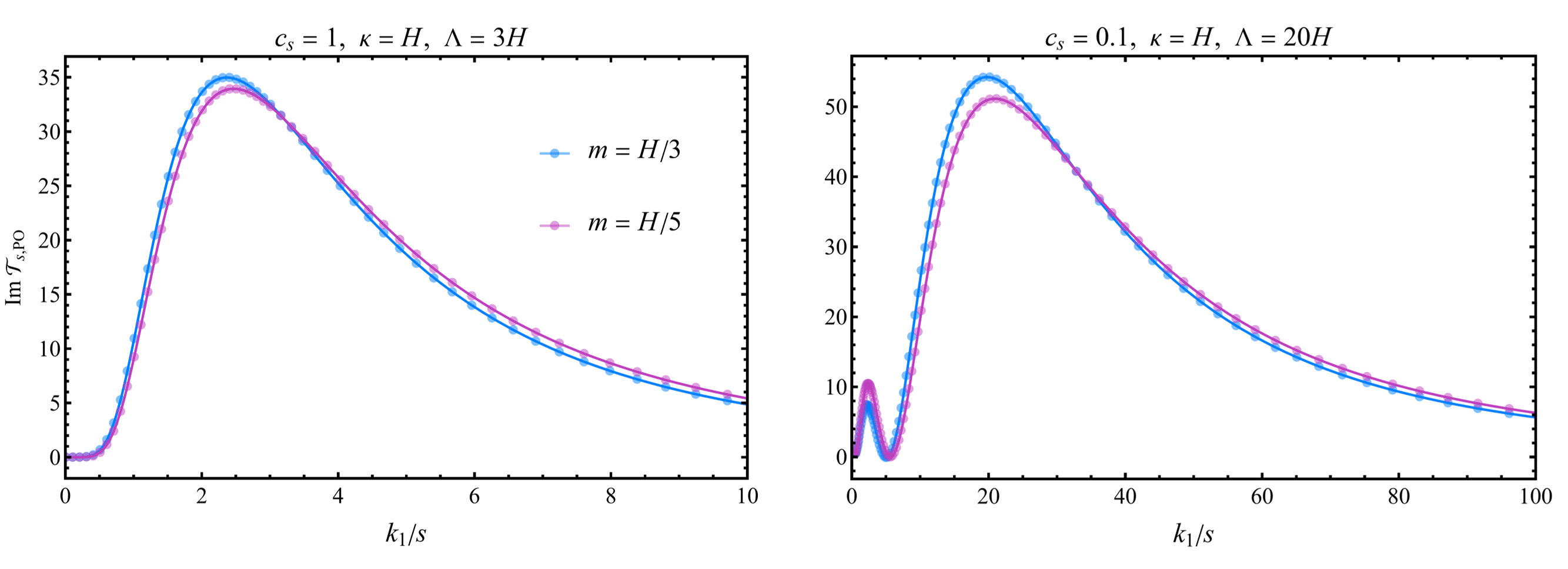}\\
	\caption{The $s$-channel dimensionless parity-odd trispectrum $\Im\mathcal{T}_{s, \text{PO}}$ as a function of the momentum ratio $k_1/s$. The kinematics is chosen as $k_1=k_3$, $k_2=k_4=\sqrt{s^2+k_1^2}$ and $\psi = \pi/3$ being the dihedral angle from the $(\mathbf{k}_1,\mathbf{k}_2)$-plane to the $(\mathbf{k}_3,\mathbf{k}_4)$-plane. The parameters are chosen as $c_s=0.1, \Lambda=3H$ (left panel) and $c_s=1, \Lambda=20H$ (right panel), together with a common chemical potential $\kappa=H$. The \textcolor{blue2}{blue} and \textcolor{magenta2}{magenta} curves show the exact solution (\ref{Spin1_light_chemical}) for vector field mass $m=H/3$ and $m=H/5$, respectively. The dots represent the numerical result computed in the UV theory, which perfectly match our exact solution.}
	\label{Spin1PPVLight}
\end{figure}

As a consistency check of this result, and therefore a check of our statement that the quartic wavefunction coefficient is purely real for this light field exchange, we also numerically computed the $s$-channel trispectrum using the in-in formalism. In Figure \ref{Spin1PPVLight} we plot the dimensionless trispectrum $\mathcal{T}_{s,\text{PO}}$, as defined in \eqref{dimlessTDef}, for both our exact expression and the result from numerics. Clearly the exact solution agrees with the numerics very well as seen from the plot. The mass dependence of $\mathcal{T}_{s,\text{PO}}$ turns out to be mild and smooth within the complementary series mass range. Interestingly, with a reduced inflaton sound speed $c_s<1$ and chemical potential $\kappa\gtrsim H$, the trispectrum has a peak roughly lying at a momentum ratio $k_1/s\sim 2\kappa/H$, which is followed by a dip to zero at $k_1/s\sim 2\kappa/(c_s^{1/2} H)$ and a second peak at $k_1/s\sim 2\kappa/(c_s H)$. The precise location and the relative height of the peaks and the dip depend on the detailed choice of mass and chemical potential, and if probed, can be utilized to break the degeneracy of parameters.

\paragraph{Heavy mass case.} We now move to the heavy field case where the mass of the vector $\Phi$ is in the principle series. We now have a complex $\psi_4$ and that will contribute to the final parity-odd trispectrum. However, as we have discussed at length in this paper, the modification of the bulk-bulk propagator by the addition/subtraction of a factorised contribution ensures that the connected component of the wavefunction coefficient is real. Then what we need to take into account are those factorised contributions to the bulk-bulk propagator which we have been denoting by $F$ as a pair of cut dashed lines, as shown diagrammatically by \eqref{BPOHeavyFormula}. For concrete calculations we choose the simplest form of $F$ which is given by
\begin{align}
    F_{1,1}^{h}(\eta_1, \eta_2, k) &= -\frac{\Phi_{1,1}^{h}(\eta_0, k)}{\Phi_{1,1}^{h \star}(\eta_0, k)}\Phi_{1,1}^{h \star}(\eta_1, k)\Phi_{1,1}^{h \star}(\eta_2, k) - \mathcal{A}_{h,1}\Phi_{1,1}^{h \star}(\eta_1, k)\Phi_{1,1}^{h \star}(\eta_2, k)~,
\end{align}
with
\begin{align}	          \mathcal{A}_{h,1}=\frac{i\pi \sech (\pi\tilde{\kappa})}{\Gamma\left(\frac{1}{2}-i\tilde{\kappa}-i\mu\right)\Gamma\left(\frac{1}{2}-i\tilde{\kappa}+i\mu\right)}.
\end{align}
This factorised contribution from $F$ is then given by
\begin{align}
	\nonumber\psi_4^{h,\text{PO}}=&\quad\begin{gathered}
		\includegraphics[width=0.18\textwidth]{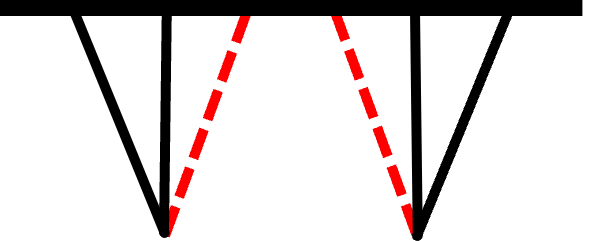}
		\end{gathered}\\
	=&~{ihc_s^4}\left(\frac{H}{\Lambda^3}\right)^2\frac{k_2^2k_4^2}{4s^8}\left(\mathbf{k}_1\cdot\mathbf{k}_2\right)\left(\mathbf{k}_3\cdot\mathbf{k}_4\right)\left[\mathbf{s}\cdot\left(\mathbf{k}_1\times\mathbf{k}_3\right)\right]\left(1-k_1\frac{\partial}{\partial k_{1}}\right)\left(1-k_3\frac{\partial}{\partial k_{3}}\right)\nonumber\\
    &\times e^{-\pi\tilde{\kappa}}\Bigg[\mathcal{A}_{h,1}+\frac{W_{i\tilde{\kappa},i\mu}(2is\eta_0)}{W_{-i\tilde{\kappa},i\mu}(-2is\eta_0)}\Bigg]\mathcal{I}^h_2(s,c_sk_{12},i\mu)\mathcal{I}^h_2(s,c_sk_{34},i\mu)+\text{3 perms}~,
\end{align}
and the corresponding parity-odd density matrix coefficient is
\begin{align}
    \rho^{\text{PO}}_4=&\,{ic_s^4}\left(\frac{H}{\Lambda^3}\right)^2\frac{k_2^2k_4^2}{4s^8}\left(\mathbf{k}_1\cdot\mathbf{k}_2\right)\left(\mathbf{k}_3\cdot\mathbf{k}_4\right)\left[\mathbf{s}\cdot\left(\mathbf{k}_1\times\mathbf{k}_3\right)\right]\left(1-k_1\frac{\partial}{\partial k_{1}}\right)\left(1-k_3\frac{\partial}{\partial k_{3}}\right)\nonumber\\
    &\times\Bigg\{\left[2\cosh(\pi \tilde{\kappa})\mathcal{A}_{+1,1}+\frac{e^{-\pi\tilde{\kappa}}W_{i\tilde{\kappa},i\mu}(2is\eta_{0})}{W_{-i\tilde{\kappa},i\mu}(-2is\eta_{0})}-\frac{e^{\pi\tilde{\kappa}}W_{i\tilde{\kappa},i\mu}(-2is\eta_{0})}{W_{-i\tilde{\kappa},i\mu}(2is\eta_{0})}\right] \nonumber \\  & ~~~~~~~~~~~~~~~~~ \times\mathcal{I}^{(+1)}_2(s,c_sk_{12},i\mu)\mathcal{I}^{(+1)}_2(s,c_sk_{34},i\mu) -\left(\tilde{\kappa}\rightarrow-\tilde{\kappa}\right)\Bigg\}+\text{3 perms}~.\label{rho4POHeavy}
\end{align}
For the light mass case the entire bracket in \eqref{rho4POHeavy} vanishes, which is consistent with $\psi_{4}$ being purely real. For the heavy mass case, $\rho^{\text{PO}}_4$ is dependent on $\eta_0$, however this dependence cancels with the contribution to the correlator coming from the cubic wavefunction coefficients thereby rendering the final correlator $\eta_0$-independent (as it should be for IR-finite interactions). The computation of the cubic wavefunction coefficients is identical as above, and is omitted here for brevity. Putting everything together and projecting on to the parity-odd sector, we arrive at
\begin{keyeqn}
\begin{align}\label{Spin1_heavy_chemical}
	\nonumber B_4^{\zeta,\mathrm{PO}}
	=&\,i\left(\frac{H}{\Lambda}\right)^6 \frac{\pi^4\Delta_\zeta^4}{2c_s^{2}}\frac{\left(\mathbf{k}_1\cdot\mathbf{k}_2\right)\left(\mathbf{k}_3\cdot\mathbf{k}_4\right)}{k_1k_2k_3k_4}\frac{\mathbf{s}\cdot\left(\mathbf{k}_1\times\mathbf{k}_3\right)}{k^2_1k^2_3s^8}\left(1-k_1\frac{\partial}{\partial k_{1}}\right)\left(1-k_3\frac{\partial}{\partial k_{3}}\right)\\
  &  \times\Bigg\{\cosh(\pi \tilde{\kappa})\mathcal{A}_{+1,1}\mathcal{I}^{+1}_2(s,c_sk_{12},i\mu)\mathcal{I}^{+1}_2(s,c_sk_{34},i\mu)+e^{\pi\tilde{\kappa}}\Re\left[\mathcal{I}^{+1}_2(s,c_s k_{12},i\mu)\mathcal{I}^{-1}_2(s,c_s k_{34},i\mu)\right]\nonumber\\
	&\qquad-\big(\tilde{\kappa}\rightarrow -\tilde{\kappa}\big)\Bigg\}+\text{3 perms} \nonumber \\
  & ~ + (t \text{-channel}) + (u \text{-channel}).
\end{align}
\end{keyeqn}
Again this result has the correct momentum scaling, and is purely imaginary. If we compare this result with that of light fields \eqref{Spin1_light_chemical}, we see that they can be converted into each other by replacing $i\mu\leftrightarrow\nu$. Hence as we expected, there is no discontinuity in the mass parameters. This property extends to other examples too: the heavy field result is always given by an analytic continuation of the light field result. Given that the calculation for light fields is less involved, for the other examples we will concentrate on light fields and then extract the heavy field result via this simple replacement rule.

\begin{figure}[ht]
	\centering
	\includegraphics[width=\textwidth]{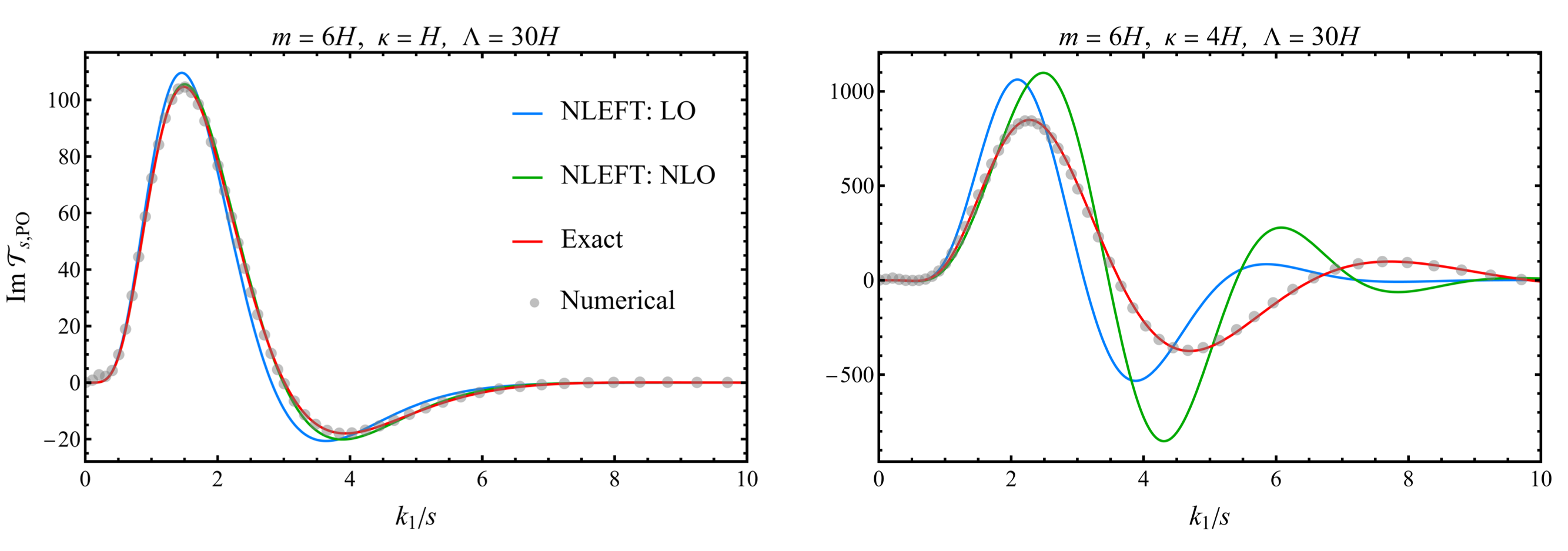}\\
	\caption{The $s$-channel dimensionless parity-odd trispectrum $\Im\mathcal{T}_{s, \text{PO}}$ as a function of the momentum ratio $k_1/s$. The kinematics is chosen as $k_1=k_3$, $k_2=k_4=\sqrt{s^2+k_1^2}$ and $\psi = \pi/3$ being the dihedral angle from the $(\mathbf{k}_1,\mathbf{k}_2)$-plane to the $(\mathbf{k}_3,\mathbf{k}_4)$-plane. The parameters are chosen as $m=6H, \kappa=H$ (left panel) and $m=6H, \kappa=4H$ (right panel), together with a common sound speed $c_s=0.1$ and $\Lambda=30H$. The \textcolor{blue2}{blue} and \textcolor{green2}{green} curves show the leading-order (LO) and the next-to-leading-order (NLO) non-local EFT results. The \textcolor{red2}{red} curve denotes the exact result (\ref{Spin1_heavy_chemical}) and the \textcolor{gray}{gray} dots represent the numerics in the UV theory. We see that the non-local EFT predictions agree with numerics in the small-$\kappa$ case, yet deviations start to appear at large $\kappa$. In principle, such deviations may be cured by adding higher order contributions which are in practice tedious to compute. In contrast, the exact result we have computed in this paper matches the numerics very well for all parameter choices.}
	\label{NLEFTvsPOE}
\end{figure}

Interestingly, for a small inflaton sound speed (i.e. $c_s\ll 1$), this model of a heavy vector field with chemical potential admits a non-local single-field EFT description in the IR, which well approximates the behaviour of the parity-odd trispectrum in the regime $c_s \kappa < c_s m < 1$ \cite{Jazayeri:2023kji}. After partially integrating out the heavy vector, parity violation resurges through \textit{emergent non-locality} in the effective vertex, and the resulting parity-odd trispectrum is neatly computed as a residue of the non-local pole in the effective vertex. Such a miraculous behaviour as seen from the non-local EFT now becomes understandable from the parity-factorisation perspective. The fact that the parity-odd trispectrum is necessarily factorised in the UV theory is precisely the reason why we only acquire a non-vanishing contribution from the non-local pole in the IR EFT.

To compare our exact result for the parity-odd trispectrum in this model with the non-local EFT prediction, and to check them against numerics, we plot the corresponding dimensionless trispectra in Figure \ref{NLEFTvsPOE}. As we can see from the plot, the exact result agrees with numerics very well, while the non-local EFT predictions start to deviate from the exact result when the chemical potential $\kappa$ is large. 

Before moving to some other examples, let us first comment on the massless case with a $U(1)$ gauge symmetry, as promised. In this case we set $m=0$ to preserve the gauge symmetry in the free theory of the vector field. We therefore have $\nu = 1/4$. Without adding any additional interactions beyond those in \eqref{MassiveVector}, a parity-odd trispectrum can be generated at $1$-loop due to the $\pi \Phi \Phi$ coupling. At tree-level we would again need to add interactions of the form $\pi \pi \Phi$ that preserve the $U(1)$ gauge symmetry. Since the field strength is anti-symmetric, this requires more derivatives than what we have studied so far. Indeed, the first non-zero operator is dimension-$8$. It would be interesting to study this class of trispectra in more detail. 


\subsection{Example 2: spin-2 exchange in CCM} \label{Example2}

We now move to a second example where we consider the exchange of a spin-$2$ field with its dynamics described by the cosmological condensed matter physics scenario. In this case we take the bulk-bulk propagator to be parity-even $\tilde{\kappa} = 0$, and source the parity-violation via the interaction vertices. We therefore need one vertex to have an even number of spatial derivatives, and for the other to have an odd number. As before, we need to ensure IR-convergence. We will therefore work with the following interactions:
\begin{align}\label{spin2CCM_int}
	S_{\text{int}}=\int d^3x d\eta\left(\frac{a}{\Lambda^2_1}\pi_c'\partial_{i}\partial_{j}\pi_c\sigma_{ij}+\frac{1}{\Lambda_2^3}\epsilon_{ijk}\partial_{i}\pi_c'\partial_{j}\partial_{l}\pi_c\sigma_{kl}\right)~,
\end{align}
where the first term is dimension-6 while the second is dimension-7. These are the only operators with those mass dimensions and are the leading ones which are IR-finite. The corresponding EFToI operators are $\delta g^{00} \delta K_{\mu\nu} \Sigma^{\mu\nu}$ and $n_{\mu} \varepsilon^{\mu \nu\alpha\beta} \nabla_{\nu} \delta g^{00} \delta K_{\alpha \gamma} \Sigma^{\gamma}{}_{\beta}$. In the CCM scenario the conformal weight of the massive field is the same as that of a massless scalar so the counting of the scale factors is simply $4 - (\text{total number of derivatives})$. As with Section \ref{sec:massivefields} we write
\begin{align}\label{Spin2_CM_Polar}
	\sigma_{ij}(\eta,\mathbf{k})=\sum_{h=-2}^{2}\sigma_h(\eta,k){\rm{e}}^{(h)}_{ij}(\mathbf{k})~,
\end{align}
and the polarisation tensors are chosen to satisfy conditions (\ref{PolStar}) and (\ref{PolNorm}), and are given by
\begin{align}
	{\rm{e}}_{ij}^{(0)} =\sqrt{3}\left(\hat{k}_{i}\hat{k}_{j} - \frac{1}{3}\delta_{ij} \right), \qquad {\rm{e}}_{ij}^{(\pm 1)} =i(\hat{k}_{i} \hat{e}_{j}^{\pm} + \hat{k}_{j} \hat{e}_{i}^{\pm}), \qquad {\rm{e}}_{ij}^{(\pm 2)} = \sqrt{2}\hat{e}_{i}^{\pm}\hat{e}_{j}^{\pm}~.
\end{align}
The mode functions for each helicity are given by
\begin{align}
	\sigma_h(\eta,k)=-\frac{H\eta}{\sqrt{2 c_{h,2}k}}W_{0,\nu}(2i c_{h,2}k\eta)~,
\end{align}
where as we mentioned before we take the chemical potential to vanish $\tilde{\kappa} = 0$. In this limit the Whittaker function recovers the Hankel function of the first kind, c.f. \eqref{WtoH}. When the mass of the spin-$2$ field is light, the only contributions to the parity-odd trispectrum come from the cubic wavefunction coefficients which are given by
\begin{align}
	\nonumber\psi_{3}^{(h)}(\mathbf{k}_1,\mathbf{k}_2,\mathbf{-s})
	=&\quad\begin{gathered}
		\includegraphics[width=0.1\textwidth]{psi3Diag}
		\end{gathered}\\
	=&\,\frac{ik_1^2}{\eta_{0} W_{0,\nu}(-2ic_{h,2}s\eta_0)}\times\Bigg[\frac{1}{s^2H\Lambda_1^2}\left(k_{2}^ik_{2}^j {\rm{e}}^{(h)}_{ij}(\mathbf{s})\right)^*\left(1-k_2\frac{\partial}{\partial{k_2}}\right)\mathcal{I}^{0}_1(c_{h,2}s,k_{12},\nu)\nonumber\\
	&\qquad+\frac{1}{s^3\Lambda_2^3}\left(\epsilon_{ijk}k_{1}^ik_{2}^jk_{2}^l{\rm{e}}^{(h)}_{kl}(\mathbf{s})\right)^*\left(1-k_2\frac{\partial}{\partial{k_2}}\right)\mathcal{I}^{0}_2(c_{h,2} s,k_{12},\nu)\Bigg]+\left(1\leftrightarrow 2\right)~.
\end{align}
In the absence of the chemical potential, the $\mathcal{I}^{h}_n$ integral is identical for each helicity (except for the sound speeds) and purely real. The helicity dependence then resides in the kinematic factors rather than the dynamical ones. The density matrix then reads
\begin{align}
	\rho_{2,1}^{(h)}(\mathbf{k}_1,\mathbf{k}_2,\mathbf{-s})&=\,\frac{k_1^2}{s^2H\Lambda_1^2}\Re\left(\frac{2i}{\eta_{0} W_{0,\nu}(-2ic_{h,2}s\eta_0)}\right)\left(k_{2i}k_{2j}\cdot {\rm{e}}^{(h)}_{ij}(\mathbf{s})\right)^*\left(1-k_2\frac{\partial}{\partial{k_2}}\right)\mathcal{I}^{0}_1(c_{h,2}s,k_{12},\nu)\nonumber\\
	&+i\frac{k_1^2}{s^3\Lambda_2^3}\Re\left(\frac{2}{\eta_{0} W_{0,\nu}(-2ic_{h,2}s\eta_0)}\right)\left(\epsilon_{ijk}k_{1i}k_{2j}k_{2l}{\rm{e}}^{(h)}_{kl}(\mathbf{s})\right)^*\left(1-k_2\frac{\partial}{\partial{k_2}}\right)\mathcal{I}^{0}_2(c_{h,2} s,k_{12},\nu)\nonumber\\
	&+\left(1\leftrightarrow 2\right)~.
\end{align}
To make the angular dependence transparent, let us decompose the $s$-channel trispectrum into two separate parts arising from the exchange of different helicity modes:
\begin{align}
	B^{\zeta}_4=B^{\zeta}_{4,h=\pm1}+B^{\zeta}_{4,h=\pm2}~,
\end{align}
where we have dropped the contribution from $h=0$ since scalar exchanges cannot yield a parity-odd contribution. For the higher helicity modes, we need to fix the polarisation sums. For spin-$1$ the form of $\sum_{h=\pm 1} \hat{e}_i(\mathbf{k})\hat{e}_j(\mathbf{-k})$ can be easily fixed without choosing any particular basis. The result should be parity-even and real given the properties of the polarisation vectors. Scale invariance further constrains it to only depend on $\delta_{ij}$ and $\hat{k}_i=k_i/k$. The free parameters can be then fixed by requiring the result to be transverse, and appropriately normalised: $\hat{e}^{\pm}_i(\mathbf{k})\hat{e}^{\pm}_i(\mathbf{-k})=1$. We then have
\begin{align}\label{PolarVec_sum}
	{\pi}_{ij}(\mathbf{k})\equiv\hat{e}^{+}_i(\mathbf{k})\hat{e}^{+}_j(\mathbf{-k})+\hat{e}^{-}_i(\mathbf{k})\hat{e}^{-}_j(\mathbf{-k})= \delta_{ij}-\hat{k}_i\hat{k}_j ~.
\end{align}
For spin-$2$, where polarisation tensors are combinations of $\hat{k}_i$ and $\hat{e}^{\pm}_i$, we can proceed in a similar way using (\ref{Spin2_CM_Polar}) and (\ref{PolarVec_sum}). We have
\begin{align}
	\sum_{h=\pm 1}{\rm{e}}^h_{ij}(\mathbf{k}){\rm{e}}^h_{mn}(-\mathbf{k})=\hat{k}_i\hat{k}_m \pi_{jn}+\hat{k}_j\hat{k}_m \pi_{in}+\hat{k}_i\hat{k}_n \pi_{jm}+\hat{k}_j\hat{k}_n \pi_{im}~,
\end{align}
and 
\begin{align}
	\sum_{h=\pm 2}{\rm{e}}^h_{ij}(\mathbf{k}){\rm{e}}^h_{mn}(-\mathbf{k})=\pi_{im} \pi_{jn}+ \pi_{in} \pi_{jm}-\pi_{ij}\pi_{mn}~.
\end{align}
By combining these polarisation sums evaluated at $\mathbf{k}=-\mathbf{s}$ and the density matrix coefficients we can extract the contribution from individual helicity exchanges to the final parity-odd trispectrum,
\begin{keyeqn}
	\begin{align}\label{CCM_Spin2_h1}
		B^{\zeta,\text{PO}}_{4,h=\pm 1}=&\,2i\pi^4\Delta_\zeta^4\cos(\pi\nu)\left(\frac{H}{\Lambda_1}\right)^2\left(\frac{H}{\Lambda_2}\right)^3\frac{\left(\mathbf{k}_2\cdot\mathbf{s}\right)\left(\mathbf{k}_4\cdot\mathbf{s}\right)}{k_1 k_2 k_3 k_4}\frac{\left[\mathbf{s}\cdot\left(\mathbf{k}_1\times\mathbf{k}_3\right)\right]}{c_{1,2}k_2^2k_4^2s^8} \nonumber\\
		&\times\left(1-k_2\frac{\partial}{\partial k_{2}}\right)\left(1-k_4\frac{\partial}{\partial k_{4}}\right)\mathcal{I}^{0}_1(c_{1,2} s,k_{12},\nu)\mathcal{I}^{0}_2(c_{1,2} s,k_{34},\nu)+\text{7 perms} \nonumber \\
  & ~ + (t \text{-channel}) + (u \text{-channel}).
	\end{align}
	\begin{align}\label{CCM_Spin2_h2}
		B^{\zeta,\text{PO}}_{4,h=\pm 2}=&\,2i\pi^4\Delta_\zeta^4\cos(\pi\nu)\left(\frac{H}{\Lambda_1}\right)^2\left(\frac{H}{\Lambda_2}\right)^3\frac{\left[s^2\left(\mathbf{k}_2\cdot\mathbf{k}_4\right)-\left(\mathbf{k}_2\cdot\mathbf{s}\right)\left(\mathbf{k}_4\cdot\mathbf{s}\right)\right]}{k_1 k_2 k_3 k_4}\frac{\left[\mathbf{s}\cdot\left(\mathbf{k}_1\times\mathbf{k}_3\right)\right]}{c_{2,2}k_2^2k_4^2s^8} \nonumber\\
		&\times\left(1-k_2\frac{\partial}{\partial k_{2}}\right)\left(1-k_4\frac{\partial}{\partial k_{4}}\right)\mathcal{I}^{0}_1(c_{2,2} s,k_{12},\nu)\mathcal{I}^{0}_2(c_{2,2} s,k_{34},\nu)+\text{7 perms} \nonumber \\
  & ~ + (t \text{-channel}) + (u \text{-channel}).
	\end{align}
\end{keyeqn}
As a consistency check, we see that the terms proportional to $\left(\mathbf{k}_2\cdot\mathbf{s}\right)\left(\mathbf{k}_4\cdot\mathbf{s}\right)$ in (\ref{CCM_Spin2_h1}) and (\ref{CCM_Spin2_h2}) differ only by a sign and the sound speeds of the different modes. These two contributions would therefore cancel once added together if the sound speeds were identical ($c_{1,2}=c_{2,2}$). In that case the total trispectrum would be independent of $s_i$, which is to be expected since in that case the three polarisation sums add up to an object that is independent of $s_i$. We also see that the result is purely imaginary, and has the correct momentum scaling. For the special cases of $\nu = 3/2$ and $\nu = 1/2$, where the mode functions simplify to exponentials, the trispectrum vanishes which is consistent with the no-go theorem of \cite{Cabass:2022rhr}. Note that here we add $+7 ~ \text{perms}$ rather than $+3 ~ \text{perms}$ which we had in example $1$ since here the two vertices on either side of a diagram are different. 

It is noteworthy that in de-Sitter/inflationary four-point functions, spin-$1$ exchange is typically characterised by linear factors of $t^2-u^2$ in $s$-channel diagrams, which originate from contractions between momenta and the polarisation sum. Exchanges of higher spin are then non-linear in $t^2-u^2$. However, here things are slightly different due to the Levi-Civita $\epsilon$-tensor, and it is easy to check no such factor arises for spin-1 exchange as indicated by (\ref{CCM_Spin2_h1}). For the exchange of $h=\pm 2$ modes, this dependence appears from the $\mathbf{k}_2\cdot\mathbf{k}_4$ factor and its corresponding permutation. Indeed we have 
\begin{align}
	\mathbf{k}_2\cdot\mathbf{k}_4=\frac{1}{4}\left(t^2-u^2\right)+\frac{1}{4}\left(k_1^2 + k_3^2 - k_2^2 - k_4^2 - s^2\right)~.
\end{align}
For general spin-$S$, it is simple to see that helicity $\pm h$ exchange will introduce contributions to the parity-odd trispectrum with a factor of $\left(t^2-u^2\right)^{|h|-1}$, from which we can read off the spin of the exchanged field. 

Here we have computed the trispectrum for light field exchange, and as we discussed above, from this result we can extract that of heavy field exchange by sending $\nu \rightarrow i \mu$. We have also checked by explicit calculation that this analytic continuation yields the correct result. 

\subsection{Example 3: spin-2 exchange in CC} \label{Example3}
In this example we again consider spin-$2$ exchange with parity-violation arising from the interactions vertices, however now we describe the dynamics of the spin-$2$ field in the cosmological collider physics set-up. As we discussed in Section \ref{sec:massivefields}, the mode functions in the CC and CCM scenarios can differ significantly, leading to a distinct parity-odd trispectrum compared to what we have just computed in the previous sub-section. We will denote the massive spin-$2$ field by $\Phi_{\mu\nu}$. Given that scalar modes do not contribute to the final trispectrum, our focus will be on the components $\Phi_{0j}$ and $\Phi_{ij}$. Again we need IR-finite interaction vertices and we choose
\begin{align}\label{int_CC_Spin2}
	S_{\text{int}}=\int d^3x d\eta\bigg(&\frac{a^{-1}}{\Lambda^2_1}\pi_c'\partial_{i}\partial_{j}\pi_c\Phi_{ij}+\frac{a^{-2}}{\Lambda_2^3}\epsilon_{ijk}\partial_{i}\pi_c'\partial_{j}\partial_{l}\pi_c\Phi_{kl}\nonumber\\
	&+\frac{a^{-2}}{\Lambda^3_3}\partial_i\pi_c'\partial_i\partial_{j}\pi_c\Phi_{0j}+\frac{a^{-3}}{\Lambda^4_4}\epsilon_{ijk}\partial_{i}\partial_{l}\pi_c\partial_{j}\partial_{l}\pi'_c\Phi_{0k}\bigg)~,
\end{align}
which arise from the following EFToI operators:
\begin{align}
\delta g^{00} \delta K_{\mu\nu} \Phi^{\mu\nu} ~ \longrightarrow ~~ &\pi_c'\partial_{i}\partial_{j}\pi_c\Phi_{ij}, \\
n_{\mu} \varepsilon^{\mu\nu\alpha\beta} \nabla_{\nu} \delta g^{00} \delta K_{\alpha \gamma} \Phi^{\gamma}{}_{\beta} ~ \longrightarrow ~~ &\epsilon_{ijk}\partial_{i}\pi_c'\partial_{j}\partial_{l}\pi_c\Phi_{kl}, \\
\nabla^{\mu} \delta g^{00} \delta K_{\mu\nu} n_{\alpha} \Phi^{\alpha \nu} ~ \longrightarrow ~~ &  \partial_i\pi_c'\partial_i\partial_{j}\pi_c\Phi_{0j}, \\
n_{\mu} \varepsilon^{\mu\nu\alpha\beta}  \delta K_{\nu \rho} 
 n_{\gamma} \nabla^{\gamma} \delta K^{\rho}{}_{\alpha} n_{\delta} \Phi^{\delta}{}_{\beta} ~ \longrightarrow ~~ &  \epsilon_{ijk}\partial_{i}\partial_{l}\pi_c\partial_{j}\partial_{l}\pi'_c\Phi_{0k}.
\end{align}
Again the scale factors are fixed by scale invariance (note that here $\Phi$ scales in the same way as $a^2(\eta) \pi_c$). To establish a parity-odd trispectrum, we need a parity-even vertex and a parity-odd vertex. For simplicity, we shall henceforth focus on operators in the first line of \eqref{int_CC_Spin2}, and the inclusion of the second line is technically analogous but tedious. We now decompose the field into the helicity basis:
\begin{align}
	&\Phi_{0j}(\eta,\mathbf{k})=\sum_{h}\Phi^h_{1,2}(\eta,k){\mathfrak{e}}^{(h)}_{j}(\hat{\bfk})~,\\
	&\Phi_{ij}(\eta,\mathbf{k})=\sum_{h}\Phi^h_{2,2}(\eta,k){\mathfrak{e}}^{(h)}_{ij}(\hat{\bfk})~,
\end{align}
and from now on we will ignore the longitudinal modes $\Phi^0_{1,2}$ and $\Phi^0_{2,2}$ since they will not contribute to the parity-odd trispectrum. This leaves us with three modes: $\Phi^{\pm 1}_{1,2}$, $\Phi^{\pm 1}_{2,2}$, and $\Phi^{\pm2}_{2,2}$. The polarization tensors for these modes, which satisfy (\ref{PolStarCC}) and (\ref{PolNormCC}), are
\begin{align}
	{\mathfrak{e}}^{(\pm 1)}_i=\sqrt{2}\,\hat{e}^{\pm }_i,\qquad	{\mathfrak{e}}^{(\pm 1)}_{ij}=\frac{3}{\sqrt{2}}\left(\hat{k}_i\hat{e}^{\pm}_j+\hat{k}_j\hat{e}^{\pm }_i\right),\qquad {\mathfrak{e}}^{(\pm 2)}_{ij}=2 \hat{e}^{\pm}_i \hat{e}^{\pm}_j~.
\end{align}
The factor $\sqrt{2}$ arises because we adhere to the convention of \cite{Lee:2016vti}, where ${\mathfrak{e}}^{(\pm 1)}_i {\mathfrak{e}}^{*(\pm 1)}_i=2$. Following our discussion in Section \ref{sec:massivefields}, we can derive the following mode functions
\begin{align}
	\Phi^{\pm 1}_{1,2}(\eta,k)&=e^{i\pi(\nu+1/2)/2}Z^{+1}_2\left(-k\eta\right)^{1/2}H^{(1)}_{\nu}(-k\eta)~,\\
	\Phi^{\pm 2}_{2,2}(\eta,k)&=e^{i\pi(\nu+1/2)/2}Z^{+2}_2\left(-k\eta\right)^{-1/2}H^{(1)}_{\nu}(-k\eta)~,\\
	\Phi^{\pm 1}_{2,2}(\eta,k)&=\frac{i}{2}e^{i\pi(\nu+1/2)/2}Z^{+1}_2\left(-k\eta\right)^{-1/2}\bigg[k\eta \left(H^{(1)}_{\nu+1}(-k\eta)-H_{\nu-1}(-k\eta)\right)-3H^{(1)}_{\nu}(-k\eta)\bigg]~,\nonumber\\
	&=-\frac{i}{2}e^{i\pi(\nu+1/2)/2}Z^{+1}_2\left(-k\eta\right)^{-1/2}\bigg[2k\eta\, H^{(1)}_{\nu-1}(-k\eta)+(3+2\nu)H^{(1)}_{\nu}(-k\eta)\bigg]~,
\end{align}
where $Z^{|h|}_s$ is given by (\ref{coeff_Zhs}), and in the last line of $\Phi^{\pm 1}_{2,2}$ we have used the recursion relation of Hankel function to simplify the expression. Given that we are only turning on the interactions in the first line, the full trispectrum is
\begin{align}
	B^{\zeta}_4=B^{\zeta}_{(2,\pm1)}+B^{\zeta}_{(2,\pm2)}~\,,
\end{align}
where we have introduced the notation $B^{\zeta}_{(n,h)}$ which tells us about the helicity and spatial spin of the exchanged mode. The computation of the remaining components is the same as what we have been through above. Once again, considering light fields buys us the privilege of limiting the focus on the cubic wavefunction coefficients only (i.e. \eqref{BPOLightFormula}), before obtaining the density matrix coefficients and summing over helicities. The result for heavy fields then follows from analytic continuation. In short, we find 
\begin{keyeqn}
\begin{align}
    B^{\zeta,\text{PO}}_{(2,\pm 1)}=\,&9i\Delta^4_{\zeta}\pi^3\cos(\pi\nu)H^2\big|Z^{+1}_2(s)|^2\left(\frac{H}{\Lambda_1}\right)^2\left(\frac{H}{\Lambda_2}\right)^3\frac{\left(\mathbf{k}_2\cdot\mathbf{s}\right)\left(\mathbf{k}_2\cdot\mathbf{s}\right)}{k_1k_2k_3k_4}\frac{{\mathbf{s}}\cdot\left(\mathbf{k}_1\times\mathbf{k}_3\right)}{k_2^2k_4^2s^9} \nonumber\\ & \times\left(1-k_2\frac{\partial}{\partial k_2}\right)
    \bigg[2\,\mathcal{I}^0_2(s,k_{12},\nu-1)+(3+2\nu)\mathcal{I}^0_1(s,k_{12},\nu)\bigg] \nonumber \\ & \times \left(1-k_4\frac{\partial}{\partial k_4}\right) \bigg[2\,\mathcal{I}^0_3(s,k_{34},\nu-1)+(3+2\nu)\mathcal{I}^0_2(s,k_{34},\nu)\bigg] +\text{7~perms} \nonumber \\
    & ~ + (t \text{-channel}) + (u \text{-channel})~,\\
    B^{\zeta,\text{PO}}_{(2,\pm 2)}=&\,2i\Delta^4_{\zeta}(2\pi)^3\cos(\pi\nu)H^2\big|Z^{+2}_2(s)|^2\left(\frac{H}{\Lambda_1}\right)^2\left(\frac{H}{\Lambda_2}\right)^3\frac{\left[s^2\mathbf{k}_2\cdot\mathbf{k}_4-\left(\mathbf{k}_2\cdot\mathbf{s}\right)\left(\mathbf{k}_4\cdot\mathbf{s}\right)\right]}{k_1k_2k_3k_4}\frac{\mathbf{s}\cdot\left(\mathbf{k}_1\times\mathbf{k}_3\right)}{k_2^2k_4^2s^9}\nonumber\\
    &\times\left(1-k_2\frac{\partial}{\partial k_2}\right)\left(1-k_4\frac{\partial}{\partial k_4}\right)\mathcal{I}^0_1(s,k_{12},\nu)\mathcal{I}^0_2(s,k_{34},\nu)+\text{7~perms} \nonumber \\
    & ~ + (t \text{-channel}) + (u \text{-channel})~.
\end{align}
\end{keyeqn}
Again we see that each component is purely imaginary and has the correct scaling with momenta (in checking this point we use the fact that $Z_{2}^{h}(s) \sim s^{1/2}$). We also see that for the special case of $\nu = 1/2$ the trispectrum vanishes. This limit corresponds to the partially-massless limit where the massive spin-$2$ field has only four propagating degrees of freedom (since the $h=0$ modes do not contribute here we don't need to worry about the divergence of the corresponding two-point function in this PM limit). In this limit the mode functions simplify to exponentials and again the no-go theorem of \cite{Cabass:2022rhr} dictates that the result should vanish. 

\subsection{Example 4: spin-0 exchange with linear mixing}\label{Example4}
In this final example we will show that, perhaps counter-intuitively, a parity-odd trispectrum can be generated in a pure scalar theory at tree-level, provided that we allow for a linear-mixing between the inflaton and a new massive scalar field which we denote as $\sigma$.\footnote{See \cite{Lee:2023jby} for a discussion of how a parity-odd trispectrum can be generated at loop level.} We consider a theory with a parity-odd vertex of the form $\pi \pi \pi \sigma$ and a parity-even linear mixing of the form $\pi \sigma$. We take $\sigma$ to be in the complementary series but can again analytically continue the final result to capture the contribution from a principle series field. In terms of wavefunction coefficients, the trispectrum is then 
\begin{align}
	\begin{gathered}
		\includegraphics[width=0.7\textwidth]{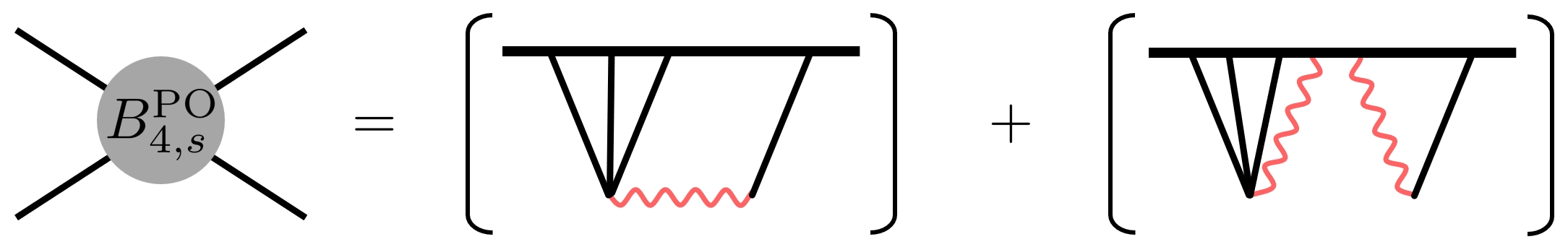}
	\end{gathered}~.\label{POlinear}
\end{align}
According to our $\psi_n$-reality theorem, the wavefunction coefficient $\psi_4$ is strictly real after Wick rotation, assuming IR convergence. Thus the first term in \eqref{POlinear} drops out when projected onto the parity-odd component. It is crucial that we only require IR convergence of the full $\psi_n$ rather than the individual vertices. Indeed, for certain couplings the mixing vertex appears to diverge as $\eta$ approaches zero using a naive power counting, yet the wavefunction remains IR finite and our reality and factorisation theorems still apply. For example, consider the lowest-order mixing operator that respects the shift symmetry of the Goldstone mode
\begin{align}
    \int d\eta\,  a^3\pi'\sigma~.
\end{align}
In the late-time limit, a naive scaling of the field operators gives $\pi'\sim \eta$ and $\sigma \sim \eta^{3/2-\nu}$. Then the total scaling of integrand is $\eta^{-1/2-\nu}$ which is superficially IR divergent for $\nu\geq 1/2$. However, since we are computing wavefunction coefficients, the vertex integrals behave slightly differently from the naive expectation using field operators, thanks to the vanishing boundary condition satisfied by the bulk-bulk propagator of the exchanged field which is given by
\begin{align}
    G_{\sigma}(\eta_{1}, \eta_{2},k) = -2 i P_{\sigma}(\eta_{0},k)\, K_{\sigma}(\eta', k) ~ \Im K_{\sigma}(\eta, k ) \theta(\eta - \eta') + (\eta \leftrightarrow \eta')~.
\end{align}
At late times the scaling of the propagator is dictated by $P_\sigma^{1/2}\Im K_\sigma$\footnote{It is $P_\sigma^{1/2}$ that contributes rather than $P_\sigma$ since its scaling is shared between the two vertices that are connected by this propagator.}, which enjoys a \textit{softer} IR scaling than $\sigma$:
\begin{align}
    P_\sigma^{1/2}(\eta_0,k)\,\Im K_\sigma(\eta,k)\sim (-\eta_0)^{3/2+\nu}\left(1-\frac{\eta}{\eta_0}+\cdots\right)~,\quad \eta,\eta_0\to 0^{-}~.
\end{align}
Therefore, the actual integrand scales as $(-\eta_0)^{-1/2+\nu}$ assuming that the linear-mixing vertex is later than the quartic vertex. This contribution to the wavefunction therefore remains convergent as $\eta_0\to 0$ for the whole range of light masses. Furthermore, as we will see below, the quartic vertices we will consider always have enough derivatives to also ensure IR convergence when the linear-mixing vertex is earlier.  We can therefore safely ignore $\psi_4$, and the only parity-odd contribution to the trispectrum comes from the factorized piece such that
\begin{align}
		\begin{gathered}
			\includegraphics[width=0.38\textwidth]{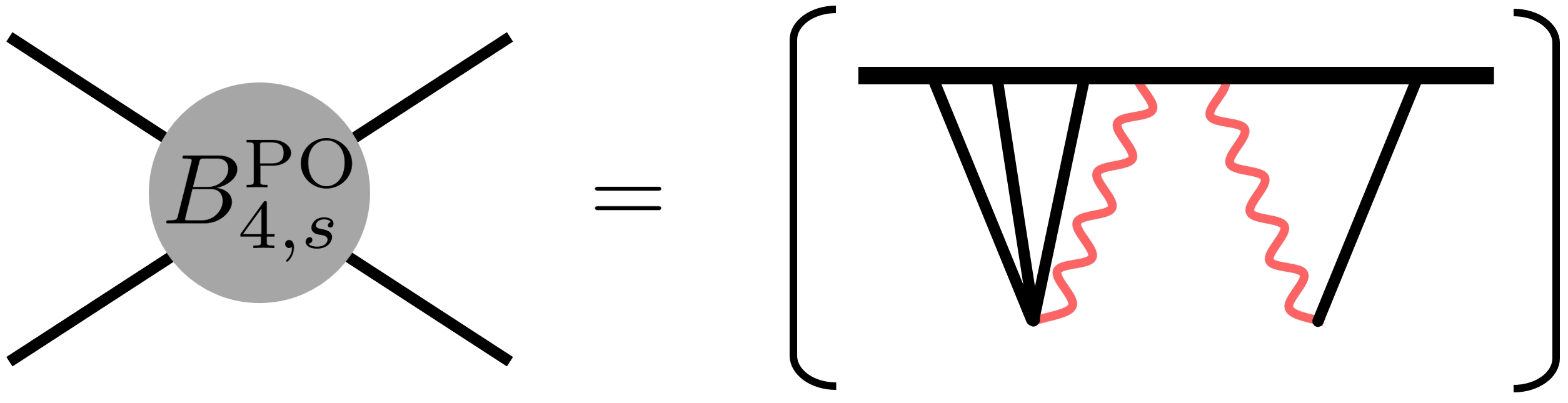}
		\end{gathered}~,\quad\text{(light fields).}\label{BPOLightFormula2}
\end{align}

To give explicit examples of a non-vanishing parity-odd trispectrum from scalar exchange, we now construct parity-odd vertices of the form $\pi \pi \pi \sigma$. Assuming dilation, rotation and translational invariance, the leading two operators are of dimension-10:
\begin{align}
& a^{-2}\sigma \epsilon_{ijk} \partial_{i} \pi '' \partial_{j} \pi' \partial_{k} \pi , \\
& a^{-2}\sigma \epsilon_{ijk} \partial_{m} \partial_{i} \pi ' \partial_{m}\partial_{j} \pi \partial_{k} \pi. \label{PO_int}
\end{align}
How can these arise in the EFToI? In general we have $B \sigma$, $BB \sigma$ and $BBB \sigma$ operators that we can consider where $B$ are building blocks that start at linear order in $\pi$ with the linear terms of the form $\pi'$ or $\partial_i \partial_j \pi$. Clearly the operators we have above cannot come from $BBB \sigma$ operators since they contain single spatial derivatives acting on a $\pi$. The $B \sigma$ option also cannot yield something parity-odd. We are therefore left with $BB \sigma$. Consider the first of the above dimension-$10$ operators. The $\partial_{k} \pi$ term needs to come from $\delta K_{ij} \sim \partial_{i} \partial_{j} \pi + \partial_{i} \pi \partial_{j} \pi + \ldots$. However since the partial derivatives in $\delta K_{ij}$ are symmetric we cannot get the anti-symmetric structure of this term. For the second we can realise this term from
\begin{align}
    \sigma\,n^{\mu}\mathbf{e}_{\mu\nu\rho\lambda}\nabla_\alpha \nabla^\nu \delta g^{00} \nabla^\rho K_{\alpha\lambda}~.
\end{align}
The leading $\pi \pi \sigma$ contribution vanishes by symmetry. 

Let us therefore consider the interaction in \eqref{PO_int}. The computation can be very easily extended to other interactions. In terms of the canonically normalised fields, the interaction part of the action is therefore
\begin{align}
	S_{\text{int}}=\int d^3x d\eta\bigg(\rho\,a^3\pi'_c\sigma+\frac{a^{-2}}{\Lambda^6}\sigma \epsilon_{ijk} \partial_{m} \partial_{i} \pi_c ' \partial_{m}\partial_{j} \pi_c \partial_{k} \pi_c\bigg)~,
\end{align}
where as always the scale factors are fixed by scale invariance. The two-point wavefunction coefficient $\psi_{1,1}$ from the quadratic mixing can be directly calculated as
\begin{align}\label{Psi11}
    \nonumber\psi_{1,1}=&\quad
    \begin{gathered}
        \includegraphics[width=0.1\textwidth]{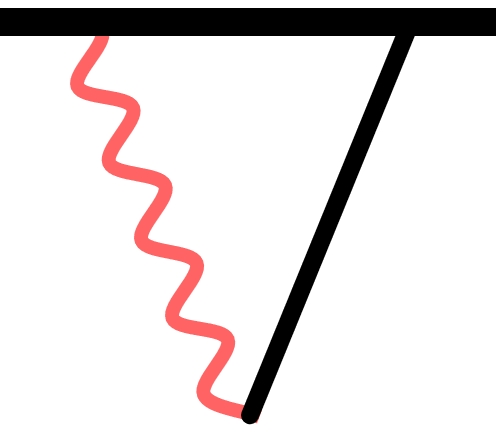}
    \end{gathered}\\
    \nonumber&=-\frac{\rho}{H^3}\frac{k^{3/2}_4}{(-\eta_0)^{3/2}H^{(2)}_{\nu}(-k_4\eta_0)}\Bigg[e^{{i\pi(\nu-1/2)}/2}\sqrt{2\pi}\sec{(\pi\nu)}\\
    &\qquad\qquad\qquad\qquad\qquad\qquad\qquad+\frac{2^{1+\nu}{(-k_4\eta_0)}^{\frac{1}{2}-\nu}\csc(\pi\nu)}{(2\nu-1)\Gamma(1-\nu)}-i\frac{2^{1+\nu}(-k_4\eta_0)^{\frac{3}{2}-\nu}\csc(\pi\nu)}{(2\nu-3)\Gamma(1-\nu)}+\cdots\Bigg]~,
\end{align}
where for simplicity we have set the speeds of sound to unity. Here, we truncate $\psi_{1,1}$ up to higher orders in the time cutoff $\eta_0$ which vanish in the limit $\eta_0\to 0^-$. For the mass range $0\le\nu<\frac{1}{2}$, the time integral is IR-finite and only the first line in the bracket contributes to the wavefunction coefficient. For the mass range $\frac{1}{2} \leq \nu \leq \frac{3}{2}$, the second line of \eqref{Psi11} actually diverges as the time cutoff is sent to zero. Despite the IR divergence in $\psi_{1,1}$, it is noteworthy that the density matrix diagonal as well as the correlator remain IR-convergent as long as $\sigma$ is massive. This is attributed to the cancellation of the IR divergence after adding the Hermitian conjugation. More explicitly, using the property $H^{(2)}_{\nu}=-H^{(1)}_\nu,~\eta\to 0^-$, we obtain the corresponding density matrix diagonal 
\begin{align}
    \rho_{1,1
}(\mathbf{k}_4)&=\psi_{1,1}(\mathbf{k}_4)+\psi^*_{1,1}(\mathbf{k}_4)\nonumber\\
    &=i\frac{\rho}{H^3}\frac{\sqrt{2\pi}\csc\left({\pi\nu}/{2}+{\pi}/{4}\right)}{H^{(2)}_{\nu}(-k_4\eta_0)}\left(-\frac{k_4}{\eta_0}\right)^{3/2}~, \qquad 0\le\nu<\frac{3}{2}~,
\end{align}
where the divergences in the wavefunction coefficient $\psi_{1,1}$ manifestly drop out. However, when the $\sigma$ field is exactly massless, the expression above becomes invalid due to the singularity of the factor $\csc\left({\pi\nu}/{2}+{\pi}/{4}\right)$ at $\nu=3/2$. We note that in the limit $\nu\to 3/2$, an additional contribution arises from the last term in the bracket of equation \eqref{Psi11} that goes like $(-k_4\eta_0)^{3/2-\nu}\sim 1+(3/2-\nu)\ln (-k_4\tau_0)$. This contribution serves to remove the pole of the $\sec(\pi\nu)$ factor in the first line of \eqref{Psi11}. Consequently, the density matrix diagonal for the massless case becomes
\begin{align}
    \rho_{1,1
    }(\mathbf{k}_4)\sim -2 k_4^3 \log\left(-k_4 \eta_0\right)~,\label{logdivrho11}
\end{align}
which contains the logarithmic term as typically expected for massless exchange \cite{Wang:2022eop}. For the other factorized component, the parity-odd wavefunction $\psi_{3,1}$ can be expressed as 
\begin{align}
    \nonumber\psi_{3,1}
    &=
    \begin{gathered}
        \includegraphics[width=0.1\textwidth]{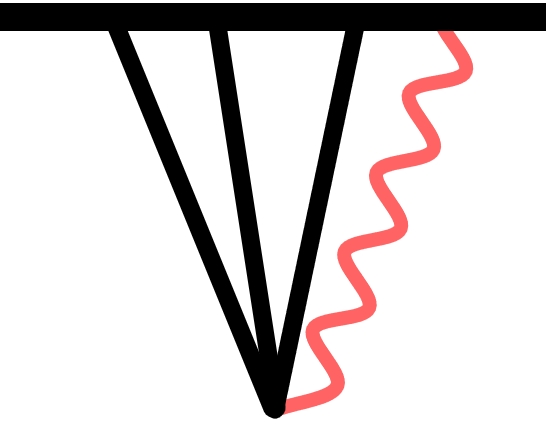}
    \end{gathered}\\
    \nonumber&=-\left(\mathbf{k}_1\cdot\mathbf{k}_2\right)\left[\mathbf{k}_1\cdot\left(\mathbf{k}_2\times\mathbf{k}_3\right)\right]\frac{H^2}{\Lambda^6}\frac{ e^{{i\pi(\nu-1/2)}/2}\,k_1^2}{k_4^{11/2}(-\eta_0)^{3/2}H^{(2)}_\nu(-k_4\eta_0)}\sqrt{\frac{2}{\pi}}\\
    &\quad\times\left(1-k_2\frac{\partial}{\partial{k_2}}\right)\left(1-k_3\frac{\partial}{\partial{k_3}}\right)\mathcal{I}^0_4(k_4,k_{123},\nu)+\text{6 perms}~.
\end{align}
For the massless case, it is apparent that the density matrix $\rho_{3,1}$ vanishes as a power-law function of the cutoff time $\eta_0$. Therefore, even with a logarithmically divergent $\rho_{1,1}$ \eqref{logdivrho11}, exchanging a massless scalar \textit{cannot} generate a parity-odd trispectrum, reinforcing the no-go theorems \cite{Liu:2019fag,Cabass:2022rhr}. As a summary, the parity-odd tripsectrum from massive scalar exchange is thus finite and scale-invariant and can be exactly computed to be
\begin{keyeqn}
\begin{align}
    B_4^{\zeta,\rm{PO}}=\,&i\,\frac{\pi^5\Delta_\zeta^4}{32}\,\bigg(\frac{\rho}{H}\bigg)\bigg(\frac{H}{\Lambda}\bigg)^6\Gamma\left(\frac{11}{2}-\nu\right)\Gamma\left(\frac{11}{2}+\nu\right)\frac{\left(\mathbf{k}_1\cdot\mathbf{k}_2\right)\left[\mathbf{k}_1\cdot\left(\mathbf{k}_2\times\mathbf{k}_3\right)\right]}{k_1 k_2^3k_3^3k_4^7}\nonumber\\
    &\times\left(1-k_2\frac{\partial}{\partial{k_2}}\right)\left(1-k_3\frac{\partial}{\partial{k_3}}\right){}_2\tilde{\rm{F}}_1\Bigg[\begin{array}{c} \frac{11}{2}-\nu, \frac{11}{2}+\nu\\[2pt] 6 \end{array}\Bigg|\,\frac{1}{2}-\frac{k_{123}}{2k_4}\Bigg]+\text{23 perms}~.
\end{align}
\end{keyeqn}
We point out that this example is, to our knowledge, the first one realising a tree-level parity-violating trispectrum in a pure scalar theory. Its analytic structure is also distinct from other examples involving triple vertices. Notably, there is only a single hypergeometric function of kinematics. This simplification arises from the fact that linear coupling reduces the other hypergeometric function to a constant. In the limit $k_T=k_{123}+k_4\to 0$, the hypergeometric function encounters a singularity at unit argument, which is precisely the fake total-energy singularity we discussed above that arises as a partial-energy singularity but turns into a total-energy one by momentum conservation at the quadratic vertex. To check this result we have also numerically computed the trispectrum and find agreement.

\section{Conclusions and outlook} \label{Conclusions}

In this paper we have derived some reality theorems relating to cosmological wavefunction coefficients of massless scalars, massless gravitons and conformally coupled scalars. We have shown that the maximally-connected part of these coefficients, which is $i)$ the most difficult to compute since it contains the maximum number of nested time integrals and $ii)$ the part that can be singular as the total-energy goes to zero $k_{T} \rightarrow 0 $, is a purely real function of the kinematics. Our results allow for the exchange of states with any mass and integer spin, and in deriving our results we considered two distinct descriptions for the dynamics of massive spinning fields during inflation: cosmological condensed matter physics (where states are representations of the group of rotations) and cosmological collider physics (where states are representations of the de Sitter group). Furthermore, if all exchanged fields are in the complementary series i.e. they have light masses, then our reality theorem extends beyond the maximally-connected part to the full wavefunction coefficient. Our results apply under the following assumptions:
\begin{itemize}
\item \textbf{Tree-level approximation}: we considered tree-level Feynman diagrams which allowed us to avoid having to analytically continue the spacetime dimensions. This meant that we could work with fixed external propagators which have simple properties in $D = 3+1$, namely that they are real after Wick rotation. In Appendix \ref{CuttingComparison} we offer an alternative proof of our theorems using Hermitian analyticity of all propagators, and scale invariance. Here we use the relation $V = I+1$ where $V$ is the number of vertices and $I$ is the number of internal propagators. This relation holds only at tree-level. By relaxing the tree-level approximation the maximally-connected parts of the wavefunction coefficients can have imaginary parts \cite{Lee:2023jby}. 
\item \textbf{Bunch-Davies vacuum conditions}: this assumption enabled us to rotate all time variables by $90^{\circ}$ in the complex plane as a tool for computing wavefunction coefficients. The fact that we are computing the vacuum wavefunction for which fields vanish exponentially fast in the far past allowed us to close the contour and drop any contributions from the arc such that integration along the real line could be replaced by integration along the imaginary line. This assumption is relevant for our proof in Appendix \ref{CuttingComparison} since Hermitian analyticity is closely tied to having Bunch-Davies vacuum conditions. If we relax this assumption, for example with the Ghost Condensate, then the maximally-connected parts of the wavefunction coefficients can have imaginary parts \cite{Cabass:2022rhr}. 
\item \textbf{Scale invariance}: this assumption ensured that the vertex operators were real after Wick rotation. Indeed, scale invariance ensures that time derivatives enter as $\eta \partial_\eta$ and spatial momenta enter as $i \eta \bfk$. This could then be combined with the reality properties of the propagators to prove that various integrands are purely real. If we relax this assumption, for example by allowing for time-dependent couplings or going to general FLRW spacetimes (except when the scale factor is 
an odd-power-law function of the conformal time), then the maximally-connected parts of the wavefunction coefficients can have imaginary parts \cite{Cabass:2022rhr,Soda:2011am}.
\item \textbf{IR-convergence}: this assumption enabled us to make the leap from proving the reality of integrands to the reality of the integrated result. Indeed, IR-convergence meant that the final result was independent of $\eta_0$ and so we did not need to worry about rotating this cut-off. In the presence of an IR-divergence we would need to also rotate $\eta_0$ and this can yield imaginary parts with a logarithmic-divergence, for example. In Appendix \ref{CuttingComparison} we combined scale invariance of the bulk interactions with IR-convergence to use the fact that the wavefunction coefficients have a fixed scaling with momenta yielding simple transformation properties as all momenta and energies flip sign. If we relax this assumption, for example by allowing for IR-divergent bulk interactions as occurs for the minimal coupling between the inflaton and the massless graviton, then the maximally-connected parts of the wavefunction coefficients can have imaginary parts \cite{Creque-Sarbinowski:2023wmb}.
\end{itemize}

The reality of the maximally-connected part of wavefunction coefficients is not just of theoretical interest, rather it can make the computation of phenomenologically relevant (\cite{Philcox:2022hkh,Cabass:2022oap,Coulton:2023oug,Hou:2022wfj,Philcox:2023ypl}) cosmological correlators a far simpler task than naively expected. We have shown this in Section \ref{ExactTrispectraSection} by considering the parity-odd trispectra of curvature perturbations due to the coupling of the inflaton with another sector with massive spinning fields and parity-violation. Since parity-odd correlators depend on the imaginary part of the maximally-connected wavefunction coefficients, these trispectra are factorised and computed by considering cubic diagrams only. We presented a number examples considering both the CCM and CC scenarios, both light and heavy fields (with the final answers related by analytic continuation), and both parity-violation arising from the free theory of the massive spinning fields and from the bulk interactions. In particular, we considered a parity-violating correction to the action of a massive vector field in Section \ref{Example1} and compared our result with that computed in \cite{Jazayeri:2023kji} using a non-local EFT arising from integrating out the massive vector field. Our result recovers the EFT result in the appropriate limit, but also gives an exact result in the regime where the EFT breaks down. This example also includes axion-$U(1)$ gauge field inflation. We also considered examples with massive spin-$2$ fields in Sections \ref{Example2} and \ref{Example3}. In cases with linear-mixing couplings, we were able to establish a first-of-its-kind example of a trispectrum in scalar theories that is parity-violating at tree level. For all spins we allowed for a chemical potential in our analysis, and for hierarchies between the speed of the inflaton and the exchanged fields, such that the size of trispectra can be enhanced. 

There are many avenues for future research directions and here we outline a few:
\begin{itemize}
    \item \textbf{Moving on to loops:} it would be interesting to see if any of our reality theorems hold at loop level. As we mentioned above, total-energy poles coming from loops can be imaginary, but perhaps the structure of such imaginary terms can be constrained given that the reality properties of bulk-bulk propagators still hold. In fact, in the original $D=3+1$ spacetime dimensions, the loop integrand for $\psi_n$ of light fields still appears to be purely real after Wick rotation, but the dimensional regulator demands an evaluation in $D=4-\epsilon$ dimensions. The Wick rotation in such a case is expected to bring factors of $(-\eta)^{\epsilon}=(-i\chi)^{\epsilon}=(1-i\pi\epsilon/2)\chi^{\epsilon}$. The $\mathcal{O}(\epsilon)$ imaginary part turns out to be cancellable by the $1/\epsilon$ divergences, giving finite imaginary contributions to $\psi_n$. This has been demonstrated for massless and conformally coupled theories in \cite{Lee:2023jby}. It would be interesting to see if such a phenomenon persists for general massive fields and at higher loops.
    \item \textbf{Distinguishing between massive spinning field set-ups}: we have considered two different descriptions of massive spinning fields during inflation (CC and CCM). It would be interesting to investigate if these two set-ups could be distinguished from each other at the level of massless scalar and massless graviton correlation functions. 
    \item \textbf{Kramers-Kronig for correlators}: it would be interesting to investigate if the parity-even part of a correlator can be constrained given the parity-odd part. This is conceivable for examples where the parity-violation is driven by a chemical potential correction to the free theory. Perhaps consistency of higher-point functions could be used to constrain lower-point ones in this regard. Since the parity-odd part is always imaginary and the parity-even part is always real, such reconstruction of a full correlator from its imaginary part, if possible, will be an interesting cosmological analogy of the well-celebrated Kramers-Kronig relations in electromagnetism.
    \item \textbf{EAdS perspective}: the Wick rotation used extensively throughout this work is in fact a contour deformation, i.e. a change of integrated bulk time which are dummy variables. However it is tempting to conceive a further Wick rotation for the boundary time $\eta_0=i\chi_0$ which is explicit. In doing so, one arrives at a quantity \textit{different} from the wavefunction of the universe. The reality of this quantity is especially transparent since the whole spacetime becomes Euclidean. In fact, a further rotation of the Hubble parameter $H=-i/L_{\mathrm{AdS}}$ yields a theory defined in Euclidean Anti-de Sitter (EAdS) space. The wavefunction then becomes the partition function of the boundary CFT of the EAdS bulk \cite{Sleight:2021plv}. The difficulty, however, is in the fact that the continuations of $\eta_0$ and $H$ do not seem to commute with the recipe of obtaining correlators from wavefunction coefficients, which relies on the unitarity of the wavefunction. It would be helpful to understand how to correctly perform this continuation, i.e. how to understand in-in correlators in Euclidean field theories.
    \item \textbf{Classifying singularities}: we have concentrated on the total-energy singularities in this work, yet they are by no means the only singularities of the wavefunction. Notably, there are also partial-energy singularities when the total energy flowing into a sub-component of a Feynman diagram approaches zero. For IR-convergent massless or conformally coupled theories in de Sitter and flat spacetime, these singularities are always poles at tree-level since the wavefunction coefficients are rational functions of momenta \cite{Arkani-Hamed:2017fdk,Salcedo:2022aal}. However, in more general theories with arbitrary mass, spin, sound speed and chemical potential, these singularities have not yet been classified even at tree-level. The classification of these singularities would crucially serve as a first step toward a complete understanding of the analytic structure of wavefunction of the universe.
\end{itemize}


\paragraph{Acknowledgements} We thank Paolo Benincasa, Zongzhe Du, Sadra Jazayeri, Austin Joyce, Hayden Lee, Arthur Lipstein, Scott Melville, Enrico Pajer, Sébastien Renaux-Petel, Denis Werth, Yi Wang, Zhong-Zhi Xianyu and Yang Zhang for helpful discussions. D.S. is supported by a UKRI Stephen Hawking Fellowship [grant number EP/W005441/1] and a Nottingham Research Fellowship from the University of Nottingham. XT and YZ are supported in part by
the National Key R$\&$D Program of China (No. 2021YFC2203100). XT is also supported by STFC consolidated grants ST/T000694/1 and ST/X000664/1. YZ is also supported by the IBS under the project code, IBS-R018-D3. D.S. thanks The Hong Kong University of Science and Technology for kind hospitality. For the purpose of open access, the authors have applied a CC BY public copyright licence to any Author Accepted Manuscript version arising.

\newpage
\appendix

\section{General solution of $\Delta G_{\sigma}^{(h)}$}\label{GeneralDeltaGhAppendix}

In this appendix we construct the general solution of $\Delta G_{\sigma}^{(h)}$ such that the connected propagator $C_{\sigma}^{(h)} = G_{\sigma}^{(h)}+\Delta G_{\sigma}^{(h)}$ is helically real after Wick rotation. We allow for a chemical potential $\kappa$ in the massive field's dispersion relation in the CCM scenario such that the relevant mode function is given by a Whittaker function, c.f. \eqref{CCMmodefunction}. Recall that we need to decompose the bulk-bulk propagator into two parts,
\begin{align}
	C_{\sigma}^{(h)}(\eta_{1}, \eta_{2},k) &=[\sigma_{h}(\eta_{1}, k)\sigma^{*}_{h}(\eta_{2},k) \theta(\eta_{1}-\eta_{2}) + (\eta_{1} \leftrightarrow \eta_{2})] + \Delta G_{\sigma}^{(h)}(\eta_{1}, \eta_{2},k)~, \\
	F_{\sigma}^{(h)}(\eta_{1}, \eta_{2},k) &= - \frac{\sigma_{h}(\eta_{0},k)}{\sigma^{*}_{h}(\eta_{0},k)}\sigma^{*}_{h}(\eta_{1},k)\sigma^{*}_{h}(\eta_{2},k) - \Delta G_{\sigma}^{(h)}(\eta_{1}, \eta_{2},k)~.
\end{align}
In order to ensure that the connected part still satisfies the propagator equation (\ref{PropagatorEoMhelicityBasisCCMPPV}), the added term must satisfy the homogeneous equation
\begin{align}
	\left(\eta_{1}^2 \frac{\partial^2}{\partial \eta_{1}^2} - 2 \eta_{1}  \frac{\partial}{\partial \eta_{1}} + c_{h,S}^2 k^2 \eta_{1}^2- 2c_{h,S}\tilde{\kappa}\, k  + \frac{m^2}{H^2} \right) \Delta G_{\sigma}^{(h)}(\eta_{1}, \eta_{2},k) =0~.
\end{align}
The UV convergence of bulk time integrals requires the Bunch-Davies initial condition at $\eta\to -\infty$, namely
\begin{align}
	\lim_{\eta_1, \eta_2 \rightarrow -\infty(1 - i \epsilon)}\Delta G_{\sigma}^{(h)}(\eta_1, \eta_2, k) = 0~.
\end{align}
Note that in contrast to (\ref{PropagatorBCs}), we do not impose any boundary condition at $\eta_0$ for $\Delta G^{(h)}$. Upon symmetrizing over $\eta_1\leftrightarrow\eta_2$, we are left with
\begin{align}
	\Delta G_{\sigma}^{(h)}(\eta_{1}, \eta_{2},k) = \mathcal{A}_h \, \sigma_{h}^{*}(\eta_1, k) \sigma_{h}^{*}(\eta_2, k)~.
\end{align}
Here $\mathcal{A}_h=\mathcal{A}_h(\kappa, \mu)$ is a helicity-dependent constant to be determined. Now we demand that this connected propagator is helically real after rotation, i.e.
\begin{align}
	\left[\tilde{C}_{\sigma}^{(h)}(\chi_{1}, \chi_{2},k)\right]^{\star}=\tilde{C}_{\sigma}^{(-h)}(\chi_{1}, \chi_{2},k)~.\label{CRealityRequirement}
\end{align}
Due to the symmetry between $\chi_1\leftrightarrow\chi_2$, we are free to pick $\chi_1<\chi_2$ without loss of generality. Thus the Wick-rotated connected propagator becomes
\begin{align}
	\nonumber\tilde{C}_{\sigma}^{(h)}(\chi_{1}, \chi_{2},k)&=\left[\sigma_h(i e^{i\epsilon}\chi_1,k)+\mathcal{A}_h \sigma_h^*(i \chi_1,k)\right]\sigma_h^*(i\chi_2,k)\\
	&=-\frac{H^2 \chi_1 \chi_2}{2 c_{h,S}k}e^{-\pi\tilde{\kappa}} \nonumber \\
 & ~~~~~~~ \times \left[W_{i\tilde{\kappa},i\mu}(-2e^{i\epsilon}c_{h,S}k\chi_1)+\mathcal{A}_h W_{-i\tilde{\kappa},-i\mu}(2c_{h,S}k\chi_1)\right]W_{-i\tilde{\kappa},-i\mu}(2c_{h,S}k\chi_2)~.
\end{align}
Notice that Whittaker functions have a branch cut along the negative real axis, which is why we have kept the $e^{i\epsilon}$ factor to ensure that the branch cut is never crossed. However, the complex conjugation of Whittaker functions is most transparent when their argument lies along the positive real axis,
\begin{align}
	\left[W_{a,b}(z)\right]^*=W_{a^*,b^*}(z)~,\quad\left[M_{a,b}(z)\right]^*=M_{a^*,b^*}(z)~,\quad z\in \mathbb{R}_+~.
\end{align}
To inspect the complex conjugation $[\tilde{C}_{\sigma}^{(h)}]^{\star}$, we can first expand the Whittaker $W$-functions in terms of Whittaker $M$-functions,
\begin{align}
	W_{a, b}(z) = \frac{\Gamma(-2 b)}{\Gamma\left(\frac{1}{2} - b - a\right)}M_{ a,b}(z)+\frac{\Gamma\left(2 b\right)}{\Gamma\left(\frac{1}{2} +  b - a\right)}M_{a,-b}(z)~,\label{WtoMeqn}
\end{align}
and then rotate away arguments lying below the branch cut using
\begin{align}
	M_{a, b}(-e^{i\epsilon} z) = -i e^{- i \pi b}M_{-a, b}(z )~.
\end{align}
This yields
\begin{align}
	\nonumber\tilde{C}_{\sigma}^{(h)}&(\chi_{1}, \chi_{2},k) =-\frac{H^2 \chi_1 \chi_2}{2c_{h,S} k}e^{-\pi\tilde{\kappa}} \nonumber \\
 & \times \Bigg\{\frac{\Gamma (-2 i \mu )^2 }{\Gamma \left(\frac{1}{2}+i \tilde\kappa -i \mu\right)^2}\left[\mathcal{A}_h-\frac{i e^{\pi  \mu } \Gamma \left(\frac{1}{2}+i \tilde\kappa -i \mu \right)}{\Gamma \left(\frac{1}{2}-i \tilde\kappa -i \mu\right)}\right]M_{-+}(1) M_{-+}(2) \nonumber \\
	\nonumber&+\frac{\pi \,\text{csch}(2 \pi  \mu ) }{2 \mu \, \Gamma \left(\frac{1}{2}+i \tilde\kappa -i \mu\right) \Gamma \left(\frac{1}{2}+i \tilde\kappa +i \mu\right)}\left[\mathcal{A}_h-\frac{i e^{-\pi  \mu } \Gamma \left(\frac{1}{2}+i \tilde\kappa +i \mu \right)}{\Gamma \left(\frac{1}{2}-i	\tilde\kappa +i \mu\right)}\right]M_{--}(1) M_{-+}(2)\\
	&+(\mu\to-\mu)\Bigg\}~,
\end{align}
where we have abbreviated
\begin{align}
	M_{\pm\pm}(j)\equiv M_{\pm i\tilde{\kappa},\pm i\mu}(2c_{h,S} k \chi_j)~,\quad M_{\pm\mp}(j)\equiv M_{\pm i\tilde{\kappa},\mp i\mu}(2c_{h,S} k \chi_j)~,\quad j=1,2~.
\end{align}
Now we can perform a complex conjugation on the Wick-rotated propagator and obtain
\begin{align}
	[\tilde{C}_{\sigma}^{(h)}&(\chi_{1}, \chi_{2},k)]^*=-\frac{H^2 \chi_1 \chi_2}{2c_{h,S} k}e^{-\pi\tilde{\kappa}} \nonumber \\
 \times &\Bigg\{\frac{\Gamma (2 i \mu )^2 }{\Gamma \left(\frac{1}{2}-i \tilde\kappa +i \mu\right)^2}\left[\mathcal{A}_h^*+\frac{i e^{\pi  \mu } \Gamma \left(\frac{1}{2}-i \tilde\kappa +i \mu \right)}{\Gamma \left(\frac{1}{2}+i \tilde\kappa +i \mu\right)}\right]M_{+-}(1) M_{+-}(2) \nonumber \\
	\nonumber&+\frac{\pi \,\text{csch}(2 \pi  \mu ) }{2 \mu \, \Gamma \left(\frac{1}{2}-i \tilde\kappa +i \mu\right) \Gamma \left(\frac{1}{2}-i \tilde\kappa -i \mu\right)}\left[\mathcal{A}_h^*+\frac{i e^{-\pi  \mu } \Gamma \left(\frac{1}{2}-i \tilde\kappa -i \mu \right)}{\Gamma \left(\frac{1}{2}+i	\tilde\kappa -i \mu\right)}\right]M_{++}(1) M_{+-}(2)\\
	&+(\mu\to-\mu)\Bigg\}~,\label{ConjugateC}
\end{align}
while a helicity flip gives
\begin{align}
	\nonumber\tilde{C}_{\sigma}^{(-h)}&(\chi_{1}, \chi_{2},k) =-\frac{H^2 \chi_1 \chi_2}{2c_{h,S} k}e^{\pi\tilde{\kappa}} \nonumber \\
 & \times \Bigg\{\frac{\Gamma (-2 i \mu )^2 }{\Gamma \left(\frac{1}{2} - i \tilde\kappa -i \mu\right)^2}\left[\mathcal{A}_{-h}-\frac{i e^{\pi  \mu } \Gamma \left(\frac{1}{2} - i \tilde\kappa -i \mu \right)}{\Gamma \left(\frac{1}{2} + i \tilde\kappa -i \mu\right)}\right]M_{++}(1) M_{++}(2) \nonumber \\
	\nonumber&+\frac{\pi \,\text{csch}(2 \pi  \mu ) }{2 \mu \, \Gamma \left(\frac{1}{2} - i \tilde\kappa -i \mu\right) \Gamma \left(\frac{1}{2} - i \tilde\kappa +i \mu\right)}\left[\mathcal{A}_{-h}-\frac{i e^{-\pi  \mu } \Gamma \left(\frac{1}{2}-i \tilde\kappa +i \mu \right)}{\Gamma \left(\frac{1}{2} + i	\tilde\kappa +i \mu\right)}\right]M_{+-}(1) M_{++}(2)\\
	&+(\mu\to-\mu)\Bigg\}~. \label{HelicityflipC}
\end{align}
By comparing the coefficients of (\ref{ConjugateC}) and (\ref{HelicityflipC}), we conclude that reality of the connected propagator requires us to satisfy:
\begin{align}
	e^{\pi \tilde{\kappa}}\mathcal{A}_{-h}-e^{-\pi \tilde{\kappa}}\mathcal{A}_{h}^*=\frac{2i\pi}{\Gamma\left(\frac{1}{2}+i	\tilde\kappa -i \mu) \Gamma(\frac{1}{2}+i	\tilde\kappa +i \mu\right)}~.\label{HelicalRealityConstraint}
\end{align}
In principle, $\mathcal{A}_{h}$ can be chosen to be an arbitrary complex constant as long as it satisfies (\ref{HelicalRealityConstraint}). However, it is often convenient to pick a symmetric choice such that
\begin{align}
	\mathcal{A}_{-h}=-\mathcal{A}_{h}^*~,
\end{align}
which yields (\ref{AhSolution}) in the main text i.e
\begin{align}
	\mathcal{A}_h = \frac{i \pi \sech(\pi \tilde{\kappa})}{\Gamma\left(\frac{1}{2} - i	\tilde\kappa -i \mu) \Gamma(\frac{1}{2} - i	\tilde\kappa +i \mu\right)}~.\label{Ahappresult}
\end{align}
This concludes the construction of the desired connected propagator in the CCM scenario.

In the CC scenario, we turn off the chemical potential, and the general constraint (\ref{HelicalRealityConstraint}) for the mode with $n=|h|$ reduces to
\begin{align}
	\mathcal{A}_{-h,|h|}-\mathcal{A}_{h,|h|}^*=2i\cosh\pi\mu_S~.
\end{align}
The simplest choice is to demand $\mathcal{A}_{-h,|h|}=\mathcal{A}_{h,|h|}=-\mathcal{A}_{h,|h|}^*$, which gives
\begin{align}
	\mathcal{A}_{h,|h|}=i\cosh\pi\mu_S~.
\end{align}
Note that this solution establishes the reality of $C^{h}_{|h|,S}$, while that of $C^{h}_{n,S},~n>|h|$ follows trivially from acting with real derivative operators on $C^{h}_{|h|,S}$, as we explained in the main text.

Finally, we point out that although the inclusion of heavy fields motivated the $\Delta G$ piece that achieves the reality of $C$, the same procedure works equally well for light fields except when $\nu$ is a half-integer. It is easy to see that all the derivations above straightforwardly go through with the replacement of $\mu\to-i\nu$. In fact, after choosing the specific choice \eqref{Ahappresult} and replacing $\mu$ by $-i\nu$, the connected propagator becomes identical to the original bulk-bulk propagator in the $\eta_0\to 0$ limit (c.f., \eqref{Canceleta0Whittaker}), i.e. $C=G$ and $F=0$. For the case of $\nu=n/2$, $n\in\mathbb{N}_+$, the equation \eqref{WtoMeqn} reaches a singularity, rendering the proof invalid. However, it is easy to check the validity of \eqref{Ahappresult} by inserting it into \eqref{CRealityRequirement}, using the same derivation that appeared in Section \ref{CCMPsubsection}. This crucially demonstrates that the proof of $k_T$-reality does not make any artificial distinction between light fields and heavy fields; we can extract the connected propagator part of all the internal lines regardless of their masses, and make use of their reality property to show the total-energy poles are always real.

\section{Proofs via the in-in/Schwinger-Keldysh formalism} \label{InIn}
In this appendix we streamline the derivation of the reality and parity-odd factorisation theorems in the language of the more conventional in-in and Schwinger-Keldysh formalisms. Both of them focus directly on the observable, i.e. the $n$-point correlation function itself, without introducing intermediate quantities such as the wavefunction representation of the quantum state of the universe. Therefore, we shall translate the $k_T$-reality theorem for wavefunction coefficients (\ref{kTRealityTheorem}) to the correlator level, and show that consequently all parity-odd correlators are factorised at tree-level. Here we will only deal with the CCM case, and the proof straightforwardly extends to the CC case as shown in the main text.

The perturbative computation of correlators can be organized using diagrammatics with slightly different Feynman rules from those of the cosmological wavefunction (see, for example, \cite{Chen:2017ryl}). In short, every $n$-point correlator is computed by a set of diagrams with coloured vertices indicating whether they are time-ordered (black or ``$+$'') or anti-time-ordered (white or ``$-$''). Thus in general, a diagram of $V$ vertices will comprise of $2^V$ coloured copies which need to be summed. A change from a black vertex to a white one (i.e. from $+$ to $-$) corresponds to a local complex conjugation plus a flip of the direction of all the 3-momenta flowing into the vertex. The internal propagators connecting these vertices are thus classified into four types according to the colour of their vertices:
\begin{subequations}
	\begin{align}
		\mathcal{G}_{++}(\eta_1,\eta_2,k)&=\theta(\eta_1-\eta_2)\varphi(\eta_1,k)\varphi^*(\eta_2,k)+\theta(\eta_2-\eta_1)\varphi^*(\eta_1,k)\varphi(\eta_2,k)~,\\
		\mathcal{G}_{+-}(\eta_1,\eta_2,k)&=\varphi^*(\eta_1,k)\varphi(\eta_2,k)~,\\
		\mathcal{G}_{-+}(\eta_1,\eta_2,k)&=\varphi(\eta_1,k)\varphi^*(\eta_2,k)~,\\ \mathcal{G}_{--}(\eta_1,\eta_2,k)&=\theta(\eta_1-\eta_2)\varphi^*(\eta_1,k)\varphi(\eta_2,k)+\theta(\eta_2-\eta_1)\varphi(\eta_1,k)\varphi^*(\eta_2,k)~.\label{SKPropDefs}
	\end{align}
\end{subequations}
The external propagators are simply obtained from sending one of the vertices to the boundary in the internal propagators,
\begin{subequations}
	\begin{align}
		\mathcal{K}_{+}(\eta,k)&=\mathcal{G}_{-+}(\eta_0,\eta,k)=\varphi(\eta_0,k)\varphi^*(\eta,k)~,\\
		\mathcal{K}_{-}(\eta,k)&=\mathcal{G}_{+-}(\eta_0,\eta,k)=\varphi^*(\eta_0,k)\varphi(\eta,k)~.
	\end{align}
\end{subequations}
All of these propagators satisfy the Bunch-Davies (or anti-Bunch-Davies for anti-time-ordered vertices) initial condition in the far past, while no boundary condition at $\eta_0$ is posed for them. Instead, they satisfy the conjugation rule
\begin{align}
	\left[\mathcal{G}_{\mathrm{ab}}(\eta_1,\eta_2,k)\right]^*&=\mathcal{G}_{\mathrm{(-a)(-b)}}(\eta_1,\eta_2,k)~,\\
	\left[\mathcal{K}_{\mathrm{a}}(\eta,k)\right]^*&=\mathcal{K}_{-\mathrm{a}}(\eta,k)~,\qquad\qquad \mathrm{a,b}=\pm~.
\end{align}
Notice that the flip of vertex colour is accompanied by a flip of momentum in the kinematic structure in addition to the complex conjugation.

\begin{figure}[ht]
	\centering
	\includegraphics[width=0.7\textwidth]{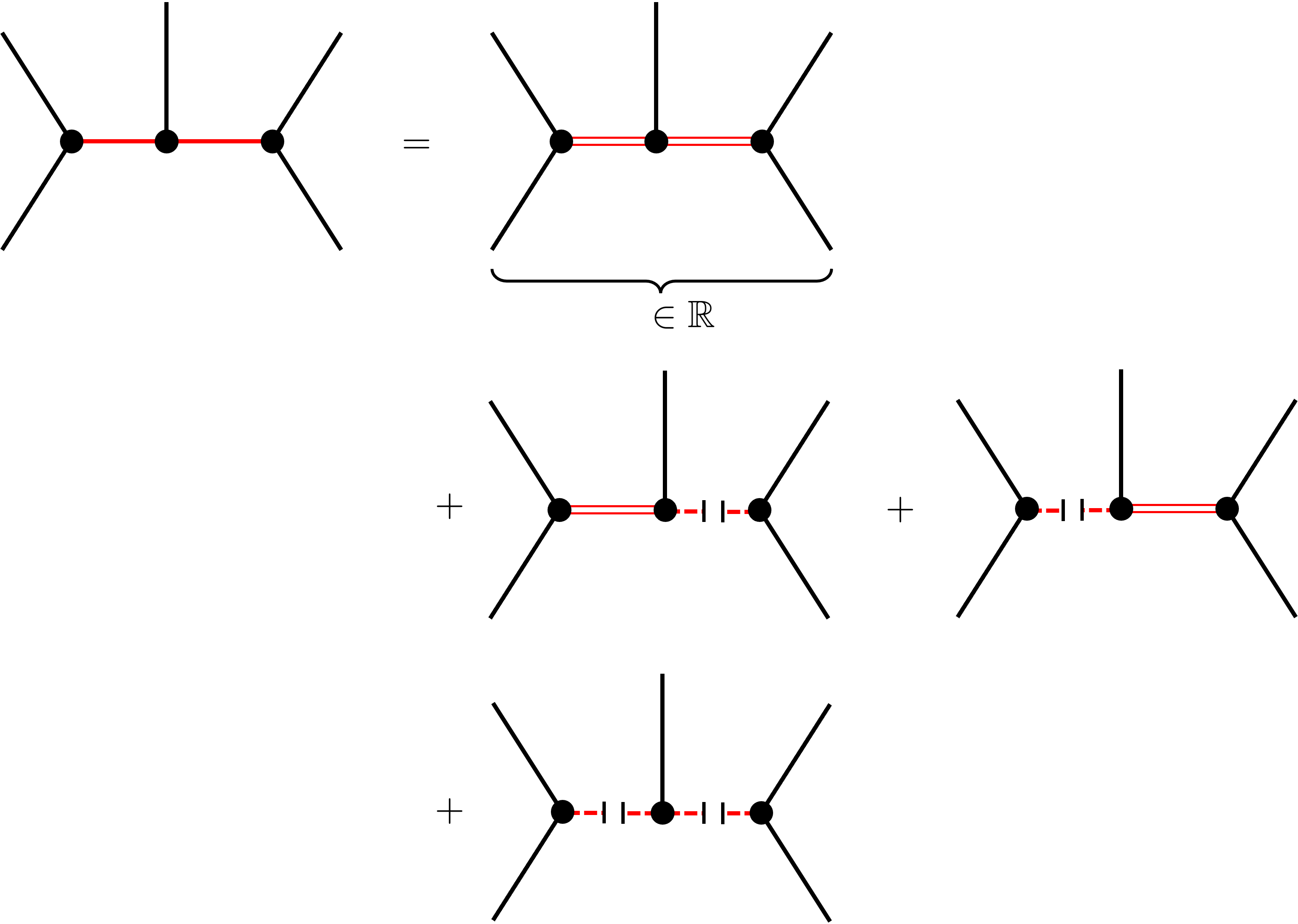}\\
	\caption{A five-point illustration of the $k_T$-reality in the in-in/Schwinger-Keldysh diagrammatics. The all-black vertices indicate the diagram is a fully time-ordered diagram $\langle\varphi^5\rangle_+$. We then expand the internal Schwinger-Keldysh propagators (solid lines) into the connected (double lines) and factorized parts (dashed lines), and use the reality of the connected propagator to conclude the reality of the maximally-connected first term and the total-energy singularities therein.}
	\label{SK5ptIllustration}
\end{figure}

We start by noticing that the total-energy singularities can only reside in \textit{monochromatic} diagrams where the vertices are either all-black or all-white. This is simply a consequence of the factorised nature of the Wightman functions $\mathcal{G}_{\pm\mp}(\eta_1,\eta_2,k)$, i.e. any polychromatic diagram is necessarily disconnected in time at the internal line with opposite colours. In the wavefunction formalism language, they correspond to contributions from the factorised third term in the bulk-bulk propagator \eqref{GdefIntro} together with sewing disconnected $\rho_n$ in \eqref{rhoToCorrelators}. Since the all-white diagram is just the complex conjugation plus momentum inversion of the all-black diagram,
\begin{align}
	\left\langle \varphi(\mathbf{k}_1)\cdots\varphi(\mathbf{k}_n)\right\rangle'_- =\left\langle \varphi(-\mathbf{k}_1)\cdots\varphi(-\mathbf{k}_n)\right\rangle_+^{\prime\,*}~,\label{blackAndwhiteConjRule}
\end{align}
we will focus on the all-black diagram without loss of generality,
\begin{align}
	\left\langle \varphi(\mathbf{k}_1)\cdots\varphi(\mathbf{k}_n)\right\rangle'_+=\int_{-\infty(1-i\epsilon)}^{0}\left[\,\prod_{v=0}^V  d\eta_v \,(+i)\lambda_v\, D_{v}\right] \left[\,\prod_{e=1}^n \mathcal{K}_{e,+}\right] \left[\,\prod_{e'=1}^I \mathcal{G}_{e',++}\right]~,
\end{align}
where $D_v$ is given by \eqref{LvCCM} for the CCM scenario. As with the derivation for the cosmological wavefunction $\psi_n$, we perform a Wick rotation $\eta=i\chi$,
\begin{align}
	\left\langle \varphi(\mathbf{k}_1)\cdots\varphi(\mathbf{k}_n)\right\rangle'_+=(-1)^V\int_0^{\infty}\left[\,\prod_{v=0}^V  d\chi_v \,\lambda_v\, \tilde{D}_{v}\right] \left[\,\prod_{e=1}^n \tilde{\mathcal{K}}_{e,+}\right] \left[\,\prod_{e'=1}^I \tilde{\mathcal{G}}_{e',++}\right]~.
\end{align}
The reality of the vertices and external propagators becomes automatic assuming scale invariance and massless external fields,
\begin{align}
	\tilde{D}_{v}^*=\tilde{D}_{v}~,\quad \tilde{\mathcal{K}}_{e,+}^*=\tilde{\mathcal{K}}_{e,+}~.
\end{align}
For the internal Feynman propagators, reality can be achieved for the connected part via adding and subtracting a homogeneous solution to the equations of motion,
\begin{align}
	\tilde{\mathcal{G}}_{e',++}=\tilde{\mathcal{C}}_{e',++}+\tilde{\mathcal{F}}_{e',++}~,\quad \left(\tilde{\mathcal{C}}_{e',++}\right)^*=\tilde{\mathcal{C}}_{e',++}~,
\end{align}
where
\begin{align}
	\tilde{\mathcal{C}}_{e',++}&\equiv \tilde{\mathcal{G}}_{e',++}+\Delta \tilde{G}~,\\
	\tilde{\mathcal{F}}_{e',++}&\equiv -\Delta \tilde{G}~.
\end{align}
The solution of $\Delta \tilde{G}$ is identical to that in the wavefunction approach (see Appendix \ref{GeneralDeltaGhAppendix}). Thus after expanding the internal propagators of the all-black diagram, we deduce that the maximally-connected contribution
\begin{align}
	\left\langle \varphi(\mathbf{k}_1)\cdots\varphi(\mathbf{k}_n)\right\rangle_+^{\prime C}=(-1)^V\int_0^{\infty}\left[\,\prod_{v=0}^V  d\chi_v \,\lambda_v\, \tilde{D}_{v}\right] \left[\,\prod_{e=1}^n \tilde{\mathcal{K}}_{e,+}\right] \left[\,\prod_{e'=1}^I \tilde{\mathcal{C}}_{e',++}\right]~,
\end{align}
is purely real (see the diagrammatic illustration in Figure \ref{SK5ptIllustration}). Consequently, all the total-energy poles inside the full tree-level correlator must also be real,
\begin{align}
	\Im \left\langle \varphi(\mathbf{k}_1)\cdots\varphi(\mathbf{k}_n)\right\rangle_+^{\prime C}=\Im \underset{k_T\to 0}\Res \left[k_T^m \,\left\langle \varphi(\mathbf{k}_1)\cdots\varphi(\mathbf{k}_n)\right\rangle^{\prime }\right]=0~,~ m,n\in\mathbb{N}~.\label{kTRealityFromSK}
\end{align}
Other total-energy singularities are also real when understood as analytically continued from the positive-$k_T$ direction, see the discussions in Section \ref{HeavyFieldkTRealitySubSect}. The factorisation of parity-odd correlators then directly follows from the $k_T$-reality,
\begin{align}
	\nonumber\left\langle\varphi(\mathbf{k}_{1}) \cdots \varphi(\mathbf{k}_{n})\right\rangle^{\prime\mathrm{PO}}&=\frac{1}{2}\left[\left\langle\varphi(\mathbf{k}_{1}) \cdots \varphi(\mathbf{k}_{n})\right\rangle'-\left\langle\varphi(-\mathbf{k}_{1}) \cdots \varphi(-\mathbf{k}_{n})\right\rangle'\right]\\
	\nonumber&=\frac{1}{2}\left[\left\langle\varphi(\mathbf{k}_{1}) \cdots \varphi(\mathbf{k}_{n})\right\rangle_+^{\prime C}+\left\langle\varphi(\mathbf{k}_{1}) \cdots \varphi(\mathbf{k}_{n})\right\rangle_-^{\prime C}\right.\\
	\nonumber&\left.\qquad-\left\langle\varphi(-\mathbf{k}_{1}) \cdots \varphi(-\mathbf{k}_{n})\right\rangle_+^{\prime C}-\left\langle\varphi(-\mathbf{k}_{1}) \cdots \varphi(-\mathbf{k}_{n})\right\rangle_-^{\prime C}\right]+\text{factorised}\\
	&=0+\text{factorised}~,
\end{align}
where we have applied \eqref{blackAndwhiteConjRule} and \eqref{kTRealityFromSK}. Finally, we comment that in the in-in/Schwinger-Keldysh formalism, there is no explicit reference to the boundary time $\eta_0$ except for the external propagator $\mathcal{K}_\pm$ which can be trivialised by sending $\eta_0\to 0$ first, and consequently the parity-odd correlators factorise in the same fashion for light fields and heavy fields. This generalises the proof of \cite{Cabass:2022rhr}, which applied to massless and conformally-coupled mode functions, allowing for general massive mode functions for the internal lines.

\section{Reality from Hermitian analyticity} \label{CuttingComparison}
In \cite{Cabass:2022rhr}, the Cosmological Optical Theorem (COT) of \cite{COT} was used to deduce that contact diagrams of massless scalars arising from IR-finite interactions are purely real, which in turn implies that such diagrams do not contribute to parity-odd trispectra. This result itself suggests that such a trispectrum is a very nice probe of exotic inflationary physics. The same result was also derived in \cite{Liu:2019fag} using Wick rotations. It is therefore tempting to wonder if the more general results we have derived in this paper, namely the inclusion of exchange processes and massive fields, can also be understood using the COT (or more generally without invoking Wick rotations). In other words, can we deduce from the COT that the imaginary part of wavefunction coefficients is factorised? Let us first review the argument for contact diagrams. The COT states that
\begin{align} \label{ContactCOT}
	\psi_{n}(\{ k \}, \{ \bfk \})+\psi^{\star}_{n}(\{ - k \}, \{ -\bfk \}) = 0~,
\end{align}
where in the second term we flip both the energies and the momenta. Note that the COT relies on our ability to analytically continue the momenta away from the physical region. The COT follows from having real coupling constants (by unitarity) and \textit{Hermitian analyticity} of the massless bulk-boundary propagators, $K(\eta, k) = K^{\star}(\eta, -k)$ c.f. \eqref{MasslessBulktoBoundary}, and the spatial momenta which enter as $i \bfk$. We refer the reader to \cite{COT,Cespedes:2020xqq,Melville:2021lst,Goodhew:2021oqg} for full details. If we have exact scale invariance then $\psi_{n} \sim k^3$, where the cubic scaling is there to cancel the scaling of the momentum-conserving delta function in three spatial dimensions, and therefore \eqref{ContactCOT} becomes
\begin{align}
	\psi_{n}(\{ k \}, \{ \bfk \}) - \psi^{\star}_{n}(\{ k \}, \{ \bfk \}) = 0 \qquad \implies \qquad \text{Im} ~ \psi_{n} = 0~.
\end{align}
Since it is the imaginary part of the wavefunction coefficient that contributes to the parity-odd correlator, c.f. \eqref{WavefunctionRho}, this tells us that contact diagrams do not contribute. Here we have used exact scale invariance of the wavefunction coefficients which of course requires scale invariance of the bulk interactions, but also IR-convergence of the time integrals otherwise scale invariance is broken by the IR cut-off $\eta_0$. In that case the wavefunction can indeed have an imaginary part \cite{Cabass:2022rhr}.

Now consider a four-point exchange diagram. For an $s$-channel diagram the COT reads \cite{COT}
\begin{align} \label{COTExchange}
	\psi_{4,s}(\{k\}, s, \{ \bfk \}) + \psi^{\star}_{4,s}( \{- k\}, s,  \{ -\bfk \}) = \text{factorised}~,
\end{align}
where the factorised RHS depends on the three-point sub-diagrams that contribute to this four-point coefficient. In addition to unitarity and Hermitian analyticity of the bulk-boundary propagators and spatial momenta, this expression holds since the real part of the bulk-bulk propagator is factorised \cite{Goodhew:2021oqg} (we remind the reader that in this paper our bulk-bulk propagator differs by a factor of $i$ from that of \cite{Goodhew:2021oqg} which is why the real part rather than the imaginary part is factorised). We can see this explicitly. Indeed, 
\begin{align}
	G(\eta_1, \eta_2, k) = \sigma(\eta_1,k)\sigma^{\star}(\eta_2,k) \theta(\eta_1 - \eta_2) + \sigma(\eta_2,k)\sigma^{\star}(\eta_1,k) \theta(\eta_2 - \eta_1) + \text{factorised}~,
\end{align}
and therefore 
\begin{align}
	G(\eta_1, \eta_2, k)+G^{\star}(\eta_1, \eta_2, k)  & = [\sigma(\eta_1,k)\sigma^{\star}(\eta_2,k)+\sigma^{\star}(\eta_1,k)\sigma(\eta_2,k)][\theta(\eta_1 - \eta_2) +  \theta(\eta_2 - \eta_1)] + \text{factorised} \nonumber \\
	& = \text{factorised}~.
\end{align}
This straightforwardly generalises to spinning fields. The LHS of the COT is therefore picking out the factorised part of the bulk-bulk propagator. The fact that the RHS is factorised suggests that such a relation could be used to derive similar results to what we have found in this paper, namely that the imaginary part of wavefunction coefficients is factorised (under the assumptions of scale invariance and IR-convergence). We would then naturally want to use scale invariance to pick out the imaginary part like we did for contact diagrams, however this leads to
\begin{align} \label{COTExchange}
	\psi_{4,s}(\{k\}, s, \{ \bfk \}) - \psi^{\star}_{4,s}( \{k\}, - s,  \{ \bfk \}) = \text{factorised}~,
\end{align}
since under a scale transformation $s$ is rescaled too. In general, wavefunction coefficients can contain both odd and even terms in $s$, since the bulk-bulk propagator does not enjoy any symmetry property under $s \rightarrow - s$, so we cannot conclude that the imaginary part is factorised from this expression alone.

However, this discussion suggests that instead we need to use a property of the bulk-bulk propagator that requires us to flip the sign of $s$. This is precisely Hermitian analyticity of the bulk-bulk propagator which for massive scalars and spinning fields within the CCM set-up was discussed in detail in \cite{Goodhew:2021oqg}. Let us explain how we can use this property to offer another perspective on the results we have found in this paper. As one might expect, the story for the CCM and CC scenarios are slightly different, and in each case light and heavy fields are slightly different. Let us therefore take each possibility in turn.

\paragraph{Cosmological condensed matter physics} For concreteness we will restrict ourselves to $\tilde{\kappa} = 0$ since Hermitian analyticity has been well-established in this case. To match the notation of \cite{Goodhew:2021oqg} we define $\sigma^{-}_{h}(\eta, k) \equiv - i  \sigma_{h}(\eta, k)$ and $\sigma^{+}_{h}(\eta, k) \equiv + i \sigma^{\star}_{h}(\eta, k)$. We then have\footnote{These two solutions are complex conjugate to each other for both light and heavy fields.}
\begin{align}
	\sigma^{-}_{h}(\eta, k) &= - i  \frac{H \sqrt{\pi}}{2} (-\eta)^{3/2}e^{i \pi (\nu+1/2)/2}H_{\nu}^{(1)}(- c_{h,S} k \eta)~, \\
	\sigma^{+}_{h}(\eta, k) &=  + i \frac{H \sqrt{\pi}}{2} (-\eta)^{3/2}e^{-i \pi (\nu+1/2)/2}H_{\nu}^{(2)}(- c_{h,S} k \eta)~, \label{SigmaMinusHA}
\end{align}
and the helical bulk-bulk propagator can be written as
\begin{align}
	G^{(h)}_{\sigma}(\eta_1, \eta_2, k) = \sigma_{h}^{+}(\eta_1, k)\sigma_{h}^{+}(\eta_2, k)\left(\frac{\sigma_{h}^{-}(\eta_1, k)}{\sigma_{h}^{+}(\eta_1, k)} -\frac{\sigma_{h}^{-}(\eta_0, k)}{\sigma_{h}^{+}(\eta_0, k)}  \right) \theta(\eta_1 - \eta_2) + (\eta_1 \leftrightarrow \eta_2)~. 
\end{align}
Since the new mode functions only differ from the old ones by a phase, the bulk-bulk propagator is unchanged. Now as shown in \cite{Goodhew:2021oqg}, the mode functions satisfy the properties:
\begin{align}
	\left[ \sigma_{h}^{+}(\eta, - k^{\star}) \right]^{\star}  &= i \sigma_{h}^{+}(\eta, k)~, \label{PhiPlusHA} \\
	\left[ \sigma_{h}^{-}(\eta, - k^{\star}) \right]^{\star}  &= i \sigma_{h}^{-}(\eta, k) + 2 \cos( \pi \nu)\sigma_{h}^{+}(\eta, k) \label{PhiMinusHA}~,
\end{align}
from which one can show that the helical bulk-bulk propagator is anti-Hermitian analytic:
\begin{keyeqn}
	\begin{align}
		\left[ G_{\sigma}^{(h)}(\eta_1, \eta_2, -k^{\star}) \right]^{\star} = - G_{\sigma}^{(h)}(\eta_1, \eta_2, k)~.
	\end{align}
\end{keyeqn}
Note that this property holds for both light and heavy fields. Here $-k^{\star}$ indicates that we must include a small negative imaginary contribution, in additional to the negative real part, such that we do not cross any branch cuts. We can now attempt to put this property into good use to conclude some reality properties of wavefunction coefficients. Consider the general diagram structure of \eqref{PsiNBeforeWick}:
\begin{align}
	\psi_{n}(\{ k \}, \{p  \}, \{\bfk \}) =\int \left[\,\prod_{v=1}^V  d\eta_v \,i\lambda_v\, D_{v}\right]\left[\,\prod_{e=1}^n K_e(k_{e})\right] \left[\,\prod_{e'=1}^I G_{e'}(p_{e'})\right]~,
\end{align}
where we have only indicated the energy dependence of the propagators. We can now use the various Hermitian analyticity properties to write
\begin{align}
	\psi^{\star}_{n}(\{- k \}, \{ - p^{\star}  \}, \{ - \bfk \}) = - \int \left[\,\prod_{v=1}^V  d\eta_v \,i\lambda_v\, D_{v}\right]\left[\,\prod_{e=1}^n K_e(k_{e})\right] \left[\,\prod_{e'=1}^I G_{e'}(p_{e'})\right]~,\label{psiAfterHA}
\end{align}
where we have used the fact that at tree-level we have $V = I+1$. For wavefunction coefficients of massless scalars that adhere to exact scale invariance, i.e. have no $\eta_0$ dependence and therefore scale as $\psi_n \sim k^3$, we can then write
\begin{align}
	\psi^{\star}_{n}(\{k \}, \{p \}, \{ \bfk \}) =  \int \left[\,\prod_{v=1}^V  d\eta_v \,i\lambda_v\, D_{v}\right]\left[\,\prod_{e=1}^n K_e(k_{e})\right] \left[\,\prod_{e'=1}^I G_{e'}(p_{e'})\right] = \psi_{n}(\{k \}, \{p \}, \{ \bfk \})~,
\end{align}
which establishes the reality of $\psi_n$. Here it was crucial that there was no $\eta_0$ dependence. If there is an $\eta_0$ dependence then flipping the signs of all energies and momenta does not simply yield an overall minus sign since $\eta_0$ itself carries a conformal weight. As we have discussed a number times in this paper, the bulk-bulk propagator is independent of $\eta_0$ only for light fields, whereas for heavy fields it does indeed depend on $\eta_0$. This proof therefore applies for light fields only, and offers a complementary proof of the result we derived in Section \ref{RandFSection} using Wick rotations. 

For heavy fields this argument does not hold (and indeed we wouldn't expect it to hold since we have already seen that for heavy fields $\psi_n$ is not real), but the discussion and the C-F decomposition we made in Section \ref{PropagatorPropertySection} suggests a clear way forward. Indeed, consider the connected bulk-bulk propagator c.f. \eqref{ConnectedHelicalProp},
\begin{align}
	C_{\sigma}^{(h)}(\eta_1, \eta_2, k) = \sigma_{h}^{+}(\eta_1, k)\sigma_{h}^{+}(\eta_2, k)\left(\frac{\sigma_{h}^{-}(\eta_1, k)}{\sigma_{h}^{+}(\eta_1, k)} -i \cos(\pi \nu)  \right) \theta(\eta_1 - \eta_2) + (\eta_1 \leftrightarrow \eta_2)~,
\end{align}
where we have taken the minimal solution of $\mathcal{A}_{h}$ which we derived in Appendix \ref{GeneralDeltaGhAppendix}. The relative minus sign between the two terms in the brackets comes from the fact we have written this propagator in terms of $\sigma^-$ and $\sigma^+$ rather than $\sigma$ and $\sigma^{\star}$. Here we have written the solution for $\mathcal{A}_{h}$ that is valid for both light and heavy fields. We can then use the Hermitian analytic properties of the mode functions to conclude that this connected bulk-bulk propagator is anti-Hermitian analytic:
\begin{keyeqn}
	\begin{align}
		\left[ C_{\sigma}^{(h)}(\eta_1, \eta_2, -k^{\star}) \right]^{\star} = - C_{\sigma}^{(h)}(\eta_1, \eta_2, k)~,
	\end{align}
\end{keyeqn}
which holds for both light and heavy fields. In Appendix \ref{GeneralDeltaGhAppendix} we saw that we could add any real term to $\mathcal{A}_{h}$ while maintaining the reality of the connected bulk-bulk propagator after Wick rotation. Here we can also add any real term to $\mathcal{A}_{h}$ and still realise the anti-Hermitian analyticity given \eqref{PhiPlusHA}. We can now run the same argument as above but with the full bulk-bulk propagator replaced by the connected one, and with the crucial difference that the connected propagator is independent of $\eta_0$ such that we can use exact scale invariance,\footnote{Note that the connected propagator has the same conformal weight as the full bulk-bulk propagator so it remains the case that $\psi_{n}^{C} \sim k^3$ by scale invariace.} to conclude that $\psi_{n}^{C}$ is real which complements the proof we derived in Section \ref{RandFSection} using Wick rotations. 

\paragraph{Cosmological collider physics} As expected, things are a little more involved for the CC scenario since the mode functions are more complicated, however here we prove anti-Hermitian analyticity of the full bulk-bulk propagator, and the connected part, for each helicity mode. To the best our of knowledge this has not been shown before.

As always we start with the $n = |h|$ modes with mode functions given in \eqref{ModeFunctionCC} which are the same as in the CCM scenario up to some real factors, and integer powers of $k$. We can therefore immediately conclude that the bulk-bulk propagator is anti-Hermitian analytic:
\begin{keyeqn}
	\begin{align}
		\left[ G^{h}_{|h|,S}(\eta_1, \eta_2, -k^{\star}) \right]^{\star} = - G^{h}_{|h|,S}(\eta_1, \eta_2, k)~.
	\end{align}
\end{keyeqn}
As we discussed at length in Section \ref{PropagatorPropertySection}, the bulk-bulk propagator for the other modes can be written in terms of $\Phi_{|h|,S}^{h}$ by iteratively using the relation \eqref{ModeFunctionCCDifferentialOperator}:
\begin{align}
	G_{n,S}^{h}(\eta_1,\eta_2,k) = &\hat{\mathcal{D}}_{h,n}^{\star}(i \eta_1,k)[\Phi_{|h|,S}^{h \star}(\eta_1,k)] \hat{\mathcal{D}}_{h,n}^{\star}(i \eta_2,k)[\Phi_{|h|,S}^{h \star}(\eta_2,k)] \nonumber \\ \times  &\left( \frac{\hat{\mathcal{D}}_{h,n}(i \eta_1,k)[\Phi_{|h|,S}^{h}(\eta_1,k)]}{\hat{\mathcal{D}}_{h,n}^{\star}(i \eta_1,k)[\Phi_{|h|,S}^{h \star}(\eta_1,k)]} -\frac{\hat{\mathcal{D}}_{h,n}( i \eta_0,k)[\Phi_{|h|,S}^{h}(\eta_0,k)]}{\hat{\mathcal{D}}_{h,n}^{\star}(i \eta_0,k)[\Phi_{|h|,S}^{h \star}(\eta_0,k)]} \right) \theta(\eta_1 - \eta_2) + (\eta_1 \leftrightarrow \eta_2 )~.
\end{align}
The main observation that allows us to make progress is that the differential operators $\hat{\mathcal{D}}_{h,n}(i \eta, k)$ are Hermitian analytic. This follows straightforwardly from the fact that the differential operator in \eqref{ModeFunctionCCDifferentialOperator} is Hermitian analytic (and so acting with it iteratively also yields something Hermitian analytic). The same Hermitian analyticity relations we used above for CCM can then be used to infer that this bulk-bulk propagator is anti-Hermitian analytic:
\begin{keyeqn}
	\begin{align}
		\left[ G^{h}_{n,S}(\eta_1, \eta_2, -k^{\star}) \right]^{\star} = - G^{h}_{n,S}(\eta_1, \eta_2, k)~.
	\end{align}
\end{keyeqn}
Note that in arriving at this conclusion we also had to make use of the fact that $\hat{\mathcal{D}}_{h,n}(i \eta, k)$ are either purely real, for even $n-|h|$, or purely imaginary, for odd $n-|h|$, as we discussed in Section \ref{PropagatorPropertySection}. This leads to
\begin{align}
	\frac{\hat{\mathcal{D}}_{h,n}(i \eta_1,k)[\Phi_{|h|,S}^{h \star}(\eta_1,k)]}{\hat{\mathcal{D}}_{h,n}^{\star}(i \eta_1,k)[\Phi_{|h|,S}^{h \star}(\eta_1,k)]} = (-1)^{n-|h|}~.
\end{align}
This is used to cancel the $\cos(\pi \nu)$ pieces that come from \eqref{PhiMinusHA}. We can now use a similar argument as above for the CCM scenario, using the general wavefunction coefficients we discussed in Section \ref{RandFSection} for the CC scenario, to conclude that the wavefunction coefficients of massless scalars exchanging such massive spinning fields are purely real, as long as the $\eta_0$ dependence cancels out. As with the CCM case, this is only the case for light fields c.f. \eqref{Eta0CancelCC}. This complements the proof we detailed in Section \ref{RandFSection} using Wick rotations. 

The situation for heavy fields now follows in the same way as for the CCM scenario: the connected bulk-bulk propagator for all modes is Hermitian analytic from which we can easily deduce that the connected part of wavefunction coefficients is purely real. In this CC scenario the connected bulk-bulk propagator is given by \eqref{ConnectedBBCC} with reality of this propagator after Wick rotation fixing 
\begin{align}
	\mathcal{A}_{h,n}(\mu) = (-1)^{n- |h|} i \cos(\pi \nu_{S})~,
\end{align}
which we have again written in a way that is valid for both light and heavy fields. We can then again use the Hermitian analyticity of $\hat{\mathcal{D}}_{h,n}(i \eta,k)$, and the fact that these differential operators are purely real for even $n-|h|$ and purely imaginary for odd $n-|h|$, to conclude that this connected bulk-bulk propagator is anti-Hermitian analytic:
\begin{keyeqn}
	\begin{align}
		\left[ C^{h}_{n,S}(\eta_1, \eta_2, -k^{\star}) \right]^{\star} = - C^{h}_{n,S}(\eta_1, \eta_2, k)~.
	\end{align}
\end{keyeqn}

\paragraph{A caution on locality and the anomaly of Hermitian analyticity}
In all of the above proofs, locality of the general form of the interactions, meaning that the vertex operator $D_v$ is composed of derivatives but not inverse derivatives, is an implicit assumption. In momentum space, locality tells us that $D_v\sim (i k)^{n}$ is a polynomial in the energy variable $k=|\mathbf{k}|$ with $n\in \mathbb{N}$. Such a local interaction vertex trivially satisfies Hermitian analyticity, i.e. $D_v^*(-k)=D_v(k)$, leading to \eqref{psiAfterHA} and the reality theorems. However, if the assumption of locality is dropped, and $D_v(k)$ is allowed to have a non-polynomial dependence on $k$, there can be an intriguing ``anomaly'' of Hermitian analyticity after performing the time integrals, thereby invalidating the conclusions about reality of wavefunction coefficients.  

To demonstrate the essential idea, consider the following toy model of a scale-invariant non-local interaction:
\begin{align}
	\mathcal{L}=\lambda\phi'^2\frac{1}{1-\bm{\nabla}^2/(a H)^2}\phi'^2~,
\end{align}
where $\bm{\nabla}^2=\delta_{ij}\partial_i\partial_j$ is the three-dimensional Laplacian in flat space. This non-local theory can be understood as describing massless $\phi$ particles interacting via a Yukawa-like force. It can be derived from a UV theory of $\phi$ and a massive scalar $\sigma$, by integrating out $\sigma$ and taking the leading order contribution from the time-derivative expansion \cite{Jazayeri:2022kjy}. The resulting $s$-channel contact wavefunction is\footnote{By ``$s$-channel" here we mean the contribution to the full four-point wavefunction coefficient that has the same symmetries as an $s$-channel exchange diagram.}
\begin{align}
	\psi_4(\{k\},s)=i\lambda (k_1 k_2 k_3 k_4)^2\int_{-\infty}^0 d\eta \frac{\eta^4}{1+s^2\eta^2} e^{ik_T\eta}~.
\end{align}
Taking the Hermitian-analytic conjugate, we obtain
\begin{align}
	\psi_4^*(\{-k^*\},-s)=-i\lambda (k_1 k_2 k_3 k_4)^2\int_{-\infty}^0 d\eta \frac{\eta^4}{1+s^2\eta^2} e^{ik_T\eta}~,
\end{align}
indicating that $\psi_4$ is anti-Hermitian analytic,
\begin{align}
	\psi_4(\{k\},s)+\psi_4^*(\{-k^*\},-s)=0~,\quad\text{(before time integration)}~,\label{naiveAHA}
\end{align}
as expected from \eqref{psiAfterHA}. We might then be tempted to use scale invariance and conclude that such a wavefunction coefficient is purely imaginary. However, this is not the case. Indeed, we can carry on and compute the time integral to obtain
\begin{align}
	\psi_4(\{k\},s)=i\lambda \frac{(k_1 k_2 k_3 k_4)^2}{s^5}\left[i \left(\text{Ci}\frac{i k_T}{s} \sinh\frac{k_T}{s}-\text{Shi}\frac{k_T}{s} \cosh\frac{k_T}{s}+\frac{s}{k_T}+\frac{2s^3}{k_T^3}\right)+\frac{\pi}{2}  \cosh \frac{k_T}{s}\right]~,\label{anomalousIntResult}
\end{align}
which, under Hermitian-analytic conjugation, becomes
\begin{align}
	\psi_4^*(\{-k^*\},-s)=i\lambda \frac{(k_1 k_2 k_3 k_4)^2}{s^5}\left[-i \left(\text{Ci}\frac{-i k_T^*}{s} \sinh\frac{k_T}{s}-\text{Shi}\frac{k_T}{s} \cosh\frac{k_T}{s}+\frac{s}{k_T}+\frac{2s^3}{k_T^3}\right)+\frac{\pi}{2}  \cosh \frac{k_T}{s}\right]~,\label{anomalousIntResultHAC}
\end{align}
where
\begin{align}
	\text{Ci}(x)=-\int_{x}^{\infty}\frac{\cos t}{t} dt~,\quad \text{Shi}(x)=\int_{0}^{x}\frac{\sinh t}{t} dt~,
\end{align}
are the cosine integral and hyperbolic sine integral functions, respectively. Adding \eqref{anomalousIntResult} and \eqref{anomalousIntResultHAC} together, we see that in contradiction to our naive expectation \eqref{naiveAHA}, (anti-)Hermitian analyticity is violated,
\begin{align}
	\psi_4(\{k\},s)+\psi_4^*(\{-k^*\},-s)=i \pi \lambda \frac{(k_1 k_2 k_3 k_4)^2}{s^5} e^{-k_T/s}\neq 0,\quad\text{(after time integration)}~.\label{AHAViolation}
\end{align}
Consequently, the wavefunction coefficient $\psi_4$ is complex in general (rather than being purely imaginary).

\begin{figure}[ht]
	\centering
	\includegraphics[width=0.8\textwidth]{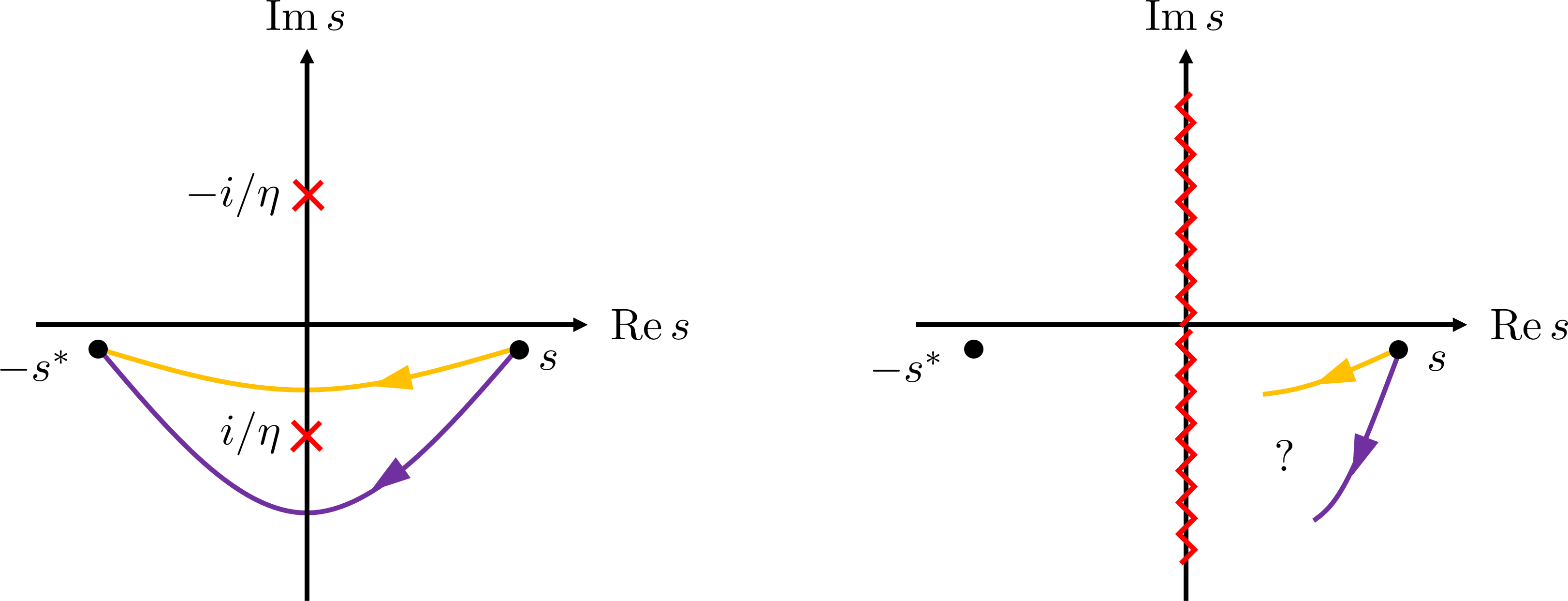}\\
	\caption{Left panel: At any fixed time $\eta$, one can choose to continue from $s$ to $-s^*$ by either passing \textcolor{orange2}{above} or \textcolor{purple2}{below} the singularity at $i/\eta$. Right panel: After finishing the time integral, the singularities merge into a branch cut that goes all the way from zero to infinity, preventing a uniform definition of the analytic continuation.}
	\label{HAAnomaly}
\end{figure}

Such an anomaly of Hermitian analyticity stems from the non-analytic behaviour of the vertex function with respect to the energy variable $s$: at any finite time $\eta$, there are poles at $s=\pm i/\eta$ which affect the definition of the Hermitian analytic image. One can choose to continue path-wise in the $s$-plane from either side of the poles, but since the integration time $\eta$ ranges from the origin all the way to infinity, there is no uniform way to perform the continuation throughout time (see Figure \ref{HAAnomaly}). The Hermitian analytic properties of the integrand therefore do not imply some simple relations for the final integrated result when there is some element of non-locality in the interactions.

Things are somewhat clearer from the Wick rotation perspective which we have primarily used in this paper: the pole at $\eta_c=i/s$ on the complex $\eta$-plane prevents the deformation of the integration contour into the Wick-rotated one. Instead, we must include half of the residue at $\eta_c$ to account for the half-circle touring around the pole, which gives exactly \eqref{AHAViolation}. In fact, this is precisely how parity violation is generated in the non-local single-field EFT model \cite{Jazayeri:2023kji}.

\section{Beyond scale invariance: reality in other FLRW spacetimes} \label{MoreRealities}

The discussion in Appendix \ref{CuttingComparison} suggests a strong connection between the reality properties we have derived in this work, namely that of wavefunction propagators after Wick rotation, and their Hermitian analyticity properties that have been discussed in the literature. Indeed, with the exact scale invariance of de Sitter space, these two properties are equivalent at the level of equations of motion. To see this equivalence more clearly, recall the equation of motion for a free bosonic field without the chemical potential term:
\begin{align}
	\hat{\mathcal{E}}(\eta,k)\sigma_{h}(\eta, k) = 0~,\quad\hat{\mathcal{E}}(\eta,k)\equiv\eta^2\frac{\partial^2}{\partial \eta^2} - 2\eta \frac{\partial}{\partial \eta} + c_{h,S}^2 k^2\eta^2 + \frac{m^2}{H^2} ~.
\end{align}
Our reality theorems rely on the fact that after Wick rotation, $\eta=i\chi$, the equation of motion remains real:
\begin{align}
	\left[\hat{\mathcal{E}}(i\chi,k)\right]^*=\hat{\mathcal{E}}(i\chi,k)~,\quad \text{or}\quad\hat{\mathcal{E}}^*(-i\chi,k)=\hat{\mathcal{E}}(i\chi,k)~.\label{RealityRequirement}
\end{align}
Hermitian analyticity, on the other hand, states that sending energies to minus energies, while doing a complex conjugation, is a unit transformation,
\begin{align}
	\left[\hat{\mathcal{E}}(\eta,-k)\right]^*=\hat{\mathcal{E}}(\eta,k)~,\quad \text{or}\quad\hat{\mathcal{E}}^*(\eta,-k)=\hat{\mathcal{E}}(\eta,k)~.\label{HARequirement}
\end{align}
For a scale-invariant free theory in de Sitter space, the equation of motion operator must be a function of the combination
\begin{align}
	\hat{\mathcal{E}}(\eta,k)=f(\eta\partial_\eta, k\eta)~,
\end{align}
which means \eqref{RealityRequirement} and \eqref{HARequirement} are equivalent (at least for a vanishing chemical potential):
\begin{align}
	\hat{\mathcal{E}}^*(-i\chi,k)=f^*(\chi\partial_\chi,-i k\chi)=f^*(\eta\partial_\eta,-k\eta)=\hat{\mathcal{E}}^*(\eta,-k)~.
\end{align}

However, in the absence of scale invariance, reality and Hermitian analyticity are drastically different notions, since they constrain the functional dependence on different variables in $\hat{\mathcal{E}}(\eta,k)$: reality constrains the \textit{time} dependence $(\eta,\cdot)$, whereas Hermitian analyticity constrains the \textit{energy} dependence $(\cdot,k)$. For fields with more complicated dispersion relations, as can appear in general FLRW spacetimes, the time and energy dependence decouples, and it is easy to find examples where one of them is satisfied but not the other. For instance, a scale-dependent mass alters the equation of motion to 
\begin{align}
	\hat{\mathcal{E}}_\alpha(\eta,k)=\eta^2\frac{\partial^2}{\partial \eta^2} - 2\eta \frac{\partial}{\partial \eta}  + c_{h,S}^2 k^2\eta^2 + \left(\frac{m^2}{H^2}+ \alpha k\right),
\end{align}
which satisfies reality but not Hermitian analyticity, while a time-dependent sound speed
\begin{align}
	\hat{\mathcal{E}}_\beta(\eta,k)=\eta^2\frac{\partial^2}{\partial \eta^2} - 2\eta \frac{\partial}{\partial \eta} + \left(c_{h,S}^2+\beta\eta\right) k^2\eta^2 + \frac{m^2}{H^2} ~,
\end{align}
satisfies Hermitian analyticity but not reality. Notice that if one assumes the usual dispersion relation $w^2=c_s^2 k_p^2+m^2$, Hermitian analyticity is valid for most theories in general FLRW spacetimes with a Bunch-Davis vacuum \cite{Goodhew:2021oqg}, whereas reality is more stringent and is only valid for certain spacetimes.

To see how far we can go without assuming scale invariance, consider theories in a power-law FLRW spacetime,
\begin{align}
	ds^2=a^2(\eta)(-d\eta^2+d\mathbf{x}^2)~,\quad a(\eta)=\left(\frac{\eta_*}{\eta}\right)^{p}~, \quad p\geq0~,
\end{align}
where $p=1$ corresponds to the case of inflation. The equation of motion operator reads
\begin{align}
	\hat{\mathcal{E}}(\eta,k)=\frac{1}{a^2(\eta)}\left(\frac{\partial^2}{\partial \eta^2} - \frac{2p}{\eta} \frac{\partial}{\partial \eta} + c_s^2 k^2\right)+ m^2~.
\end{align}
This operator is apparently Hermitian analytic for any $p\in\mathbb{R_+}$, and under some assumptions the corresponding propagators are Hermitian analytic \cite{Goodhew:2021oqg}. How about reality after Wick rotation? Replacing $\eta\to i\chi$, we find
\begin{align}
	\hat{\mathcal{E}}(i\chi,k)=i^{2p}\frac{1}{a^2(\chi)}\left(-\frac{\partial^2}{\partial \chi^2} + \frac{2p}{\chi} \frac{\partial}{\partial \chi} + c_s^2 k^2\right)+ m^2~,
\end{align}
which is real only for $p$ being an integer. Thus the propagator realities for $K_e, G_{e'}$ will continue to hold for $p\in \mathbb{N}$. To further check the $\psi_n$-reality and $k_T$-reality, we need to examine how the vertex $D_v$ transforms under Wick rotation:\footnote{We stress that the power of the scale factor in $D_v$ is fixed by diffeomorphism invariance (gauge redundancy) rather than scale invariance (isometry), and so is the same for any FLRW spacetime.}
\begin{align}
	\nonumber D_v&= a^{4-k_v-l_v}(\eta)\left[\left(\delta_{ij}\right)^{p_v}\left(\epsilon_{ijk}\right)^{q_v}\left(\partial_\eta\right)^{k_v} \left(i \, k_i\right)^{l_v} \right]_{\text{partially contract}}\\
	\nonumber&=i^{(-4+k_v+l_v)p}~ i^{-k_{v}}~ i^{l_v}\times a^{4-k_v-l_v}(\chi)\left[\left(\delta_{ij}\right)^{p_v}\left(\epsilon_{ijk}\right)^{q_v}\left(\partial_\chi\right)^{k_v} \left(k_i\right)^{l_v} \right]_{\text{partially contract}}\\
	&=i^{(k_v+l_v)(p+1)}\times \text{real}~,
\end{align}
which is real for arbitrary couplings (i.e. all $k_v,l_v\in\mathbb{N}$) only if $p$ is an odd integer.\footnote{Or if $p$ is an even integer, the sum $\sum_v(k_v+l_v)$ must be even in a diagram, in which case the reality and factorisation theorems also hold in a diagram-dependent fashion.} Therefore, we conclude with 
\begin{keythrm}
	\begin{corollary}
		In \textit{odd-power-law} FLRW spacetimes with a Bunch-Davies vacuum and IR convergence, $\psi_n$-reality, $k_T$-reality and parity-odd factorisation theorems are still valid even in the absence of scale invariance.
	\end{corollary}	
\end{keythrm}

\bibliographystyle{JHEP}
\bibliography{refsPOExact}

\end{document}